\documentclass[a4paper,11pt]{article}

\pdfoutput=1

\usepackage{amsmath,amssymb,mathtools}
\usepackage{color}
\usepackage{graphicx}
\usepackage{subfigure}
\usepackage{cite}
\usepackage[colorlinks=true,linkcolor=blue, citecolor=red, urlcolor=blue, bookmarks]{hyperref}
\usepackage{multirow,makecell}
\usepackage{textcomp}
\usepackage{wasysym}

\usepackage[text={17.1cm,24.6cm},centering]{geometry} 

\numberwithin{equation}{section}

\def \be {\begin{equation}}
\def \ee {\end{equation}}
\def \ba {\begin{array}}
\def \ea {\end{array}}
\def \bea{\begin{eqnarray}}
\def \eea{\end{eqnarray}}
\def \nn {\nonumber}

\def \a {\alpha}

\def \g {\gamma}

\def \G {\Gamma}
\def \d {\delta}
\def \D {\Delta}
\def \e {\epsilon}
\def \ve {\varepsilon}
\def \m {\mu}

\def \l {\lambda}
\def \L {\Lambda}
\def \s {\sigma}

\def \r {\rho}

\def \th {\theta}

\def \vph {\varphi}

\def \cD {\mathcal D}

\def \cR {\mathcal R}
\def \cS {\mathcal S}
\def \cT {\mathcal T}

\def \cX {\mathcal X}

\def \rC {\mathrm C}

\def \rR {\mathrm R}

\def \mD {\mathcal D}

\def \mR {\mathcal R}
\def \mS {\mathcal S}
\def \mT {\mathcal T}

\def \mX {\mathcal X}
\def \mY {\mathcal Y}

\def \p {\partial}
\def \es {\emptyset}
\def \f {\frac}

\def \lt {\left}
\def \rt {\right}

\def \lra {\leftrightarrow}
\def \sr {\sqrt}
\def \td {\tilde}

\def \inf {\infty}

\def \lag {\langle}
\def \rag {\rangle}

\def \ep {\mathrm{e}}
\def \ii {\mathrm{i}}

\def \tr {\textrm{tr}}

\def \and {{~\textrm{and}~}}

\def \CFT {{\textrm{CFT}}}

\def \NS {{\textrm{NS}}}

\begin{document}

\title{\textbf{Subsystem trace distance in low-lying states of $(1+1)$-dimensional conformal field theories}} 
\author{
Jiaju Zhang$^{1}$, 
Paola Ruggiero$^{1}$, 
and
Pasquale Calabrese$^{1,2}$ 
}

\maketitle
 \vspace{-10mm}
\begin{center}
{\it
$^{1}$SISSA and INFN, Via Bonomea 265, 34136 Trieste, Italy\\\vspace{1mm}
$^{2}$International Centre for Theoretical Physics (ICTP), Strada Costiera 11, 34151 Trieste, Italy
}
\vspace{10mm}
\end{center}

\begin{abstract}
We report on a systematic replica approach to calculate the subsystem trace distance for a quantum field theory.
This method has been recently introduced in [J. Zhang, P. Ruggiero, P. Calabrese, Phys. Rev. Lett. 122, 141602 (2019)], of which this work is a completion.
The trace distance between two reduced density matrices $\rho_A$ and $\sigma_A$ is obtained from the moments $\tr (\rho_A-\sigma_A)^n$
and taking the limit $n\to1$ of the traces of the even powers.
We focus here on the case of a subsystem consisting of a single interval of length $\ell$ embedded in the low lying eigenstates of a one-dimensional critical
system of length $L$, a situation that can be studied exploiting the path integral form of the reduced density matrices of two-dimensional conformal field theories.
The trace distance turns out to be a {\it scale invariant universal function} of $\ell/L$.
Here we complete our previous work by providing detailed derivations of all results and further new formulas
for the distances between several low-lying states in two-dimensional free massless compact boson and fermion theories.
Remarkably, for one special case in the bosonic theory and for another in the fermionic one, we obtain the exact trace distance, as well as the Schatten $n$-distance,
for an interval of arbitrary length,
while in generic case we have a general form for the first term in the expansion in powers of  $\ell/L$.
The analytical predictions in conformal field theories are tested against exact numerical calculations in XX and Ising spin chains, finding perfect agreement.
As a byproduct, new results in two-dimensional CFT are also obtained for other entanglement-related quantities, such as the relative entropy and the fidelity.
\end{abstract}

\baselineskip 18pt
\thispagestyle{empty}
\newpage


\tableofcontents

\section{Introduction}

The characterisation of the entanglement content of extended quantum system has become a crucial theme in modern
physics \cite{Amico:2007ag,calabrese2009entanglement,Laflorencie:2015eck} at the level that a few experimental protocols to measure such entanglement
have been already set up \cite{islam2015measuring,kaufman2016quantum,elben2018renyi,lukin2018probing,brydges2018probing}.
The reason of this very large and diversified interest in the entanglement of many body quantum systems is manyfold.
On the one hand, entanglement became a standard and powerful tool to characterise the phases of matter, especially
in connection with criticality \cite{hlw-94,Vidal:2002rm,Calabrese:2004eu,Calabrese:2009qy,rm-04} and topological order \cite{top-ent,hzs-07,lh-08}.
Furthermore, entanglement is also a key feature to design new numerical algorithms based on tensor network states \cite{tns}.
More generically, characterising subsystems is essential to understand the phenomenon of equilibration and thermalisation of an isolated non-equilibrium
quantum systems \cite{cc-05,cramer2008exact,barthel2008dephasing,deutsch2013microscopic,Gogolin:2016hwy,calabrese2016introduction,Essler:2016ufo,vidmar2016generalized,alba2017entanglement,rigol2008thermalization,Lashkari:2016vgj},
and the entanglement dynamics is also related to the black hole information loss paradox \cite{Hawking:1974sw,Mathur:2009hf,Maldacena:2013xja}
through gauge/gravity duality \cite{Maldacena:1997re,Fitzpatrick:2016ive,Chen:2017yze}.
For this reason, the entanglement entropy in holographic theories and its relation to quantum gravity have also been extensively studied \cite{Ryu:2006bv,Hubeny:2007xt,VanRaamsdonk:2010pw,Lewkowycz:2013nqa,Faulkner:2013ana,Faulkner:2013ica,Dong:2016fnf,Dong:2016hjy,Rangamani:2016dms}.

Nonetheless the information that the entanglement provides about a given subsystem may not be enough for some applications.
Specifically, it can be equally important to develop tools enabling to distinguish between subsystems in different states, i.e. to distinguish reduced density matrices (RDMs).
The problem of measuring the distance between density matrices has been intensively considered in quantum information theory,
where several different measures have been introduced and analysed, see e.g.  \cite{nielsen2010quantum,watrous2018theory} as reviews.
A proper measure of the difference should be a {\it metric} in a mathematical sense, meaning it should be nonnegative, symmetric in its inputs, equal to zero if and
only if its two inputs are exactly the same, and should obey the triangular inequality.
Given two normalised density matrices $\r$ and $\s$ (i.e. with $\tr\r=\tr\s=1$), an important family of distances is given by
\be \label{DnDef}
D_n(\r,\s) = \f{1}{2^{1/n}} \| \r - \s \|_n,
\ee
which depends on the real parameter $n \geq 1$.
These distances are known as (Schatten) $n$-distances, and are defined in terms of
the (Schatten) $n$-norm (of a general matrix $\L$) \cite{watrous2018theory}
\be
\|\L\|_n = \Big( \sum_i \l_i^n \Big)^{1/n},
\ee
with $\l_i$ being the nonvanishing singular values of $\L$, i.e.\ the nonvanishing eigenvalues of $\sr{\L^\dag \L}$. When $\L$ is Hermitian, $\l_i$ are just the absolute values of the nonvanishing eigenvalues of $\L$.
The  normalisation in \eqref{DnDef} is fixed so that $0 \leq D_n(\r,\s) \leq 1$.
As long as we are dealing with finite dimensional Hilbert spaces, all distances (including $D_n (\r, \s)$ in \eqref{DnDef}) are equivalent, in the sense that they bound each other
\be
c_{nm} \, D_n (\r, \s) \leq D_m (\r, \s) \leq c_{mn}\, D_n (\r, \s),
\ee
for some constants $c_{nm}$. However this ceases to be the case for infinite dimensional Hilbert spaces, because the constants $c_{nm}$ depend on the Hilbert space
dimension.
For this same reason, it is not obvious how to compare distances between RDMs associated to subsystems of different size, which is one of our main goals in this paper.
Given this state of affairs, it is natural to wonder whether one distance is on a special foot compared to the others.
In this respect, it is well known that the \emph{trace distance}
\be \label{TDDef}
D(\r,\s) = \f12 \| \r - \s \|_1,
\ee
(i.e. \eqref{DnDef} for $n=1$) has several properties that made it more effective than the others \cite{nielsen2010quantum,watrous2018theory,fagotti2013reduced}.
In particular, an important feature of such metric is that it provides an upper bound for the difference between the expectation values of observables
in the two states $\r$ and $\s$, i.e.
\be
\label{TD-bound}
\left| \tr (\r- \s) O \right| \leq \| \r - \s \|_1 \| O \|_{\infty}=
2 D (\r, \s) \| O \|_{\infty}.
\ee
It is clear that the bound \eqref{TD-bound} does not depend on the Hilbert space dimension, while it would not be the case if one uses $n\neq 1$ in \eqref{DnDef}.
This means that if $\r$ and $\s$ are ``close'', also the expectation values of an arbitrary observable $O$ (of finite norm) are ``close''.
We will provide in this paper important examples of how choosing the ``wrong distance'' could lead to misleading results.

In an extended quantum system, especially in a quantum field theory (QFT), it is extremely difficult to evaluate the trace distance \eqref{TDDef}, as,
for example, discussed in \cite{fagotti2013reduced}.
This is one of the reason why in the literature there has been an intensive investigation of the relative entropy
\cite{Blanco:2013joa,Lashkari:2014yva,Balasubramanian:2014bfa,Lashkari:2015dia,Jafferis:2015del,Sarosi:2016oks,Casini:2016udt,Sarosi:2016atx,Ruggiero:2016khg,Nakagawa:2017fzo,Murciano:2018cfp}, defined as \cite{kullback1951}
\be
\label{rel-ent}
S(\r\|\s) = \tr(\r\log\r)-\tr(\r\log\s).
\ee
$S(\r\|\s) $ bounds the trace distance according to the Pinsker's inequality \cite{watrous2018theory}
\be \label{pinsker}
D(\r,\s) \leq \sr{\f12 S(\r\|\s)}.
\ee
It is definitely a useful tool in quantum information theory, but is not a metric: indeed it is not symmetric in its inputs, it may be infinite for some density matrices,
and does not satisfy the triangle inequality \cite{watrous2018theory}.

Another quantity, already studied in literature, that provides an indication of the difference of two states is the fidelity \cite{Lashkari:2014yva},
defined as \cite{nielsen2010quantum,watrous2018theory}
\be
F(\r,\s) = \tr \sr{\sr{\s}\r\sr{\s}} = \tr \sr{\sr{\r}\s\sr{\r}}.
\ee
Although not obvious by definition, the fidelity is  symmetric in $\rho$ and $\s$.
(Notice that often in the literature the square of $F(\r,\s)$ is called fidelity, generating some confusion.)
By definition $0\leq F(\r,\s)\leq1$: in particular, for two close states $F(\r,\s)$ approaches 1, and for two far away states $F(\r,\s)$ approaches 0.
Trace distance and fidelity also satisfy the inequalities \cite{nielsen2010quantum}
\be
1 - F(\r,\s) \leq D(\r,\s) \leq \sr{1 - F(\r,\s)^2}.
\ee
Unfortunately, neither the fidelity is a metric,
and therefore does not provide a proper distance for extended quantum systems.
However, it can be used to define a metric \cite{nielsen2010quantum} as $ \arccos (F(\r,\s)$).

Recently, we developed a systematic method to calculate the trace distance between two RDMs in generic QFTs in Ref.~\cite{Zhang:2019wqo}.
The present paper is an extension of the Letter \cite{Zhang:2019wqo}.
Here, on top of providing many details of the calculations that were not reported in  \cite{Zhang:2019wqo} for lack of space, we also produced
new results  for subsystem trace distances in 2D free massless compact boson and fermion theories.
Furthermore, as a byproduct of our analysis, we provide new results for the relative entropy and fidelity, which, as mentioned above,
have both already been largely studied in literature.

Our approach to compute the trace distance is based on the path integral representation of the RDMs and an \emph{ad hoc} replica trick.
As detailed in the following, one first needs to compute $D_{n_e}(\r,\s)$ with $n_e$ being an even integer, and then consider its analytical continuation to arbitrary real values.
The trace distance is then given by the following \emph{replica limit}
\be
D(\r,\s)= \lim_{n_e \to 1} D_{n_e}(\r,\s).
\ee
This strategy closely resembles the calculation of the entanglement negativity (an entanglement measure for generic mixed states)
in \cite{Calabrese:2012ew,Calabrese:2012nk,ctt-13}.
%
%
The method can be applied to many different situations, but in \cite{Zhang:2019wqo} we focused on one-dimensional (1D) systems described by a
2D Conformal Field Theory (CFT), with the subsystem consisting of an interval of length $\ell$ embedded in a circle of length $L$.
In such setting, entanglement measures as the R\'enyi and the von Neumann entropy have been considered.
In particular, using the \emph{twist operators} \cite{Calabrese:2004eu,Cardy:2007mb,Calabrese:2009qy} and their operator product expansion (OPE)
\cite{Headrick:2010zt,Calabrese:2010he,Rajabpour:2011pt,Chen:2013kpa,rtc-18,Lin:2016dxa,Chen:2016lbu,He:2017vyf}, a universal short interval expansion has been derived.
This expansion also generalises to subsystem trace distances between the low-lying excited states in 2D CFT.

The remaining part of the paper is arranged as follows.
In Section \ref{review} we review the path integral approach to entanglement in QFT and in particular in CFT.
In Section~\ref{secGen}, after presenting in details the replica trick for the trace distance, we derive the universal formula of the leading order trace distance of one interval in
the short interval expansion and exact results for a special class of states.
In Section~\ref{secBoson}  we consider the 2D free massless compact boson theory and calculate trace distance and several $n$-distances.
We test our analytic predictions against exact numerical calculations for the XX spin chain.
We also provide some further results on relative entropy and fidelity.
The same quantities for the 2D free massless fermion theory  are investigated in Section \ref{secFermion} and tested against exact numerical
calculations in the critical Ising spin chain.
We conclude with discussions in Section~\ref{secCnD}.
In Appendix~\ref{appRXY}, we review the needed information about the XY spin chain, of which the XX and Ising models represent special cases.
In Appendix~\ref{appAID} and Appendix~\ref{appREF} we give some identities that are useful to the calculations of relative entropies in the
boson and fermion theory, respectively.
In Appendix \ref{appAC} we provides some details of the analytic continuation.

\section{Entanglement in QFT: an overview}
\label{review}

In this section we present an overview of the path integral approach to the entanglement entropy and introduce all concepts that will be used in the
following sections to calculate the trace distances between RDMs of the low-lying eigenstates in CFT.

\paragraph{Replica tricks.}

The most useful measure of bipartite entanglement in a pure state is the entanglement entropy, defined in terms of the RDM of a quantum state.
For a generic state $|\psi\rangle$ with density matrix $\rho=|\psi\rangle\langle\psi|$,  the RDM of a subsystem $A$ is $\rho_A = \tr_{\bar{A}} \rho$
($\bar{A}$ being the complement of $A$) and its entanglement entropy is the corresponding von Neumann entropy $S_A = - \tr \rho_A \log \rho_A$.
 In the replica approach, it is obtained from $\tr \rho_A^n$, computed at first for $n$ integer and then analytically continued to real values, through the limit
\be
S_A = - \lim_{n \to 1} \frac{\partial}{\partial n} \tr \rho_A^n.
\ee
When $n$ is an integer, $\tr \rho_A^n$ may be computed in the path integral formalism.  In fact,
in 1D systems described by 2D QFTs,  this path integral representation is the partition function on a $n$-sheeted Riemann surfaces,
in which the $j$-th sheet represents $\rho_{A, j}$, the $j$-th copy of the state $\rho_A$.

A generalisation of this replica trick has been then introduced in Refs.~\cite{Lashkari:2014yva, Lashkari:2015dia} for the relative entropy, cf. Eq.~\eqref{rel-ent}.
It relies on the path integral representation of $\tr \left( \rho_A^n \sigma_A^m \right)$, with $\rho_A, \sigma_A$ being two RDMs, and reads
\be
S(\rho_A \| \sigma_A ) = - \lim_{n \to 1}  \frac{\partial}{\partial n} \frac{\tr \left( \rho_A \sigma_A^{n-1} \right)}{\tr \rho_A^n}.
\ee

\paragraph{Twist fields.}

Moreover, still in 1D systems, for a subsystem $A$ consisting of $m$ disjoint intervals, $\tr \rho_A^n$ can be expressed (for integer $n$) as a $2m$-point correlation function
of some special fields $\mT$ and $\bar\mT$ known as twist fields  \cite{Calabrese:2004eu,Cardy:2007mb,Calabrese:2009qy}.
This correlation is evaluated in the state ${\boldsymbol\rho}_n = \otimes_{j=1}^n \rho_{j}$ of the corresponding $n$-fold theory, denoted as CFT$^n$, as shown
in Figure \ref{replica} (left). For the simple case of a single interval ($m=1$)
\be
\label{ttbar-renyi}
\tr \rho_A^n = \langle \mT (\ell, \ell) \bar\mT (0, 0) \rangle_{\boldsymbol\rho_n}
\ee
The above relation and twist fields in general can be defined in any 2D QFT but turn out to be particularly useful when dealing with a CFT, where twist fields are primary operators in CFT$^n$,  with conformal weights \cite{Calabrese:2004eu}
\be \label{hnbhn}
h_n = \bar h_n = \f{c(n^2-1)}{24n},
\ee
$c$ being the central charge of the single copy CFT.
In the case when $\rho_A$ corresponds to the ground state (vacuum) of the CFT, the moments of the reduced density matrix for $A$ being a single interval
in an infinite system are fixed by global conformal invariance to be
\be
\tr \rho_A^n = \langle \mT (\ell, \ell) \bar\mT (0, 0) \rangle_{\boldsymbol\rho_n}= c_n \left(\frac{\ell}{\epsilon}\right)^{-2(h_n + \bar h_n) },
\ee
where $c_n$ (with $c_1=1$) is the normalisation of the twist operators (related to the boundary conditions induced by the twist operators at
the entangling surface \cite{Ohmori:2014eia,Cardy:2016fqc,act-17}) and $\epsilon$ is an ultraviolet  cutoff.

Similarly, also $\tr \left( \rho_A^n \sigma_A^m \right)$ can be expressed in terms of correlation functions of twist fields, this time evaluated in the state $\otimes_{j=1}^n \rho_{ j} \otimes_{k=1}^m \sigma_{ k} $ in CFT$^n$. This is indeed nothing but the generalisation of \eqref{ttbar-renyi} to the case where the replicas of the CFT are in different states.

\paragraph{Short interval expansion.}

Hereafter we specialise to a 1+1 dimensional CFT in imaginary time $\tau$. The two dimensional geometry can be parametrised by a
complex coordinate $z=x+ \ii \tau$, where $\tau \in \mathbb{R}$, the spatial coordinate $x \in [0, L]$ and we consider periodic boundary conditions (PBC).

The OPE of twist operators \cite{Headrick:2010zt,Calabrese:2010he,Rajabpour:2011pt,Chen:2013kpa} can be used to write down and asymptotic expansion of
the multipoint correlation functions of the twist operators.
For example, in terms of CFT$^n$ quasiprimary operators and their derivatives, the OPE of twist operators takes the form \cite{Chen:2013kpa}
\be
\label{OPEtt}
\mT(z,\bar z)\bar\mT(0,0) = \f{c_n \e^{2(h_n+\bar h_n)}}{z^{2h_n}\bar z^{2\bar h_n}} \sum_K d_K \sum_{r,s\geq0} \f{a_K^r}{r!}\f{\bar a_K^s}{s!}
                                                                  z^{h_K+r}\bar z^{\bar h_K+s}
                                                                  \p^r \bar \p^s \Phi_K(0,0).
\ee
 The summation $K$ is over all the orthogonal quasiprimary operators $\Phi_K$ in CFT$^n$, with conformal weights $(h_K,\bar h_K)$, and they can be constructed from the orthogonal quasiprimary operators in the original one-fold CFT.
In Eq.~\eqref{OPEtt} the following constants have been defined
\be
a_K^r \equiv \f{C_{h_K+r-1}^r}{C_{2h_K+r-1}^r}, \qquad \bar a_K^s\equiv\f{C_{\bar h_K+s-1}^s}{C_{2\bar h_K+s-1}^s},
\qquad {\rm with} \quad C_x^y=\f{\G(x+1)}{\G(y+1)\G(x-y+1)}.
\ee
The OPE coefficients, moreover, can be calculated as \cite{Calabrese:2010he}
\be \label{dK}
d_K=\f{1}{\a_K\ell^{h_K+\bar h_K}} \lim_{z\to\inf}z^{2 h_K}\bar z^{2\bar h_K}\lag \Phi_K(z,\bar z) \rag_{\cR_{n}},
\ee
with $\a_K$ being the normalisation of $\Phi_K$ and $\cR_{n}$ being the $n$-fold Riemann surface for one interval $A=[0,\ell]$ on the complex plane. The expectation value on $\cR_{n}$ can be calculated by mapping to the complex plane \cite{Calabrese:2010he}.

For a general translationally invariant state $\r$, in the OPE of twist operators we only need to consider CFT$^n$ quasiprimary operators that are direct products of the quasiprimary operators $\{\cX\}$ of the original CFT \cite{Chen:2016lbu, rtc-18}
\be
\Phi_K^{j_1,j_2,\cdots,j_k} = \mX_1^{j_1}\cdots\mX_k^{j_k}.
\ee
From the OPE coefficient of $\mX_1^{j_1}\cdots\mX_k^{j_k}$, which we denote by $d_{\mX_1\cdots\mX_k}^{j_1\cdots j_k}$, one can define the coefficient
\cite{Chen:2016lbu}
\be \label{bX1dddXk}
b_{\mX_1\cdots\mX_k} = \sum_{0 \leq j_1, \cdots, j_k \leq n-1} d_{\mX_1\cdots\mX_k}^{j_1\cdots j_k},
\ee
where the sum is constrained in order to avoid overcounting.
For examples, for $\mX_{j_1}\mX_{j_2}$ one has $0\leq j_1 < j_2 \leq n-1$, and for $\mX_{j_1}\mX_{j_2}\mY_{j_3}$ with $\mX\neq\mY$ one has $0\leq j_1,j_2,j_3 \leq n-1$ with constraints $j_1<j_2$, $j_1 \neq j_3$, $j_2 \neq j_3$.
For the RDM $\r_A$ of such states, one finds the following expansion \cite{Chen:2016lbu,Lin:2016dxa,rtc-18,He:2017vyf}
\be
\label{short-exp}
\tr \r_A^n =
c_n \Big(\f{\ell}{\e}\Big)^{-4h_n}
\Big[ 1+
\sum_{k=1}^n
\sum_{\{\mX_1,\cdots,\mX_k\}}
\ell^{\D_{\mX_1}+\cdots+\D_{\mX_k}}
b_{\mX_1\cdots\mX_k}
\lag \mX_1 \rag_\r
\cdots
\lag \mX_1 \rag_\r \Big],
\ee
with the summation being over all the sets of orthogonal nonidentity quasiprimary operators $\{\cX\}$. This allows to derive the short interval behaviour  of the R\'enyi and entanglement entropies.

Similarly, given two RDMs $\r_{A}$, $\s_{A}$ associated to translationally invariant states, one can derive the universal leading order of the relative entropy in short interval expansion \cite{Sarosi:2016oks,Sarosi:2016atx,He:2017vyf}
\be \label{SUHLZ}
S(\r_A\|\s_A) =  \f{\sr\pi\G(\D_\phi+1)\ell^{2\D_\phi}}{2^{2(\D_\phi+1)}\G(\D_\phi+\f32)} \f{(\lag\phi\rag_\r - \lag\phi\rag_\s)^2}{\ii^{2s_\phi}\a_\phi} + o(\ell^{2\D_\phi}).
\ee
Here $\phi$ is one of the quasiprimary operators with the smallest scaling dimension among the ones that satisfy
\be
\label{condition}
\lag\phi\rag_\r \neq \lag\phi\rag_\s.
\ee
This is strictly true when there is a single operator satisfying \eqref{condition};
in the degenerate case, we need just to sum all the quasiprimary operators $\phi$ satisfying the constraint~\eqref{condition}.
The operator $\phi$ has conformal weights $(h_\phi,\bar h_\phi)$, scaling dimension $\D_\phi=h_\phi+\bar h_\phi$, and spin $s_\phi=h_\phi-\bar h_\phi$.
We choose $\phi$ to be Hermitian and so the normalisation  factor is $\a_\phi>0$.
Note that $\phi$ can only be bosonic, i.e. $s_\phi$ is an integer, otherwise $\lag\phi\rag_\r = \lag\phi\rag_\s =0$.
When $s_\phi$ is an even integer $\lag\phi\rag_\r$, $\lag\phi\rag_\s$ are real, and when $s_\phi$ is an odd integer $\lag\phi\rag_\r$, $\lag\phi\rag_\s$ are pure imaginary.
Moreover, we only consider unitary CFTs, so that $h_\phi>0$, $\bar h_\phi>0$.
As required by definition, $S(\r_A\|\s_A)\geq0$.
For later reference, it's important to note that \eqref{SUHLZ} applies to both the cases with and without degeneracy at scaling dimension $\D_\phi$.

\paragraph{Some exact results for excited states entanglement and relative entropy.}

We now consider excited CFT states  obtained by acting on the ground state with a field $\mX$ (i.e. $|\mX \rangle\equiv \mX(-\ii\infty)|0\rangle$).
Here, once again, $A$ is an interval of length $\ell$ in a finite, periodic, system of length $L$.
The path integral representation of the corresponding density matrix $|\mX \rangle \langle \mX |$ presents two fields insertions at $\pm \ii \infty$.
The RDM $\rho_{\mX}$ relative to the subsystem $A$ is obtained by closing cyclically
$| \mX \rangle \langle \mX |$ along $\bar A$ and leaving an open cut along $A$.
Then ${\tr} \rho^n_{\mX}$ is given by $n$ copies of the RDM $\rho_{\mX}$ sewed
cyclically along $A$.
Following this standard procedure, we end up in a world-sheet which is the $n$-sheeted Riemann surface ${\mathcal{R}_n}$,
and the moments of $\rho_{\mX}$ are \cite{Alcaraz:2011tn, Berganza:2011mh}
\begin{equation}
\label{moms}
{\tr} \rho_{\mX}^n = \frac{ Z_n (A)}{Z_1^n} \frac{\langle \prod_{k=1}^n \mX (z_k) \mX_k^{\dagger} (z'_k) \rangle_{\mathcal{R}_n}}{\langle\mX (z_1) \mX^{\dagger} (z'_1) \rangle_{\mathcal{R}_1}^n},
\end{equation}
where $Z_n (A) \equiv \langle \mathbb{I} \rangle_{\mathcal{R}_n} $ (i.e. the $n$-th moment of the RDM of the ground state)
and $z_k= \ii \infty, z'_k = - \ii \infty$ are points where the operators are inserted in the $k$-th copy.
In \eqref{moms}, the normalisation  is properly taken into account.

For convenience, one usually introduces the universal ratio between the moment of the RDM in the
state $\mX$ and the one of the ground state, i.e.,
\begin{equation} \label{ratio-F}
F_{\mX}^{(n)}\left(\frac{\ell}L\right) \equiv \frac{{\tr} \rho_{\mX}^n}{{\tr} \rho_{\mathbb{I}}^n}
=\frac{\langle \prod_{k=1}^n \mX (z_k) \mX_k^{\dagger} (z'_k) \rangle_{\mathcal{R}_n}}{\langle\mX (z_1) \mX^{\dagger} (z'_1) \rangle_{\mathcal{R}_1}^n},
\end{equation}
in which  the factors coming from the partition functions cancel out and so the ratio is a universal function solely of $\ell/L$.

In the case of $A$ being a single interval, in order to calculate the correlators appearing in \eqref{ratio-F}, one could either introduce twist fields (as mentioned above) or consider a conformal transformation mapping the Riemann surface to the complex plane, where the correlators themselves can be explicitly evaluated.
While the representation in terms of twist field is a powerful tool to get the short-interval expansion,  this second method allows in some cases to get the full analytic result for an interval of arbitrary length, at least in the case when $\mX$ is a primary field and the mapping to the complex plane has no
anomalous terms \cite{Alcaraz:2011tn, Berganza:2011mh}.
The above results have been generalised in the literature to many other situations, e.g., states
generated by descendant fields \cite{p-14,p-16}, boundary theories \cite{txas-13,top-16}, and systems with disorder \cite{rrs-14}.

The traces ${\tr }\left( \rho_{\mY}^m \rho_{\mX}^n \right)$, for two given fields $\mX, \mY$, are obtained by a simple generalisation  of
${\tr} \rho_{\mX}^n$ discussed above.
In this case, in fact, instead of $n$ copies of the RDM $\rho_{\mX}$ only, one considers further $m$ copies of $\rho_{\mY}$ and
joins them cyclically as before.
The final result is a path integral on a Riemann surface with $(m+n)$ sheets with the insertion
of $\mX, \mX^{\dagger}$
on $n$ sheets and $\mY , \mY^{\dagger}$ on the remaining $m$ sheets.
Keeping track of the normalization we get \cite{Lashkari:2015dia}
\begin{equation}
{\tr} \left(  \rho_{\mY}^m \rho_{\mX}^{n} \right) =
\frac{Z_{n+m} (A)}{Z_1^{m+n}}
\frac{ \langle \prod_{k=1}^{m} \mY (w_k) \mY^{\dagger} (w'_k) \prod_{i= 1+m}^{n+m} \mX  (w_i) \mX^{\dagger} (w'_i) \rangle_{\mathcal{R}_n}  }{
\langle \mY (w_1) \mY^{\dagger} (w'_1) \rangle_{\mathcal{R}_1}^{m} \langle \mX  (w_1) \mX ^{\dagger} (w'_1) \rangle_{\mathcal{R}_1}^{n} }.
\end{equation}
Also in this case, a universal ratio is usually introduced
\begin{equation} \label{ratio-G}
G^{(n)} ( \rho_{\mY} \| \rho_{\mX} ) \equiv \frac{{\tr} \left(  \rho_{\mY} \rho_{\mX}^{n-1} \right) }{{\tr} \left( \rho_{\mY} ^n \right)} =
\frac{ \langle \mY(w_1) \mY^{\dagger} (w'_1) \prod_{i= 2}^{n} \mX  (w_i) \mX^{\dagger} (w'_i)
\rangle_{\mathcal{R}_n}
\langle \mY (w_1) \mY^{\dagger} (w'_1) \rangle_{\mathcal{R}_1}^{n-1}  }{
\langle \prod_{i=1}^n  \mY (w_i ) \mY^{\dagger} (w'_i) \rangle_{\mathcal{R}_n} \langle \mX  (w_1) \mX ^{\dagger} (w'_1) \rangle_{\mathcal{R}_1}^{n-1} }.
\end{equation}
A similar strategy will be applied to the trace distances in the following sections.

\section{Subsystem trace distance in QFT} \label{secGen}


In this section we report on the construction of the replica trick for the trace distance (\ref{TDDef})
introduced in our previous Letter \cite{Zhang:2019wqo}.
The problem in the calculations of the trace distance (\ref{TDDef}) resides in the presence of the absolute value of the eigenvalues of $\rho_A-\sigma_A$.
Because of this absolute value, the only way to directly get the desired quantity would be by explicitly diagonalising $\rho_A-\sigma_A$, a problem that is
made even more complicated by the fact that the two RDMs generically do not commute.
Absolute values of matrices can be anyhow tackled with a replica trick, an idea first introduced, to the best of our knowledge, by Kurchan \cite{k-91} and
later applied to many different situations \cite{g-vari,cs-08,Calabrese:2012ew,Calabrese:2012nk}, including to the entanglement
negativity \cite{Calabrese:2012ew,Calabrese:2012nk}.
This trick  for the trace distance, and more generically for all the $n$-distance for arbitrary real $n$, works as follows.
Given two (Hermitian) density matrices $\r$ and $\s$, we have by definition
\be
\| \r - \s \|_n^n = \tr|\r-\s|^n = \sum_i |\l_i|^n,
\ee
with $\l_i$ being the eigenvalues of $(\r-\s)$. Note that, for $n_e$ being an \emph{even} integer, it holds
\be
\tr|\r-\s|^{n_e} = \tr(\r-\s)^{n_e}.
\ee
Therefore, if we compute $ \tr(\r-\s)^{n_e}$ for generic even integer $n_e=2,4,\cdots$, we can then consider its analytical continuation to any real number.
In case we manage to work out such an analytic continuation, the trace distance is then simply obtained as
\be
\label{eq:replica}
D (\r_A, \s_A)= \f12\lim_{n_e \to 1} \tr (\r_A-\s_A)^{n_e}.
\ee
The calculation of $\tr (\r_A-\s_A)^{n}$ for general integer $n$ is instead a relatively simple issue.
Indeed, expanding the power of the difference, one just has to compute a sum of the traces of products of $\rho_A$'s and $\sigma_A$'s;
and we know how to get each of these products, as explained in the previous section.
For example for $n=2$ we have  $\tr (\r_A-\s_A)^{2}= \tr \rho_A^2+\tr \sigma_A^2- 2 \tr (\rho_A \sigma_A)$ and so on for larger $n$
(but keep in mind that $\rho_A$ and $\sigma_A$ do not commute).
Incidentally, this simplicity is the main reason why in the literature  the (Schatten) 2-distance has been largely studied in many applications,
instead of the more physical trace norm.
We stress that for odd $n=n_o$, $\tr (\r_A-\s_A)^{n_o}$ does {\it not} provide the $n_o$-distance (because of the absence of the absolute value).
Also, the limit $n_o\to 1$ gives the trivial result $\tr(\r_A-\s_A)=0$ (in full analogy with what happens for the negativity \cite{Calabrese:2012ew,Calabrese:2012nk}).

\begin{figure}[tbp]
  \centering
  \includegraphics[width=0.7\textwidth]{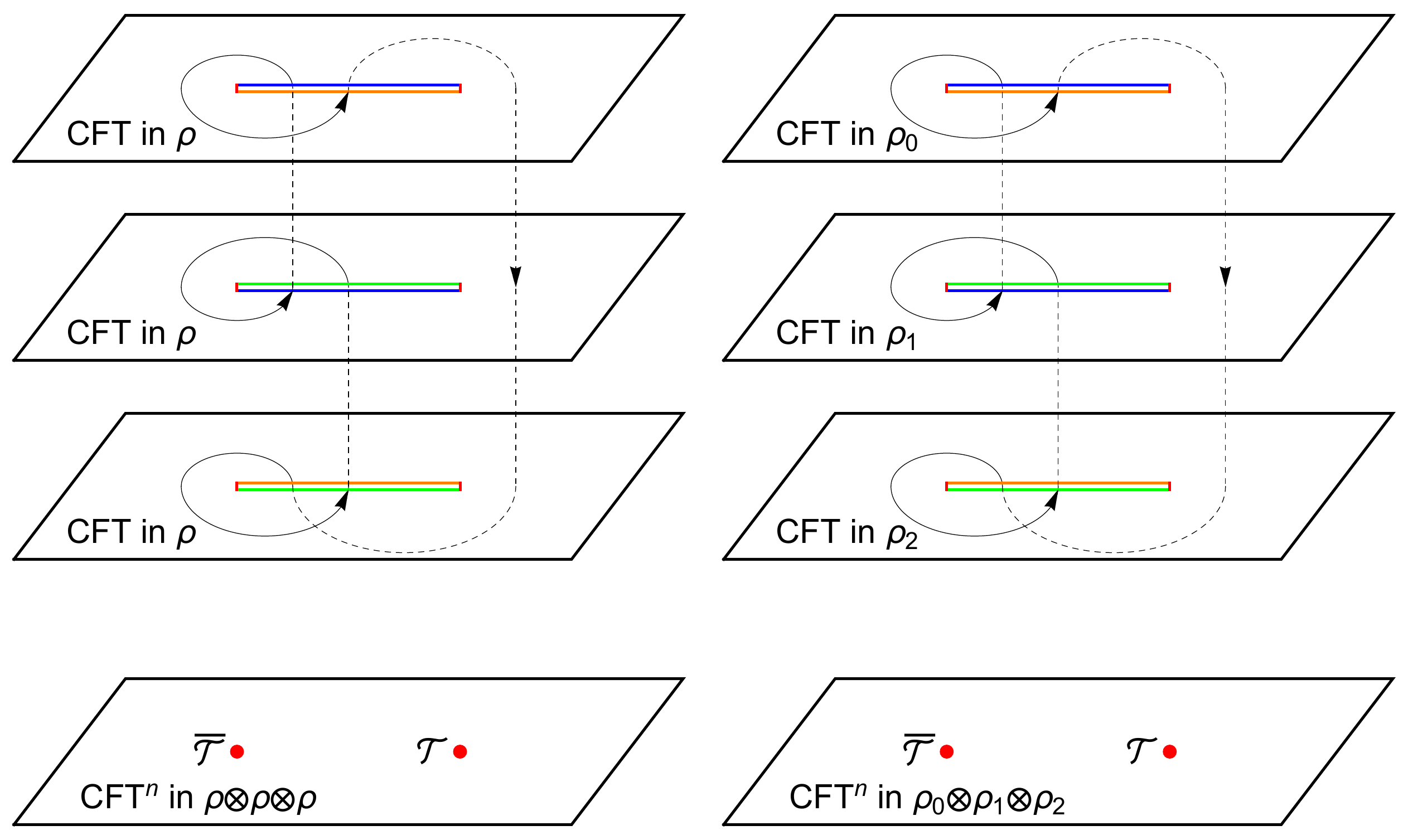}\\
  \caption{The replica trick to calculate $\tr_A \r_A^n$, Eq.~(\ref{ttbar-renyi}), (left) and $\tr_A (\r_{A,0}\r_{A,1}\cdots\r_{A,n-1})$, Eq.~(\ref{tr-rel}), (right).
  Top: path integral  in terms of Riemann surfaces.
  Bottom: equivalent representation in terms of the twist operators in CFT$^n$.
  We show the case $n=3$ as an example.}\label{replica}
\end{figure}

Therefore, in the context of a general QFT, the quantity we need to evaluate is $\tr (\r_A-\s_A)^{n}$, which may be expanded as
\be \label{expansion}
\tr (\r_A-\s_A)^{n} = \sum_{\mathcal{S}} (-)^{|\mathcal{S}|} \tr \left( \rho_{0_{\mathcal{S}}} \cdots \rho_{(n-1)_{\mathcal{S}}} \right)  ,
\ee
where the summation $\mathcal{S}$ is over all the subsets of $\mathcal{S}_0 = \{ 0, \cdots, n-1 \}$, $| \mathcal{S}|$ is the cardinality of
$\mathcal{S}$ and $\rho_{j_{\mathcal{S}}} =  \sigma_A $ if $j \in \mathcal{S}$ and $\rho_A$ otherwise.
Crucially, each term in the sum appearing in the r.h.s. of Eq.~\eqref{expansion}, in a 2D QFT, is related to a partition function on an $n$-sheeted Riemann
surface (see Fig. \ref{replica}, right) and may still be seen as a two-point function of twist fields (cfr., e.g., \cite{dei-18})
\be
\label{tr-rel}
\tr (  \rho_{0_{\mathcal{S}}} \cdots \rho_{{(n-1)}_{\mathcal{S}}}  ) = \lag \cT(\ell,\ell)\bar\cT(0,0) \rag_{\otimes_j \rho_{j_{ \mathcal{S} }}}.
\ee
Such objects already appeared in the replica trick for the relative entropy mentioned above \cite{Lashkari:2014yva, Lashkari:2015dia},
and, in some cases, they have been explicitly computed
\cite{Lashkari:2014yva, Lashkari:2015dia,Sarosi:2016oks,Sarosi:2016atx,Ruggiero:2016khg}.
Still, performing the sum in Eq.~\eqref{expansion} and obtaining its analytic continuation is not an easy task.
We stress that Eqs. \eqref{expansion} and \eqref{tr-rel} are very general in the sense they apply to generic situations for one-dimensional systems, even if
in the following we just focus on eigenstates of CFTs.

\subsection{The trace distance between primary states in CFT} \label{secExact}

In this section,  we specialise to the case when the RDMs $\r_A$ and $\s_A$ correspond to low lying eigenstates of a 2D CFT;
we focus on periodic systems of total length $L$ and on a subsystem being an interval of length $\ell$.
Similarly to the discussion in Section \ref{review}, analytical results for an interval of arbitrary length can be obtained by looking to a special class of states in a 2D CFT.
%
We study the distance between RDMs of orthogonal eigenstates associated to primary operators;
as we shall see, while the distance between the entire states is maximal, subsystems may be rather close
and they distance has different functional form depending on the considered states.

For a general primary operator $\mX$, let $(h_{\mX}, \bar{h}_{\mX})$ be its conformal weights and $\Delta_{\mX} = h_{\mX} + \bar{h}_{\mX}$ and $s_{\mX} =h_{\mX}- \bar{h}_{\mX}$ its scaling dimension and spin, respectively.
We exploit Eq.~\eqref{expansion} to compute $\tr \left( \r_{\mX} - \r_{\mY} \right)^n$ for two RDMs associated to two primary operators $\mX$ and $\mY$.
For such states, each term of the sum in the r.h.s. corresponds to a correlation function with insertions of the fields $\mX$ and $\mY$
on the Riemann surface \cite{Alcaraz:2011tn, Berganza:2011mh}, which, as mentioned above, can be mapped to the complex plane by the map
\be
f(z)= \left( \frac{z - e^{ 2 \pi \ii \ell/L}}{z-1} \right)^{1/n}.
\ee

The final result for the entire sum in Eq.~\eqref{expansion} can be then written as a sum of such correlation functions as follows
\begin{multline}
 \label{trAXY}
 \tr( \r_\mX - \r_\mY )^n = c_n \Big( \f{L}{\pi\e}\sin\f{\pi\ell}{L} \Big)^{-4h_n} \\
                             \times  \sum_{\mS} \Big\{
                             (-)^{|\mS|}
                             \ii^{2( |\bar\mS|s_\mX + |\mS|s_\mY )}
                             \Big( \f{2}{n}\sin\f{\pi\ell}{L} \Big)^{2(|\bar\mS|\D_\mX+|\mS|\D_\mY)} \\
\times \Big\lag
   \Big[ \prod_{j \in \bar \mS} \Big( f_{j,\ell}^{h_\mX} \bar f_{j,\ell}^{\bar h_\mX} f_j^{h_\mX} \bar f_j^{\bar h_\mX}  \mX(f_{j,\ell},\bar f_{j,\ell}) \mX^\dag(f_{j},\bar f_{j}) \Big)\Big] \\
\times \Big[ \prod_{j \in \mS} \Big( f_{j,\ell}^{h_\mY} \bar f_{j,\ell}^{\bar h_\mY} f_j^{h_\mY} \bar f_j^{\bar h_\mY} \mY(f_{j,\ell},\bar f_{j,\ell}) \mY^\dag(f_{j},\bar f_{j}) \Big)\Big]
   \Big\rag_{\mathbb{C}} \Big\}.
\end{multline}
Here $\bar\mS = \mS_0/\mS$, $f_j = \ep^{\f{2\pi\ii j}{n}} $ and $ f_{j,\ell} = \ep^{\f{2\pi\ii}{n}(j+\f{\ell}{L})}$.
Eq. \eqref{trAXY} relates the even (Schatten) $n$-distances between the RDM of two primary states $|\mX\rangle$ and $|\mY\rangle$ to the $2n$-point correlation function
of the corresponding primary fields on the complex plane.
Such correlation functions may be calculated in some specific cases, as we shall see, but in general it is not possible to work them out in a closed form as function of $n$
in order to obtain the analytic continuation for the trace distance.
Anyway, even if too complicated to extract direct information, from Eq. \eqref{trAXY} we can already draw one very important conclusion.
Indeed, in the limit $n \to 1$ (independently of the parity of $n$) the dependence on the ultraviolet cutoff $\epsilon$ washes out.
Importantly, this means that the trace distance is a {\it universal, cutoff independent} (i.e., UV-complete),
scale invariant  function of $\ell/L$ (i.e., it does not separately depend on $\ell$ and $L$).
This is another very important property that puts the trace distance on a special foot compared to the other Schatten distances that instead are
cutoff dependent and not scale invariant (but only scale covariant since they have non zero dimension).

It is clear from the explicit form of Eq. \eqref{trAXY} that the $n$-distance $D_n(\r_A,\s_A)$ of two RDMs $\r_A,\s_A$ with $n\neq 1$ is dependent on the UV cutoff $\e$.
As for other quantities discussed above (cfr. Eqs.~\eqref{ratio-F}, \eqref{ratio-G}), it is worth and useful to introduce a \emph{scale-invariant}
and \emph{cutoff-independent} ratio for the $n$-distance as
\be
\label{Gn}
\mD_n (\r_A,\s_A) = \f{1}{2} \f{\tr|\r_A-\s_A |^n}{\tr\r_0^n},
\ee
in which $\rho_0$ is the RDM of the subsystem $A$ in the CFT ground state.
Within this normalisation by $\tr\r_0^n$, the function $\mD_n$ in Eq. (\ref{Gn}) has the simpler expression in CFT.
Note also that this definition is slightly different compared to the one given in \cite{Zhang:2019wqo}: the present form gives a quantity between $[0,1]$,
rather than $[0,2]$ as the one in the Letter and has a normalisation that is independent from the inputs of the distance.
When there is no ambiguity, we will also call $\mD_n(\r,\s)$ loosely as the $n$-distance, but the true $n$-distance is instead
\be
D_n(\r_A,\s_A)= [\mD_n (\r_A,\s_A)\, \tr\rho_0^n  ]^{1/n}\,.
\label{DD}
\ee
The replica limit \eqref{eq:replica} now takes the form
\be
\label{eq:replica2}
D (\r_A, \s_A) =  \lim_{n_e \to 1} \mD_{n_e} (\r_A, \s_A).
\ee

Eq. \eqref{DD} also highlights one of the main reasons why the trace distance is better than all other $n$-distances.
In CFT $\cD_n(\r_A,\s_A)$ is always a smooth function of $\ell/L$ in the interval $[0,1]$ and so it is its replica limit $D(\r_A, \s_A)$.
Conversely, since $\tr\rho_0^n$ goes to zero as $L\to\infty$, irrespective of the value of $\ell/L$, the $n$-distance always vanishes in the
thermodynamic limit. Consequently, a study of $n$-distance may artificially signal the closeness of two RDMs that actually are very different.

\subsection{Short interval expansion}

Although Eq. \eqref{trAXY} is model dependent and generically complicated to be worked out analytically,
it is possible to use the OPE of twist fields (cf. Eq.~\eqref{OPEtt}) to obtain a general result in the limit $\ell\ll L$.

Let $\r_A$, $\s_A$ be the RDMs of two  CFT eigenstates $\r$, $\s$, not only primary and quasiprimary states, but also descendents or even thermal states.
The OPE of twist fields, Eq.~\eqref{OPEtt}, leads to
\be \label{trArAmsAn}
\tr_A(\r_A-\s_A)^n = c_n \Big(\f{\ell}{\e}\Big)^{-4h_n}
\sum_{\{\mX_1,\cdots,\mX_n\}}
\ell^{\D_{\mX_1}+\cdots+\D_{\mX_n}}
b_{\mX_1\cdots\mX_n}
\big(\lag \mX_1 \rag_\r - \lag \mX_1 \rag_\s\big)
\cdots
\big(\lag \mX_n \rag_\r - \lag \mX_n \rag_\s\big).
\ee
%
For two different states $\r$, $\s$, quasiprimary operators $\phi$ such that
\be
\lag\phi\rag_\r - \lag\phi\rag_\s \neq 0,
\label{condition2}
\ee
should exist (as mentioned in Eq.~\eqref{condition}).
Among these, we select the operator $\phi$ with the smallest scaling dimension $\D_\phi$.
In this section, for simplicity, we only consider the case when only one of such operators exists
with the smallest dimension $\D_\phi$  (non-degenerate case).
As mentioned in the section for the relative entropy, $s_\phi$ has to be integer.
Hence, for a general even integer $n_e$, we get
\be
\label{expansion-td}
\tr(\r_A-\s_A)^{n_e} = c_{n_e} \Big(\f{\ell}{\e}\Big)^{-4h_{n_e}}
                     \big[ \ell^{{n_e}\D_\phi} b_{\phi^{n_e}} \big(\lag \phi \rag_\r - \lag \phi \rag_\s\big)^{n_e}
                          +o(\ell^{{n_e}\D_\phi})  \big],
\ee
with $\phi^{n_e}$ denoting the direct product of $n_e$ $\phi$'s. Note that $b_{\phi^{n_e}}=d_{\phi^{n_e}}^{0\cdots(n_e-1)}$.
Then, we consider the analytical continuation in $n_e$ and get $\tr|\r_A-\s_A|^{n_e}$ for a general real number.
In particular, for $n_e \to1$, this leads to the desired universal leading order term of the trace distance in short interval expansion
\be \label{DrAsA}
D(\r_A,\s_A) = \f{x_\phi\ell^{\D_\phi}}{2} \Big| \f{\lag\phi\rag_\r - \lag\phi\rag_\s}{\sr{\a_\phi}} \Big|
               + o(\ell^{\D_\phi}).
\ee
Here $\a_\phi$ is the normalisation of the field $\phi$ (in most of the cases $\a_\phi=1$), and
the to-be-determined coefficient $x_\phi$ is given by the replica limit ($n_e = 2p, \; p=1, 2, \cdots$)
\be \label{xphi}
x_\phi = \lim_{p\to1/2} \ii^{2ps_\phi} \a_\phi^p d_{\phi^{2p}}^{0\cdots(2p-1)}
       = \lim_{p \to 1/2} \f{\ii^{2 p s_\phi}}{\a_\phi^{p}(2p)^{2p\D_\phi}}
                          \Big\lag \prod_{j=0}^{2p-1} \big[ f_j^{h_\phi} \bar f_j^{\bar h_\phi} \phi(f_j,\bar f_j) \big] \Big\rag_{\mathbb{C}}
, ~~
f_j = \ep^{\f{\pi\ii j}{p}}.
\ee
We stress however that, differently from the corresponding result \eqref{SUHLZ} for the relative entropy,  Eq.~(\ref{DrAsA}) only applies to the case with
no degeneracy at scaling dimension $\D_\phi$.
We will see in the next section how to relax this condition, while applying to the specific case of the free boson.

Note that in \eqref{xphi}, we did not keep track of the Schwarzian derivative part in the conformal transformation of the quasiprimary operator $\phi$
because it just cancels out in the limit $n_e\to1$ (i.e.\ $p\to1/2$), when using (\ref{dK}) to calculate the OPE coefficient $d_{\phi^{n_e}}^{0\cdots(n_e-1)}$.

From Eq. (\ref{xphi}) and for an integer $p=1,2,\cdots$, the replica limit can be obtained using the function $F_\phi^{(p)}(\ell/L)$ defined in Eq. (\ref{ratio-F}) and rewritten as
 \be \label{Fphipell}
F_\phi^{(p)}\Big(\frac\ell{L}\Big) = \f{\ii^{2 p s_\phi}}{\a_\phi^{p}} \Big( \f2p \sin\f{\pi\ell}{L} \Big)^{2p\D_\phi}
                          \Big\lag \prod_{j=0}^{p-1} \big[
                          f_{j,\ell}^{h_\phi} \bar f_{j,\ell}^{\bar h_\phi}
                          f_j^{h_\phi} \bar f_j^{\bar h_\phi}
                          \phi(f_{j,\ell},\bar f_{j,\ell})
                          \phi^\dag(f_j,\bar f_j) \big] \Big\rag_{\mathbb C}, \nn\\
\ee
where $f_{j,\ell} = \ep^{\f{2\pi\ii}{p}(j + \f{\ell}{L} )}, f_j = \ep^{\f{2\pi\ii j}{p}}$.
In fact, when $\phi$ is Hermitian we have $\phi^\dag=\phi$ and so
\be \label{xphiFphipell}
x_\phi = \f{F_\phi^{(1/2)}(1/2)}{2^{2\D_\phi}}.
\ee
Furthermore, if $\phi$ is a primary operator, $ F_\phi^{(p)}(\ell/L)$ is related to the $p$-th order R\'enyi entropy $S_{A,\phi}^{(p)}(\ell)$
for $A=[0,\ell]$ in the state $|\phi\rangle$
\be \label{ABS}
F_\phi^{(p)}(\ell/L) = \ep^{-(p-1) [ S_{A,\phi}^{(p)}(\ell) - S_{A,0}^{(p)}(\ell) ]} .
\ee
In Refs.~\cite{Alcaraz:2011tn,Berganza:2011mh}, Eq.~\eqref{ABS} has been explicitly evaluated for several operators and in some cases also the analytic continuation is available \cite{elc-13}.
Note that (\ref{ABS}) only applies to the case that $\phi$ is a primary operator, while (\ref{Fphipell}) also applies to the case that $\phi$  is a quasiprimary operator.

Finally we mention that from the inequality \eqref{pinsker} and from the universal leading order of the relative entropy (\ref{SUHLZ}),
one can get a universal upper bound to the coefficient $x_{\phi}$, solely depending on the scaling dimension
\be \label{bound}
x_\phi \leq x_{\rm{max}}(\D_\phi) = \sr{\f{\sr\pi\G(\D_\phi+1)}{2^{2\D_\phi+1}\G(\D_\phi+\f32)}}.
\ee
We will check such bound for various examples in the boson and fermion theories.

\subsection{Exact general result for the $2$-distance from the ground state}

We mention that the second (Schatten) norm can be straightforwardly obtained between ground state $\r_0$ and a primary state $\r_\phi$.
Indeed, we have
\be
\mD_2 (\r_0,\r_\phi)=\frac12 \frac{\tr (\r_0-\r_\phi)^2}{\tr \r_0^2}= \frac12 \Big(1+ \frac{\tr \r_\phi^2}{\tr \r_0^2}- \frac{2 \tr \r_0\r_\phi}{\tr \r_0^2}\Big),
\label{d2}
\ee
on which the second term is just the universal function $F^{(2)}_\phi(\ell/L) $ in \eqref{Fphipell} (and calculated for many $\phi$'s in \cite{Alcaraz:2011tn,Berganza:2011mh}),
while the last term is just a two point function in a two-sheeted surface given by (see also \eqref{ratio2})
\be
\f{\tr(\r_\phi\r_0)}{\tr\r_0^2} = \Big( \f{\sin\f{\pi\ell}{L}}{2\sin\f{\pi\ell}{2 L}} \Big)^{2\D_\phi},
\ee
where $\Delta_\phi$ is scaling dimension of $\phi$.
Hence, we finally have
\be
\mD_2 (\r_0,\r_\phi)=\frac{ 1+ F^{(2)}_\phi(\ell/L)}2- \Big( \cos \f{\pi\ell}{2 L}  \Big)^{2\D_\phi}.
\label{d2f}
\ee
If in \eqref{d2} we replace $\r_0$ with a primary state, the only difference is that $\tr \r_{\phi_1}\r_{\phi_2}$ is a four-point function in the 2-sheeted Riemann surface.
The latter can be easily calculated on a case by case basis, but it has not a simple expression as \eqref{d2f}.
Notice that the property $F^{(2)}_\phi(\ell/L)\geq 1$ \cite{Berganza:2011mh} ensures that the rhs of \eqref{d2f} is non negative, as it should.

\section{Free massless compact boson} \label{secBoson}

In this section, we consider the 2D free massless compact boson theory (i.e. with the target space being a circle of finite radius)
defined on an infinite cylinder of circumference $L$.  The model is a $c=1$ CFT.
In condensed matter, such a theory is usually denoted as a Luttinger liquid and describes the continuum limit of many relevant 1D systems,
among which the XX spin chain that we will explicitly consider.
We first compute the trace distance and more generally the $n$-distances (with $n=2,3,4,5$) between several low-lying excited states in the CFT,
and derive some new results for relative entropies and fidelities.
All the CFT results are then checked against numerical calculations in the XX spin chain (but we stress that our results apply to a larger class
of critical systems even interacting ones like XXZ spin chains and Bose gases).



The boson field has diffeomorphic part $\phi$ and anti-diffeomorphic part $\bar\phi$.
The states in which we are interested are those generated by the action of the following operators:
the identity operator $\mathbb{I}$, with conformal weights $(0,0)$, and its descendent at the second level,
the stress tensors $T$ and $\bar T$ with conformal weights $(2,0)$, $(0,2)$;
the currents $J=\ii\p\phi$, $\bar J=\ii\bar\p\bar\phi$ and $J\bar J$ whose conformal weights are given by $(1,0)$, $(0,1)$ and $(1,1)$, respectively;
the vertex operators $V_{\a,\bar\a}=\exp(\ii\a\phi+\ii\bar\a\bar\phi)$ with conformal weights $({\a^2}/{2},{\bar\a^2}/{2})$.
While $T$ and $\bar T$ are quasiprimary operators, all the others are primary operators.
Details of the 2D free massless compact boson theory can be found in \cite{DiFrancesco:1997nk,Blumenhagen:2009zz}.

We denote the ground state as $|0 \rangle, $ and the low energy excited states are constructed by acting on it with a primary operator, obtaining the following set of states: $|V_{\a,\bar\a}\rag$, $|J\rag$, $|\bar J\rag$, $|J\bar J\rag$.
We denote the RDMs of $A$ in these states, respectively, as $\r_{\a,\bar\a}$, $\r_{J}$, $\r_{\bar J}$, $\r_{J\bar J}$ and  $\r_{0}$ for the ground state.
Note that $\r_{0,0}=\r_0$.
One should beware to distinguish the density matrices of the entire system from the RDMs of the subsystem $A$.

\subsection{Short interval results}

In this subsection we report the explicit form of the short distance expansion for all the states we consider for the free boson.
The general form is always given by Eq. \eqref{DrAsA} with $x_\phi$ in \eqref{xphi} or equivalently \eqref{xphiFphipell}.
Here we identify the leading operator $\phi$ contributing to each distance and explicitly provide the analytic continuation for $x_\phi$.

\subsubsection{Vertex-Vertex distance: non-degenerate case}

We first consider the distance between two states generated by the a vertex operators, namely $|V_{\a,\bar\a}\rag$ and $|V_{\a',\bar\a'}\rag$.
The leading operators entering in the OPE are the primaries $J$ and $\bar{J}$ with expectation values
\be \label{JbarJ1}
\lag J \rag_{\a,\bar\a}=\f{2\pi\ii\a}{L}, \qquad \lag \bar J \rag_{\a,\bar\a}=-\f{2\pi\ii\bar\a}{L}.
\ee
They are both operators with minimal dimension $\Delta_{J}=\Delta_{\bar J}=1$ and we use the normalisation $\a_J=1$. 
The CFT formula (\ref{DrAsA}) only applies to the case with no degeneracies in the sense of Eq. \eqref{condition2}:
this implies for the vertex operator that either $\alpha= \alpha'$ or $\bar{\alpha}= \bar{\alpha}'$, else both $J$ and $\bar J$ would contribute.
We first consider the non degenerate case and in a following subsection the degenerate one.

At this point, for the non-degenerate case, the only missing factor is $x_J$ (or $x_{\bar J}$).
This can be read off Eqs. \eqref{xphiFphipell} and \eqref{ABS}. Indeed the R\'enyi entropies in the current state have been derived in the form of
a determinant in \cite{Alcaraz:2011tn,Berganza:2011mh} and analytically continued in \cite{elc-13}.
The final result reads \cite{elc-13}
\be \label{FJpell}
F_J^{(p)}(\ell/L) = F_{\bar J}^{(p)}(\ell/L) = \Big( \f2p \sin\f{\pi\ell}{L} \Big)^{2p}
\f{\Gamma^2\big(\f{1+p+p\csc\f{\pi\ell}{L}}{2}\big)}
  {\Gamma^2\big(\f{1-p+p\csc\f{\pi\ell}{L}}{2}\big)}.
\ee
Using such result and plugging $F_J^{(1/2)}(1/2)$ in Eq. (\ref{xphiFphipell}), we get
\be
x_J = x_{\bar J} = \f1\pi.
\label{xJ}
\ee
Notice that they satisfy the bound (\ref{bound}) with $x_{\rm{max}}(1) = 1/\sr{6}$.

Finally, putting all pieces together  in Eq. \eqref{DrAsA}  we get the leading orders of the trace distances:
\be \label{tdvv1}
D(\r_{{\a,\bar\a}},\r_{{\a',\bar\a}}) = \f{|\a-\a'|\ell}{L} + o\Big(\f{\ell}{L}\Big), \qquad
D(\r_{{\a,\bar\a}},\r_{{\a,\bar\a'}}) = \f{|\bar\a-\bar\a'|\ell}{L} + o\Big(\f{\ell}{L}\Big).
\ee

\subsubsection{Vertex-Current distance: non-degenerate case}

Then we consider the trace distance between a vertex state $|V_{\a,\bar\a}\rag$ and one of the three current states $|J\rag$, $|\bar J\rag$, $|J\bar J\rag$.
The OPE is again dominated by the current operator, so to apply Eq.~(\ref{DrAsA}) we need the expectation value of the current in the vertex state, as in Eq. \eqref{JbarJ1},
and also  the expectation values of $J, \bar{J}$ in the current states $|J\rag$, $|\bar J\rag$, $|J\bar J\rag$.
They are simply given by
\be \label{JbarJ2}
\lag J \rag_J = \lag J \rag_{\bar J} = \lag J \rag_{J\bar J}=
\lag \bar J \rag_J = \lag \bar J \rag_{\bar J} = \lag \bar J \rag_{J\bar J}=0.
\ee
In this case, to apply Eq.~(\ref{DrAsA}), the non degeneracy condition implies either $\alpha = 0$ or $\bar{\alpha} =0 $,
for which we simply get (using also Eq. \eqref{xJ})
\bea \label{tdjv1}
&& D(\r_{J},\r_{{\a,0}}) = D(\r_{\bar J},\r_{{0,\a}}) = \f{|\a|\ell}{L} + o\Big(\f{\ell}{L}\Big), \qquad
   D(\r_{J},\r_{{0,\bar\a}}) = D(\r_{\bar J},\r_{{\bar\a,0}})= \f{|\bar\a|\ell}{L} + o\Big(\f{\ell}{L}\Big), \nn\\
&& D(\r_{J\bar J},\r_{{\a,0}}) = 
   D(\r_{J\bar J},\r_{{0,\a}}) = \f{|\a|\ell}{L} + o\Big(\f{\ell}{L}\Big).
\eea
When both $\a$ and $\bar\a$ are non zero, we are in the degenerate case which will be considered in the following.


Instead, if $\alpha=\bar \a=0$ (i.e. for the distance between the current and the ground state), the leading term vanishes and we have to
go to the next operator in the OPE which is the stress energy tensor.
This can be obtained as follows. The expectation values of the stress tensors in a general primary state $|\mX\rag$ with conformal weights $(h_\mX,\bar h_\mX)$ are given by
\be \label{TXTbX}
\lag T \rag_\mX= \f{\pi^2 c}{6L^2} -\f{4\pi^2 h_\mX}{L^2}, ~~ \lag \bar T \rag_\mX = \f{\pi^2 c}{6L^2} -\f{4\pi^2 \bar h_\mX}{L^2}.
\ee
This result together with Eq.~(\ref{DrAsA}), with the minimal dimension quasiprimary being one of the stress tensors, eventually leads to
\bea \label{TDXXA31}
&& D(\r_{0},\r_{J}) =  D(\r_{0},\r_{\bar J})   =x_T \f{ 2\sr{2}\pi^2\ell^2}{L^2} + o\Big(\f{\ell^2}{L^2}\Big),
\eea
In this case, obtaining an analytic result for the coefficient $x_T$ is much more complicated because $T$ is not primary and Eq. \eqref{ABS} does not apply.
The general expression for $x_T$ may be written as
\be
x_T = \lim_{p \to 1/2} \Big( \f{2}{c} \Big)^p \Big\lag \prod_{j=0}^{2p-1} [ f_j^2 T(f_j) ]  \Big\rag_{\rC}, \qquad
f_j=\ep^{\f{\pi\ii j}{p}}.
\label{xT}
\ee
This result may seem, at first, quite surprising because in the mapping from the Riemann surface to the complex plane anomalous terms are present
since $T$ is not primary. This is indeed the case for the OPE coefficient for $n\neq1$.
However, since for $n=1$ the transformation from the $n$-sheeted surface to the plane is in $SL(2,\rC)$, then the Schwarzian derivative vanishes,
and so all the anomaly terms are at least of order $(n-1)$ \cite{no-anomaly}. Thus  they cancel in the $n\to1$ limit, i.e. in the $p\to1/2$ limit.
Anyhow, getting a general closed form for Eq. \eqref{xT} is rather difficult and
hence, an approximate value for this unknown coefficients $x_T$  will be extracted from the numerical results in the XX spin chain later on.

Finally let us notice that from the decoupling of the holomorphic and anti-holomorphic sectors we simply have
\be
\f{\tr(\r_J-\r_{J\bar J})^n}{\tr\r_0^n} = \f{\tr \r_J^n}{\tr\r_0^n} \f{\tr(\r_0-\r_{\bar J})^n}{\tr\r_0^n}.
\ee
In the limit $n\to1$, this decoupling leads to
\be
D(\r_J,\r_{J\bar J})=D(\r_0,\r_{\bar J}),
\ee
and, using also Eq. \eqref{TDXXA31}, we get the OPE
\be
D(\r_{\bar J},\r_{J\bar J}) = D(\r_{J},\r_{J\bar J}) =x_{ T} \f{ 2\sr{2}\pi^2\ell^2}{L^2} + o\Big(\f{\ell^2}{L^2}\Big).
\ee

\subsubsection{Vertex-Vertex and Vertex-Current distances: degenerate case}

In the 2D free massless boson theory, we can actually generalise formula (\ref{DrAsA}) to the degenerate case.
In fact, consider two states $\r$, $\s$ such  that both $\lag J \rag_\r \neq \lag J \rag_\s$ and $\lag \bar J \rag_\r \neq \lag \bar J \rag_\s$ hold.
Using Eq.~(\ref{dK}), we get the OPE coefficient of the $\CFT^n$ operator $J_{j_1}\cdots J_{j_{2k}}$
\be
d_{J^{2k}}^{j_1\cdots j_{2k}} = \f{1}{(4\ii p)^{2k}} \Big[ \f{1}{(\sin\f{\pi j_{12}}{2p}\cdots\sin\f{\pi j_{(2k-1)(2k)}}{2p})^2} + {\rm permutations} \Big]_{(2k-1)!!},
\ee
where $n=2p$ so that $p=1,2,\cdots$ and we used $J^{2k}$ to denote the direct product of $2k$ $J$'s from different replicas of the CFT. We also defined the shorthand $j_{i_1i_2} = j_{i_1} - j_{i_2}$. On the RHS of the above equation we have a sum of the permutations of all possible pairwise contractions, and the total number of terms is $(2k-1)!!$. Similarly, we get the OPE coefficient of $J_{j_1}\cdots J_{j_{2k_1}}\bar J_{j'_1}\cdots \bar J_{j'_{2k_2}}$
\bea
&& d_{J^{2k_1}\bar J^{2k_2}}^{j_1\cdots j_{2k_1}j'_1\cdots j'_{2k_2}} =
\f{1}{(4\ii p)^{2(k_1 + k_2)}} \Big[ \f{1}{(\sin\f{\pi j_{12}}{2p}\cdots\sin\f{\pi j_{(2k_1-1)(2k_1)}}{2p})^2} + {\rm permutations} \Big]_{(2k_1-1)!!} \nn\\
&& \phantom{d_{J^{2k_1}\bar J^{2k_2}}^{j_1\cdots j_{2k_1}j'_1\cdots j'_{2k_2}} =} \times
\Big[ \f{1}{(\sin\f{\pi j'_{12}}{2p}\cdots\sin\f{\pi j'_{(2k_2-1)(2k_2)}}{2p})^2} + {\rm permutations} \Big]_{(2k_2-1)!!},
\eea
which is a sum of $(2k_1-1)!!(2k_2-1)!!$ terms.
Using the definition (\ref{bX1dddXk}), we find the coefficient $b_{J^{2p}}=d_{J^{2p}}^{0\cdots(2p-1)}$, which is a sum of $(2p-1)!!$ terms.
A sum of $C_{2p}^{2k}$ number of $d_{J^{2(p-k)}\bar J^{2k}}^{j_1\cdots j_{2(p-k)}j'_1\cdots j'_{2k}}$ gives $b_{J^{2(p-k)}\bar J^{2k}}$, which is in turn a sum of totally $C_{2p}^{2k}[2(p-k)-1]!!(2k-1)!!$ terms. Since in each sum we add up all the possible permutations, eventually we simply get
\be
b_{J^{2(p-k)}\bar J^{2k}} = C_{2p}^{2k}[2(p-k)-1]!!(2k-1)!! \f{b_{J^{2p}}}{(2p-1)!!} = C_p^k b_{J^{2p}}.
\ee
The operator $J_{j_1}\cdots J_{j_{2k_1-1}}\bar J_{j'_1}\cdots \bar J_{j'_{2k_2-1}}$, instead, has a vanishing OPE coefficient
\be
d_{J^{2k_1-1}\bar J^{2k_2-1}}^{j_1\cdots j_{2k_1-1}j'_1\cdots j'_{2k_2-1}} = 0,
\ee
and we get the vanishing coefficient
\be
b_{J^{2(p-k)+1}\bar J^{2k-1}} = 0.
\ee
Now, by specifying Eq.~(\ref{trArAmsAn}), which is valid in a general 2D CFT, to the case of the 2D free massless boson theory, one finds
\bea
\label{tr-beyond}
&& \tr_A(\r_A-\s_A)^{2p} = c_{2p} \Big(\f{\ell}{\e}\Big)^{-4h_{2p}}
\Big[
\ell^{2p} \sum_{k=0}^p b_{J^{2(p-k)}\bar J^{2k}} \big(\lag J \rag_\r - \lag J \rag_\s\big)^{2(p-k)} \big(\lag \bar J \rag_\r - \lag \bar J \rag_\s\big)^{2k}
+o(\ell^{2p})
\Big] \nn\\
&& \phantom{\tr_A(\r_A-\s_A)^{2p}} =
c_{2p} \Big(\f{\ell}{\e}\Big)^{-4h_{2p}}
\Big\{
\ell^{2p} b_{J^{2p}} \big[ \big(\lag J \rag_\r - \lag J \rag_\s\big)^2 + \big(\lag \bar J \rag_\r - \lag \bar J \rag_\s\big)^2 \big]^p
+o(\ell^{2p})
\Big\}.
\eea
In particular we can apply Eq.~\eqref{tr-beyond} to two generic vertex operators and finally obtain their trace distance as
\be \label{TDXXA1}
D(\r_{{\a,\bar\a}},\r_{{\a',\bar\a'}}) = \sr{(\a-\a')^2+(\bar\a-\bar\a')^2} \f{\ell}{L} + o\Big(\f{\ell}{L}\Big),
\ee
generalising (\ref{tdvv1}) to  the degenerate cases.

Similarly, from \eqref{tr-beyond} we straightforwardly get also the distance between the generic current and generic vertex states as
\bea \label{TDXXA2}
&& D(\r_{J},\r_{{\a,\bar\a}}) = D(\r_{\bar J},\r_{{\bar\a,\a}})= \sr{\a^2+\bar\a^2}\f{\ell}{L} + o\Big(\f{\ell}{L}\Big), \nn\\
&& D(\r_{J\bar J},\r_{{\a,\bar\a}}) = \sr{\a^2+\bar\a^2}\f{\ell}{L} + o\Big(\f{\ell}{L}\Big),
\eea
which are the generalisations of  (\ref{tdjv1}) to the degenerate cases.

\subsubsection{Numerical results in the XX spin chain}

In this subsection we test the results for the short length expansion of the trace distance against exact numerical calculations
in the XX spin chain at half filling.
Actually our results apply more generically to all models described by a free boson with arbitrary  compactification radius (i.e. a Luttinger
liquid with arbitrary Luttinger parameters $K$), including, e.g., XXZ spin chains, repulsive Lieb Liniger model, etc.
We focus onto the XY spin chain in transverse field, of which the XX spin chain is a special case, because it can be mapped in a free fermionic model
for which the RDM can be written in terms of the two-point correlation function exploiting of the Wick theorem
\cite{chung2001density,Vidal:2002rm,peschel2003calculation} and
the construction of the excited states is  discussed \cite{Alba:2009th,Alcaraz:2011tn,Berganza:2011mh}.
The required details of this approach based on correlation functions are briefly reviewed in Appendix~\ref{appRXY}, with particular emphasis
to the CFT-XX states correspondence.

Within this approach, the RDM is a $2^\ell\times 2^\ell$ matrix whose $2^\ell$ eigenvalues are related the the $2\ell$  eigenvalues (for the generic XY chain) of the
correlation function. In this way, the entanglement entropy, as well as many entanglement  related quantities are easily extracted just by diagonalising
a matrix which is linear and not exponential in $\ell$.
Clearly this approach cannot be used for the trace distance because this requires the diagonalisation of the difference $\rho_A-\sigma_A$ and, usually,
the two RDMs do not commute.
For this reason, we rely on a brute-force approach that consists in explicitly constructing $\rho_A$ and $\sigma_A$ as $2^\ell\times 2^\ell$ matrices
as a Gaussian matrix (see the Appendix~\ref{appRXY} for details). Since RDMs are exponentially large in $\ell$ we can only access relatively small
subsystem sizes (up to $\ell\sim7$). Anyhow, compared to exact diagonalisation methods, we can consider arbitrarily large systems sizes $L$.

We will also consider the (Schatten) $n$-distances. When $n$ is even, this amounts just to consider products of RDM that can be manipulated with
standard correlation matrix techniques (cf. Ref. \cite{Fagotti:2010yr}).
Consequently, in this case we can very easily access subsystem of very large lengths. See Appendix \ref{appRXY} for details.
We stress that this methods cannot be applied to the (Schatten) $n$-distances with $n$ odd.

\begin{figure}[t]
  \centering
  \includegraphics[width=0.99\textwidth]{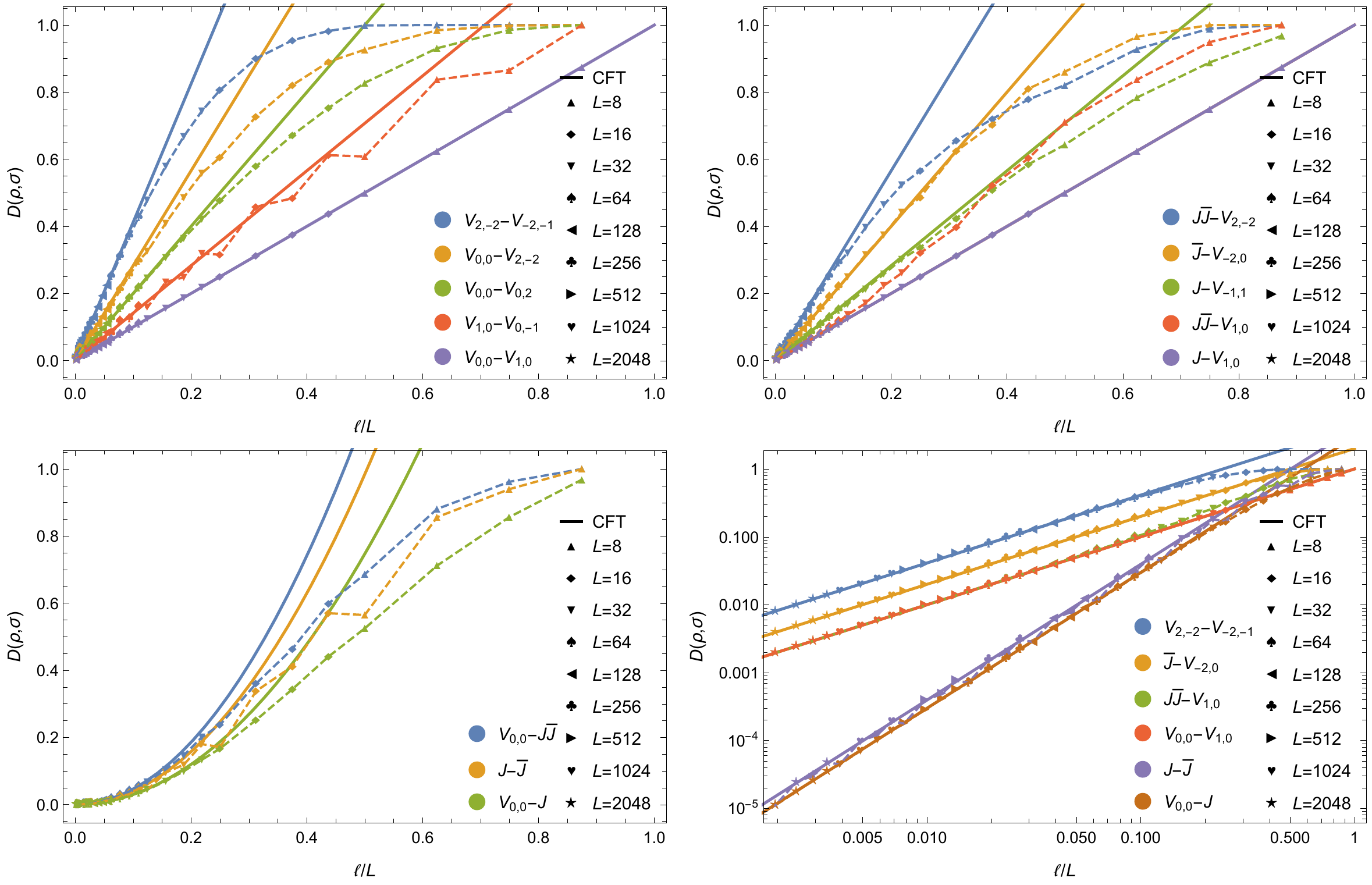}\\
  \caption{Trace distance $D(\r, \s)$ between the RDMs in several low-lying states as a function of the ratio between the subsystem $\ell$ and the system size $L$
  in the XX spin chain.
  The solid lines denote the leading order CFT prediction in the limit of short interval, Eqs.~(\ref{TDXXA1}), (\ref{TDXXA2}) and (\ref{TDXXA3}).
 The symbols joined by dashed lines represent numerical data, obtained with the method in Appendix \ref{appRXY}.
 Different symbols correspond to different $L$ and different colours correspond to different pairs of states.
  }
  \label{TDXX}
\end{figure}

In the remaining of this section we present our results for the trace distances among the RDMs of several low-lying excited states and
discuss their behaviour for $\ell\ll L$, comparing with the universal CFT prediction just obtained.
Our results for several representative states are reported in  Figure~\ref{TDXX}.
The various numerical data for the XX chain (symbols in the figure) perfectly match the  leading order CFT results obtained above (and full lines in the figure) for $\ell\ll L$.
Such agreement is highlighted in the fourth panel where the data are reported in log-log plots to make more evident the power law behaviour at small $\ell$.
Notice that in some (few) cases the first term in the short length expansion are equal, but the numerics surely rule out  the possibility that the entire scaling functions are the same.
This for example happens for the distances $D(\r_{J\bar J},\r_{{1,0}})$ and $D(\r_{J},\r_{{1,0}})$.

The numerical data reported in Figure~\ref{TDXX} can be used to complement the analytic CFT results obtained above.
Indeed for a few distances we have not been able to perform the analytic continuation to calculate exactly the amplitude $x_\phi$
appearing in the short length expansion \eqref{DrAsA}.
In such cases, matching Eq. \eqref{DrAsA} with the numerical results, we get approximately
\bea \label{TDXXA3}
&& D(\r_{0},\r_{J}) =
   D(\r_{0},\r_{\bar J}) =
    D(\r_{J},\r_{J\bar J}) =
   D(\r_{\bar J},\r_{J\bar J}) \approx 0.107 \f{ 2\sr{2}\pi^2\ell^2}{L^2} + o\Big(\f{\ell^2}{L^2}\Big), \nn\\
&& D(\r_{0},\r_{J\bar J}) \approx 0.166 \f{ 2\sr{2}\pi^2\ell^2}{L^2} + o\Big(\f{\ell^2}{L^2}\Big), \nn\\
&&   D(\r_{J},\r_{\bar J}) \approx 0.141 \f{ 2\sr{2}\pi^2\ell^2}{L^2} + o\Big(\f{\ell^2}{L^2}\Big).
\eea
Some of these results are also shown in Figure~\ref{TDXX}.
Comparison with (\ref{TDXXA31}) leads to $x_T\approx0.107$ (which
satisfies the bound (\ref{bound}) with $x_{\rm max}(2)=1/\sr{30}\approx 0.183$).

\subsection{$n$-distances for arbitrary subsystem size and analytic continuation} \label{secBND}

In this subsection, we consider the calculation of the $n$-distances for arbitrary $n$ and for arbitrary values of the ratio $\ell/L$, specialising the general approach
in Section \ref{secExact} to a few primary operators in the 2D free massless boson theory.
In a specific case we have also been able to obtain the analytic continuation in $n$ and find the exact expression of the trace distance for an interval of arbitrary length.
%

\subsubsection{Distances between vertex states}

\begin{figure}[tbp]
  \centering
  \includegraphics[width=0.99\textwidth]{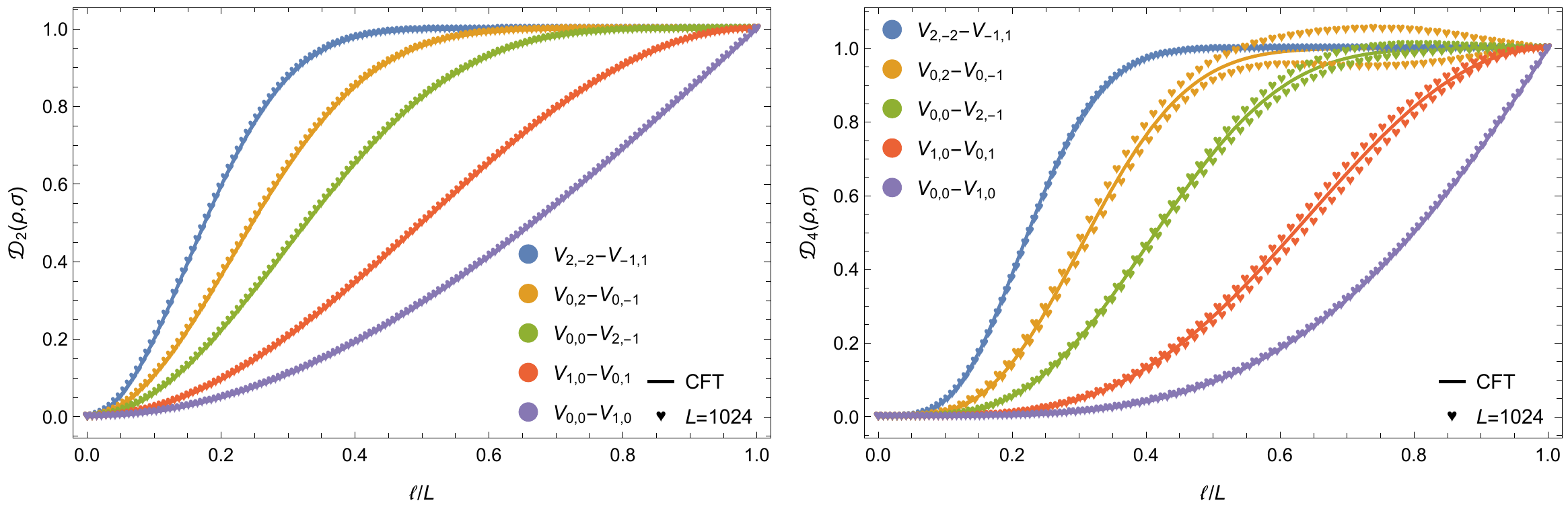}\\
  \caption{Even $n$-distance $\mD_n [\Delta \alpha]$ for $n=2,4$ as a function of the ratio between the subsystem $\ell$ and the system size $L$ in the free compact boson theory. The solid lines are the exact CFT predictions in (\ref{G2XXA1}).
The symbols are the numerical data for a system of size $L=1024$ and arbitrary $\ell$.
Different colours correspond to different pairs of vertex states $\r$ and $\s$ (several values of $\Delta \alpha$ are considered).}
  \label{XXD2D4}
\end{figure}

We first consider the distances between two states generated by vertex operators for which we can obtain many analytical results.
Specialising Eq.~\eqref{trAXY} to vertex operators and using the explicit form of the multipoint correlation functions of the vertices (see e.g.  \cite{DiFrancesco:1997nk}),
we straightforwardly obtain the general form for the $n$-distance with $n$ even
\be
\label{Hn}
\mD_n [\Delta \alpha] \equiv \mD_n (\r_{\a,\bar\a} , \r_{\a',\bar\a'} ) =
\frac{1}{2}                                               \sum_{k=0}^n (-)^k
                                               \sum_{0\leq j_1 < \cdots < j_k \leq n-1}      h_n(\{j_1,\cdots,j_k\})^{\Delta \alpha},
\ee
where we defined
\be
\Delta \alpha \equiv (\alpha - \alpha')^2 + (\bar{\alpha} - \bar{\alpha}' )^2,
\ee
and the function $h_n(\{j_1,\cdots,j_k\})$ of the set $\{j_1,\cdots,j_k\}$  as
\be \label{FnS}
h_n(\mS) = \Big( \f{\sin{\f{\pi\ell}{L}}}{n\sin{\f{\pi\ell}{n L}}} \Big)^{|\mS|}
         \prod^{j_1 < j_2}_{j_1,j_2 \in \mS}
         \f{\sin^2 \f{\pi (j_1 - j_2)}{n}}{\sin\f{\pi (j_1 - j_2 + \ell/L)}{n}\sin\f{\pi (j_1 - j_2 - \ell/L)}{n}}.
\ee
We stress that for an odd integer, $n=n_o$, Eq. \eqref{Hn} does not provide the $n_o$-distance.
Indeed, when  $n=n_o$ is an odd integer, using the identity (\ref{id1}) in Appendix~\ref{appAID}, we immediately have
the rhs of Eq. \eqref{Hn} vanishes identically. 
%
For even integer $n= n_e$, \eqref{Hn} is the (Schatten) $n_e$-distance  $\mD_{n_e}$.
Anyhow, the expression \eqref{Hn} as a sum of products of terms is not in the right form to be manipulated for the analytic continuation, but
it can be considerably simplified for the smallest even integers, leading to the compact expressions
\bea \label{G2XXA1}
 \mD_2 [\Delta \alpha] &=& 1 - \Big(\cos\f{\pi\ell}{2L}\Big)^{\Delta \alpha}, \nn\\
  \mD_4  [\Delta \alpha] &=& 1 + \Big(\cos^2\f{\pi\ell}{2L}\Big)^{\Delta \alpha}
                                            +2\Big(\cos^2\f{\pi\ell}{4L}\cos\f{\pi\ell}{2L}\Big)^{\Delta \alpha} ,
\eea
where we defined $  \mD_n [\Delta \alpha] \equiv \mD_{n}(\r_{\a,\bar\a},\r_{\a',\bar\a'})$.
These two expressions are consistent with the leading order results in short interval expansion obtained from (\ref{trArAmsAn}) and \eqref{tr-beyond}
\bea \label{G2XXA2}
&& \mD_2 [\Delta \alpha]= \Delta \alpha \f{\pi^2\ell^2}{8L^2}
                                                + o\Big(\f{\ell^2}{L^2}\Big) , \nn\\
&& \mD_4[\Delta \alpha] = (\Delta \alpha)^2 \f{9\pi^4\ell^4}{512L^4}
                                                + o\Big(\f{\ell^4}{L^4}\Big).
\eea
To get the above leading order results we used the coefficients $b_{JJ}=b_{\bar J\bar J}=-\f{1}{16}$ for $n=2$,
and $b_{JJJJ}=b_{\bar J\bar J\bar J\bar J}=\f{9}{4096}$, $b_{JJ\bar J\bar J}=\f{9}{2048}$ for $n=4$, which are easily read off from the results in \cite{Li:2016qbo}.
The predictions (\ref{G2XXA1}) can be tested against numerical computation in the XX spin chain for very large $\ell$ and $L$,
using the method of composition of Gaussian operators  \cite{Fagotti:2010yr} (see Appendix \ref{appRXY}).
The results are reported in Fig.~\ref{XXD2D4}: the agreement is excellent, although some oscillating subleading corrections to the scaling affect the data,
but the presence of such deviations is not unexpected since their presence is well known for entanglement related quantities \cite{ccen-10,ce-10,fc-11,cmv-11,cc-10}.
We checked carefully, by considering several values of $L$ and performing extrapolations, that indeed such pronounced oscillations go to zero in the thermodynamic limit.

The (Schatten) $n$-distance $\mD_n [\Delta \alpha]$ for general $n$ (also odd or non-integer) is obtained from the analytic continuation of $\mD_{n_e} [\Delta \alpha]$
from $n_e \in 2 \mathbb{N}$ to an arbitrary real number.
To obtain this analytic continuation, we need to rewrite \eqref{Hn} in such a way to remove the sum over the permutations.
We managed to do this only  for the special case $ \Delta \alpha = 1$.
Indeed, for $ \Delta \alpha = 1$, the scaling function \eqref{Hn} for an even integer $n_e$ may be rewritten (after some work) as
\be
\mD_{n_e}[1] = 2^{n_e -1} \prod_{j=1}^{n_e/2} \Big[ \sin\f{\pi(2j-1)x}{2n_e} \Big]^2,
\label{nnorm_prod}
\ee
where $x=\ell/L$.
This product formula is of the right form to obtain the analytic continuation.
Indeed, using the identity
\be
\log \sin(\pi s) = \log\pi - \int_0^{\inf} dt \f{\ep^{-t}}{t} \Big[ \f{\ep^{st}+\ep^{(1-s)t}-2}{1-\ep^{-t}} -1 \Big],
\ee
we get the analytic continuation to arbitrary $n$
\be \label{anacon}
\log 2\mD_{n}[1] =  n\log(2\pi) -  2\int_0^{\inf} dt \f{\ep^{-t}}{t}
\Big\{
\f{1}{1-\ep^{-t}}
\Big[
\f{(\ep^{\f{t x}{2}}-1)\big[ \ep^{\f{t x}{2{n} }} + \ep^{\big( 1 - \f{({n}-1)x}{2{n}} \big)t} \big]}
  {\ep^{\f{t x}{{n} }}-1}
- {n} \Big]
-\f{{n}}{2}
\Big\}.
\ee
%
In particular, for $n=1$, Eq. \eqref{anacon} simplifies dramatically to  (see Appendix \ref{appAC})
\be
D [\Delta \alpha= 1] = \f{\ell}{L}.
\label{lin}
\ee
Such simple expression tells us that the trace distance in this case is entirely determined by the leading OPE in Eq. (\ref{tdvv1}).
Then all the contributions from operators different from $J$ have OPEs with amplitudes that must vanish in the limit $n_e\to1$.
It is rather natural to wonder whether there is a deeper and general explanation of this fact and if there are other non trivial implications of this property
(not only for trace distances, but also for other quantities determined by the same OPE coefficients).
Other OPE amplitudes in fact vanish in the limit $n\to1$ \cite{Calabrese:2010he} and this has important consequences, e.g., for the entanglement
negativity \cite{Calabrese:2012ew,Calabrese:2012nk}.
Finally notice that the data for the trace distance $D [\Delta \alpha= 1]$ in Fig. \ref{TDXX} are perfectly compatible with the simple linear behaviour of Eq. (\ref{lin}).

\begin{figure}[tbp]
  \centering
  \includegraphics[width=0.99\textwidth]{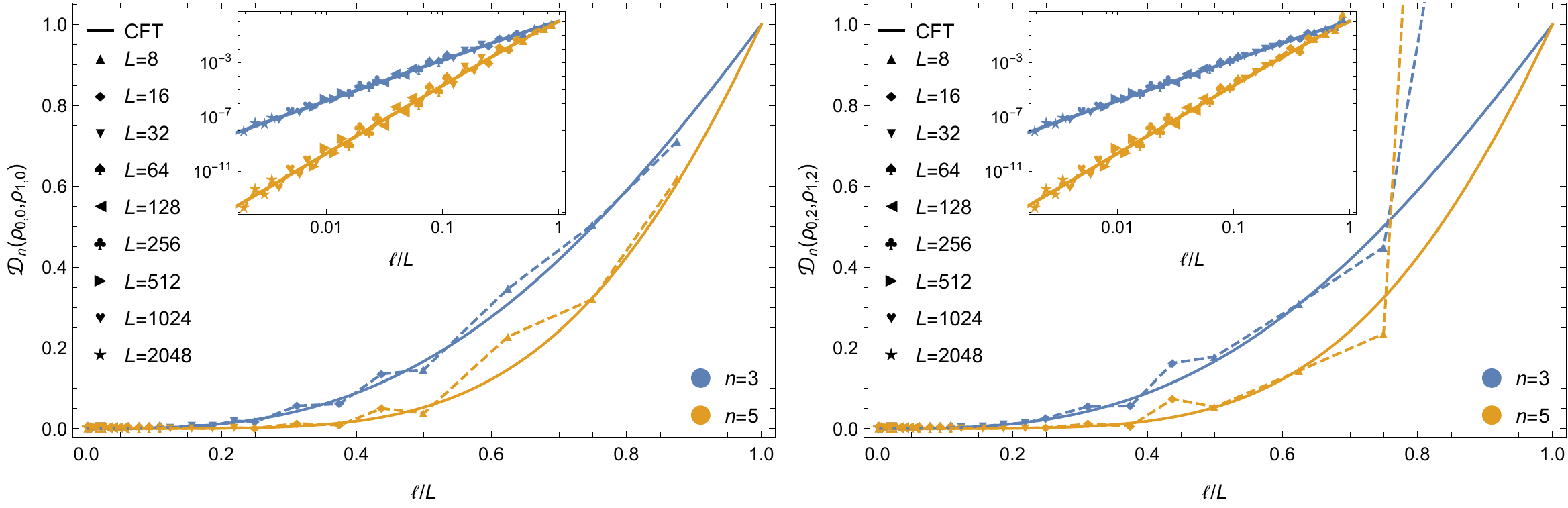}\\
  \caption{Odd $n$-distance $\mD_n (\r,\s)$ for $n=3$ (blue) and $n=5$ (yellow) as a function of the ratio between the subsystem $\ell$ and the system size $L$
  in the free compact boson theory.
  The solid lines are the analytic continuation (\ref{anacon}) of the CFT prediction.
  The symbols (joined by dashed lines) are numerical data, 
  with different symbols corresponding to different $L$.
  The two panels show two different pairs of vertex states $\r$ and $\s$, but  both with $\Delta \alpha =1$.
  Insets: Zoom in log-log scale of the region $\ell\ll L$.
    }
  \label{G35NXXN}
\end{figure}

The analytic continuation (\ref{anacon}) provides also non-trivial predictions for the $n$-distance $\mD_n [1]$, for arbitrary $n$.
We can test this prediction against the XX results which we obtained from the full RDMs as in Eq.~\eqref{bruteforce}.
In Figure~\ref{G35NXXN}, we report the results we obtained for  $n=3,5$.
The spin chain calculations and the analytic continuation (\ref{anacon}) match rather well, in spite of the presence of the
oscillating correction to the scaling \cite{ccen-10,ce-10,fc-11,cmv-11,cc-10}.
They can appear larger than those reported for even $n$ in Fig. \ref{XXD2D4}, but this is only due to the smaller values of $\ell$
we can access from the diagonalisation of the entire density matrix, compared to the correlation matrix technique used for even $n$.

\begin{figure}[tbp]
\centering
  \includegraphics[width=0.6\textwidth]{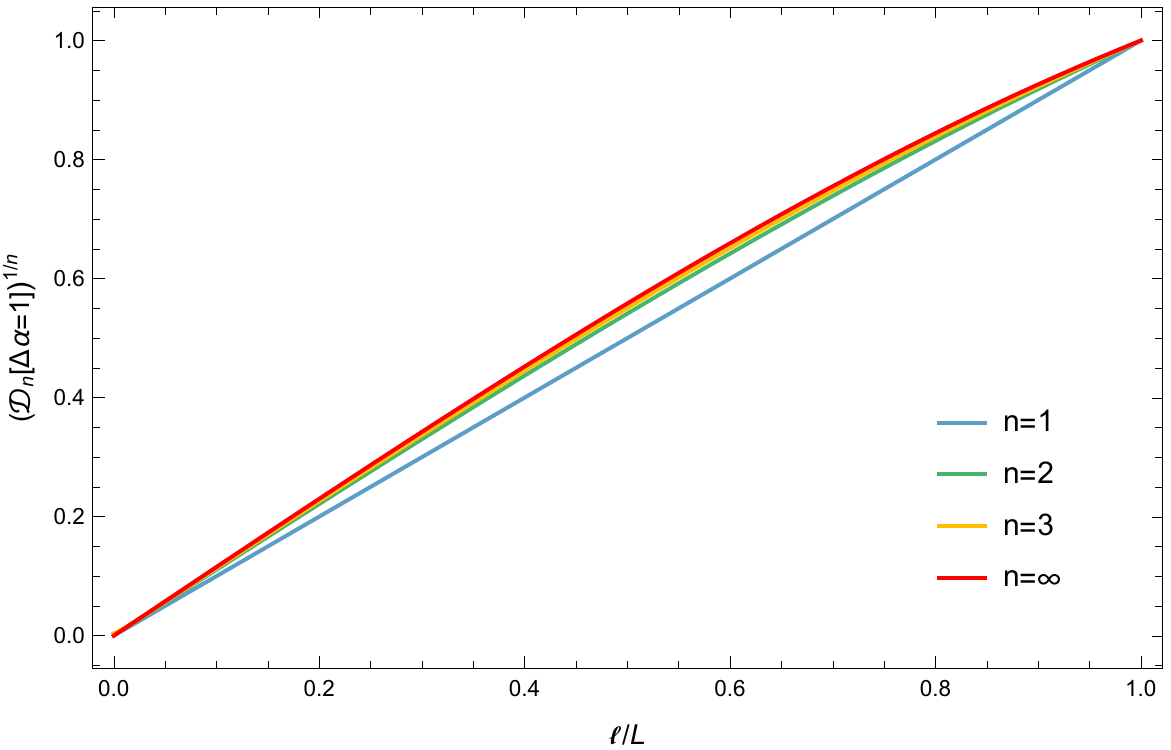}
  \caption{The scaling functions of the $n$-distances $ (\mD_{n}[1])^{1/n}$ as function of $\ell/L$ for $n=1,2,3,\infty$ for two vertex operator states
  with $\Delta\alpha=1$. Notice the monotonicity of the curves in both $n$ and $\ell/L$.}
  \label{Dnn}
\end{figure}

Finally, another limit in which \eqref{anacon} simplifies is for $n\to\infty$ when we get (see Appendix \ref{appAC}, Eq. \eqref{Dinf})
\be
\lim_{n\to\infty}\f{\log \mD_{n}[1]}{n} =\frac{2 \left(\zeta'\left(-1,1-\frac{x}{2}\right)-\zeta'\left(-1,\frac{x}{2}\right)\right)}{x},
\label{Dinft}
\ee
where $\zeta'(z,y)\equiv \partial_z\zeta(z,y)$ denotes the derivative of the generalised $\zeta$ function with respect to the first argument and $x=\ell/L$.
We plot $ (\mD_{n}[1]^{1/n})$ as function of $x$ for various $n$ in Figure \ref{Dnn}.
It is clear that the various curves are very close to each other, but always different.
Furthermore they are monotonous functions of $n$, i.e.  $ (\mD_{n'}[1]^{1/n'})> (\mD_{n}[1]^{1/n})$ if $n'>n$ and $x$.
By no means this implies that the various $n$-distances are equivalent:
the true $n$ norm, cf. Eq. \eqref{DD}, is obtained by multiplying $(\mD_{n}[1])^{1/n}$ by $(\tr \rho_0^n)^{1/n}$ that in the thermodynamic limit $L\to\infty$
goes to zero for any $n>1$. Hence all these $n$-distances are zero unless $n=1$.
Once again this fact shows that the trace distance is the most appropriate distance when one needs to compared subsystems of different sizes.

When $\Delta\a\neq 1$, we are not able to simplify the general expression \eqref{Hn} to a form useful for the analytic continuation without the sum over the permutation.
We only obtained few specific formulas. For example, for $\Delta\a=2$ and $n=6$, the general expression can be simplified to ($x=\ell/L$)
\begin{multline}
{\cal D}_6 [\Delta \alpha= 2]= \frac{4}{9} \Big(\sin\left(\frac{\pi  x}{6}\right) \Big)^6 \left(1+2 \cos \left(\frac{\pi x}{3}\right)\right)^2 \\ \times
   \left( 4\left( \sin \left(\frac{\pi  x}{3}\right)+2 \sin   \left(\frac{2 \pi  x}{3}\right)\right)^2
   +9 \left(1+2 \cos \left(\frac{\pi  x}{3}\right)+2 \cos \left(\frac{2 \pi  x}{3}\right)\right)^2\right),
\end{multline}
and a more cumbersome expression can be found for $n=8$, but the general structure (if it exists) is not understood yet.

\subsubsection{Distances involving current states}

The replicated distances involving current states have a much more complicated structure compared to the vertex states.
For this reason, we briefly focus here on the distance $D(\r_J, \r_{1,0})$ because the numerical data in Figure \ref{TDXX} strongly suggest
that this distance is exactly equal to $\ell/L$, as the distance between vertex operator with $\Delta\a=1$.

Using the correlation functions between current and vertex states, after long but straightforward algebra one arrives to
\bea
&& \f{\tr(\r_J - \r_{1,0})^n}{\tr\r_0^n} = \sum_{\cS\subseteq\cS_0}\sum_{\cR\subseteq\td\cS}
   \Big[ (-)^{|\bar\cS|}
         \f{(\f{1}{n}\sin\f{\pi\ell}{L})^{2n-|\bar\cS|}}{(\sin\f{\pi\ell}{nL})^{|\bar\cS|-|\bar\cR|}}
         \Big( \det_{r_1,r_2\in\cR}^\sim \f{1}{\sin\f{\pi(r_1-r_2)}{n}} \Big) \nn\\
&& \phantom{\f{\tr(\r_J - \r_{1,0})^n}{\tr\r_0^n} =} \times
         \Big( \prod_{\bar s_1,\bar s_2\in\bar\cS}^{\bar s_1 < \bar s_2}
               \f{\sin^2 \f{\pi (\bar s_1 - \bar s_2)}{n}}{\sin\f{\pi (\bar s_1 - \bar s_2 + \ell/L)}{n}\sin\f{\pi (\bar s_1 - \bar s_2 - \ell/L)}{n}} \Big) \nn\\
&& \phantom{\f{\tr(\r_J - \r_{1,0})^n}{\tr\r_0^n} =} \times
        \Big( \prod_{\bar r \in \bar \cR} \sum_{\bar s \in \bar \cS} \f{1}{\sin\f{\pi (\bar r - \bar s)}{n}\sin\f{\pi (\bar r - \bar s - \ell/L)}{n}} \Big)
   \Big].
 \label{DJr}
\eea
Note that the sum of the set $\mS$ is over all the subsets of $\mS_0=\{0,1,\cdots,n-1\}$, and the complement set is $\bar \cS = \cS_0/\cS$.
The sum of $\mR$ is over all the subsets of $\td \cS = \cS \cup ( \cS + \f{\ell}{L} )$, and the complement set is $\bar \cR = \td\cS/\cR$.
The determinant $\overset{\sim}{\det}$ is for the matrix whose diagonal entries are vanishing.

Unfortunately, further simplifications appear very difficult.
However, this general form is enough to rule out that the Schatten $n$-distances ${\cal D}_n(\r_J, \r_{1,0})$ and ${\cal D}_n(\r_0 , \r_{1,0})$ are equal:
it is  enough to calculate the two distances for some  even integer $n$.
For example, for $n_e=2$ we have ($x=\ell/L$)
\be
{\cal D}_2 [J, V_{1,0}]= \frac{1}{2} \left(1-\sin ^3\left(\frac{\pi  x}{2}\right) \sin (\pi  x)-2 \cos
   ^3\left(\frac{\pi  x}{2}\right)+\frac{1}{64} (\cos (2 \pi  x)+7)^2\right),
\ee
which is different from \eqref{G2XXA1} with $\Delta\a=1$.
However, the differences between these distances are rather small and of higher order in $x$ (e.g.  ${\cal D}_2 [J, V_{1,0}]-{\cal D}_2 [V_{0,0}, V_{1,0}]=O(x^6)$).
The same seems true for higher $n$. Given the present state of affairs we are not able to distinguish whether $D(\r_J, \r_{1,0})$ is equal to $\ell/L$ or just very close to it:
the analytic continuation of Eq. \eqref{DJr} seems too complicated to solve this issue.

\subsection{Application of the OPEs to Relative entropies and fidelities}

The OPEs of twist fields that we employed for the trace distances can be used also to derive some new results for the relative entropies and fidelities
that can be tested against exact computations in the XX spin chain (generalising the results in  \cite{Ruggiero:2016khg,Nakagawa:2017fzo}).

\subsubsection{Relative entropy} \label{secBRE}

The relative entropies between different CFT states have been already considered in the literature.
In particular the relative entropy between vertex operators is   \cite{Lashkari:2014yva,Lashkari:2015dia}
\be \label{REXXA1}
S(\r_{\a,\bar\a}\|\r_{\a',\bar\a'}) = [(\a-\a')^2+(\bar\a-\bar\a')^2]\Big( 1- \f{\pi\ell}{L}\cot\f{\pi\ell}{L} \Big),
\ee
while the one between current and vertex is \cite{Ruggiero:2016khg}
\bea \label{REXXA2}
&& S(\r_J\|\r_{\a,\bar\a}) = S(\r_{\bar J}\|\r_{\a,\bar\a})
                           =  (2 + \a^2+\bar\a^2)\Big( 1- \f{\pi\ell}{L}\cot\f{\pi\ell}{L} \Big) \nn \\
&& \phantom{S(\r_J\|\r_{\a,\bar\a}) = S(\r_{\bar J}\|\r_{\a,\bar\a})=}
                            + 2 \Big[ \sin \f{\pi\ell}{L}
                            + \log\Big( 2 \sin \f{\pi\ell}{L} \Big)
                            + \psi\Big(\f12 \csc\f{\pi\ell}{L}\Big) \Big],
\eea
with $\psi$ denoting the digamma function.
Actually the same correlation functions already derived in \cite{Ruggiero:2016khg} also determine the relative entropy between $J\bar J$ and the vertex as
\be \label{REXXA3}
S(\r_{J\bar J}\|\r_{\a,\bar\a}) = (4 + \a^2+\bar\a^2)\Big( 1- \f{\pi\ell}{L}\cot\f{\pi\ell}{L} \Big)
                            + 4 \Big[ \sin \f{\pi\ell}{L}
                            + \log\Big( 2 \sin \f{\pi\ell}{L} \Big)
                            + \psi\Big(\f12 \csc\f{\pi\ell}{L}\Big) \Big].
\ee
We also obtain
\be \label{REXXA3x}
S(\r_{J \bar J}\|\r_{J}) = S(\r_{J \bar J}\|\r_{\bar J})
                           =  2 \Big( 1- \f{\pi\ell}{L}\cot\f{\pi\ell}{L} \Big)
                            + 2 \Big[ \sin \f{\pi\ell}{L}
                            + \log\Big( 2 \sin \f{\pi\ell}{L} \Big)
                            + \psi\Big(\f12 \csc\f{\pi\ell}{L}\Big) \Big].
\ee
The leading order of the relative entropies (\ref{REXXA1}), (\ref{REXXA2}), (\ref{REXXA3}) and (\ref{REXXA3x}) are
\bea \label{REXXA4}
S(\r_{\a,\bar\a}\|\r_{\a',\bar\a'}) &=& [(\a-\a')^2+(\bar\a-\bar\a')^2]\f{\pi^2\ell^2}{3L^2} + o\Big( \f{\ell^2}{L^2} \Big),\\
 \label{REXXA5}
S(\r_{J}\|\r_0) &=& S(\r_{\bar J}\|\r_0) = S(\r_{J \bar J}\|\r_J) = S(\r_{J\bar J}\|\r_{\bar J}) =  \f{8\pi^4\ell^4}{15L^4} + o\Big( \f{\ell^4}{L^4} \Big), \\
S(\r_{J\bar J}\|\r_0) &=& \f{16\pi^4\ell^4}{15L^4} + o\Big( \f{\ell^4}{L^4} \Big),\\
 \label{REXXA6}
 S(\r_{J}\|\r_{\a,\bar\a}) &=& S(\r_{\bar J}\|\r_{\a,\bar\a})=  (\a^2+\bar\a^2)\f{\pi^2\ell^2}{3L^2} + o\Big( \f{\ell^2}{L^2} \Big),\\
 S(\r_{J\bar J} \| \r_{\a,\bar\a}) &=& (\a^2+\bar\a^2)\f{\pi^2\ell^2}{3L^2} + o\Big( \f{\ell^2}{L^2} \Big)
\eea
and they coincide with the general prediction in (\ref{SUHLZ}).
There are other cases in which we do not know the exact form of the relative entropies, but nevertheless we can use (\ref{SUHLZ}) to
get leading order results in short interval expansion
\bea \label{REXXA7}
&& S(\r_0\|\r_{J}) = S(\r_0\|\r_{\bar J}) =S(\r_J\|\r_{J \bar J}) = S(\r_{\bar J}\|\r_{J\bar J}) = \f{8\pi^4\ell^4}{15L^4} + o\Big( \f{\ell^4}{L^4} \Big), \nn\\
&&   S(\r_0\|\r_{J\bar J}) = \f{16\pi^4\ell^4}{15L^4} + o\Big( \f{\ell^4}{L^4} \Big),
~~ S(\r_{\a,\bar\a}\|\r_{J}) = S(\r_{\bar\a,\a}\|\r_{\bar J})= (\a^2+\bar\a^2)\f{\pi^2\ell^2}{3L^2} + o\Big( \f{\ell^2}{L^2} \Big), \nn\\
&& S(\r_{\a,\bar\a}\|\r_{J\bar J}) = (\a^2+\bar\a^2)\f{\pi^2\ell^2}{3L^2} + o\Big( \f{\ell^2}{L^2} \Big), ~~
   S(\r_J\|\r_{\bar J}) = S(\r_{\bar J}\|\r_J) = \f{16\pi^4\ell^4}{15L^4} + o\Big( \f{\ell^4}{L^4} \Big).
\eea

\begin{figure}[tbp]
  \centering
  \includegraphics[width=0.99\textwidth]{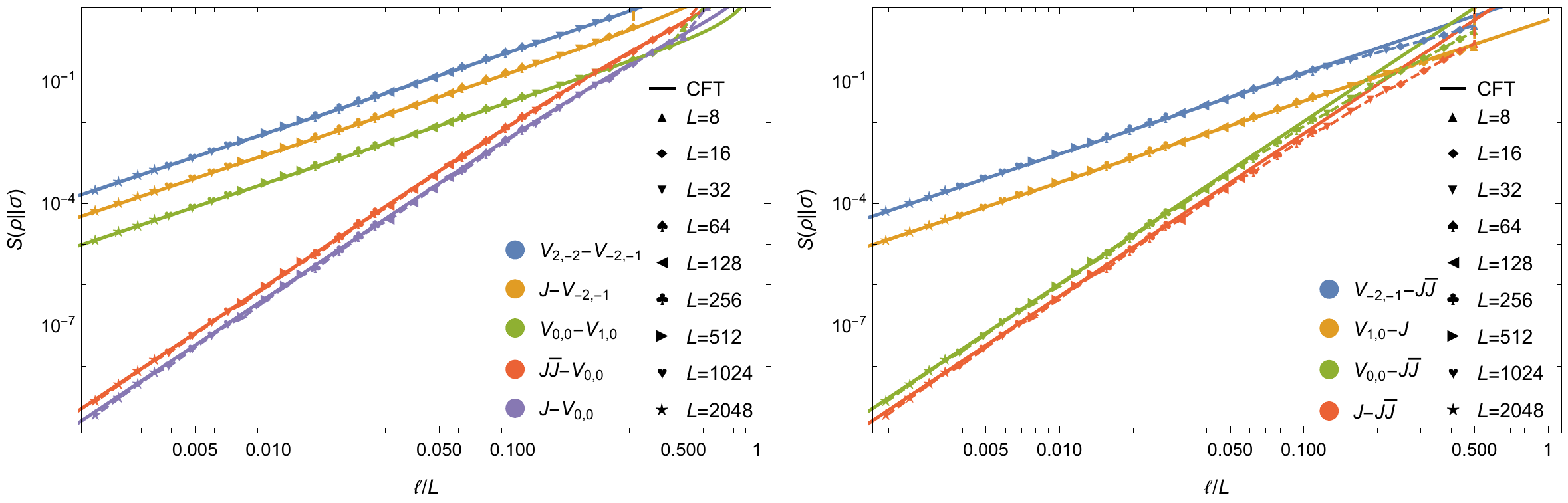}\\
  \caption{Relative entropy $S(\r \| \s)$ as a function of the ratio between the subsystem $\ell$ and the system size $L$ in the free compact boson.
  Solid lines are the CFT  prediction for short distance, Eqs.~(\ref{REXXA4}), (\ref{REXXA5}) and (\ref{REXXA6}).
  The symbols joined by dashed lines represent numerical data (obtained using Eq.~\eqref{bruteforce}),
    with different symbols corresponding to different $L$.
    Different colours correspond to different pairs of states $\r$ and $\s$.
  }
  \label{XXRE}
\end{figure}

On the spin chain side, we calculate the relative entropies directly from the RDMs using Eq.~\eqref{bruteforce}.
The obtained results are reported in Figure~\ref{XXRE} and they perfectly match the CFT predictions.

\subsubsection{Fidelity} \label{secBFD}

The fidelity $F(\r,\s)$ between RDMs of low-lying states in 2D CFT has been already investigated in \cite{Lashkari:2014yva} where it has been shown that
it can be rewritten as
\be
F(\r,\s) = \ep^{-\f12 S_{1/2}(\r\|\s)},
\label{FvsS}
\ee
in terms of the  R\'enyi relative entropy \cite{Wilde:2014eda,muller2013quantum}
\be
\label{renyiRel}
S_p(\r\|\s) = \f{1}{p-1} \log \tr \Big[ \Big( \s^{\f{1-p}{2p}} \r \s^{\f{1-p}{2p}} \Big)^p \Big].
\ee

\begin{figure}[t]
  \includegraphics[width=0.97\textwidth]{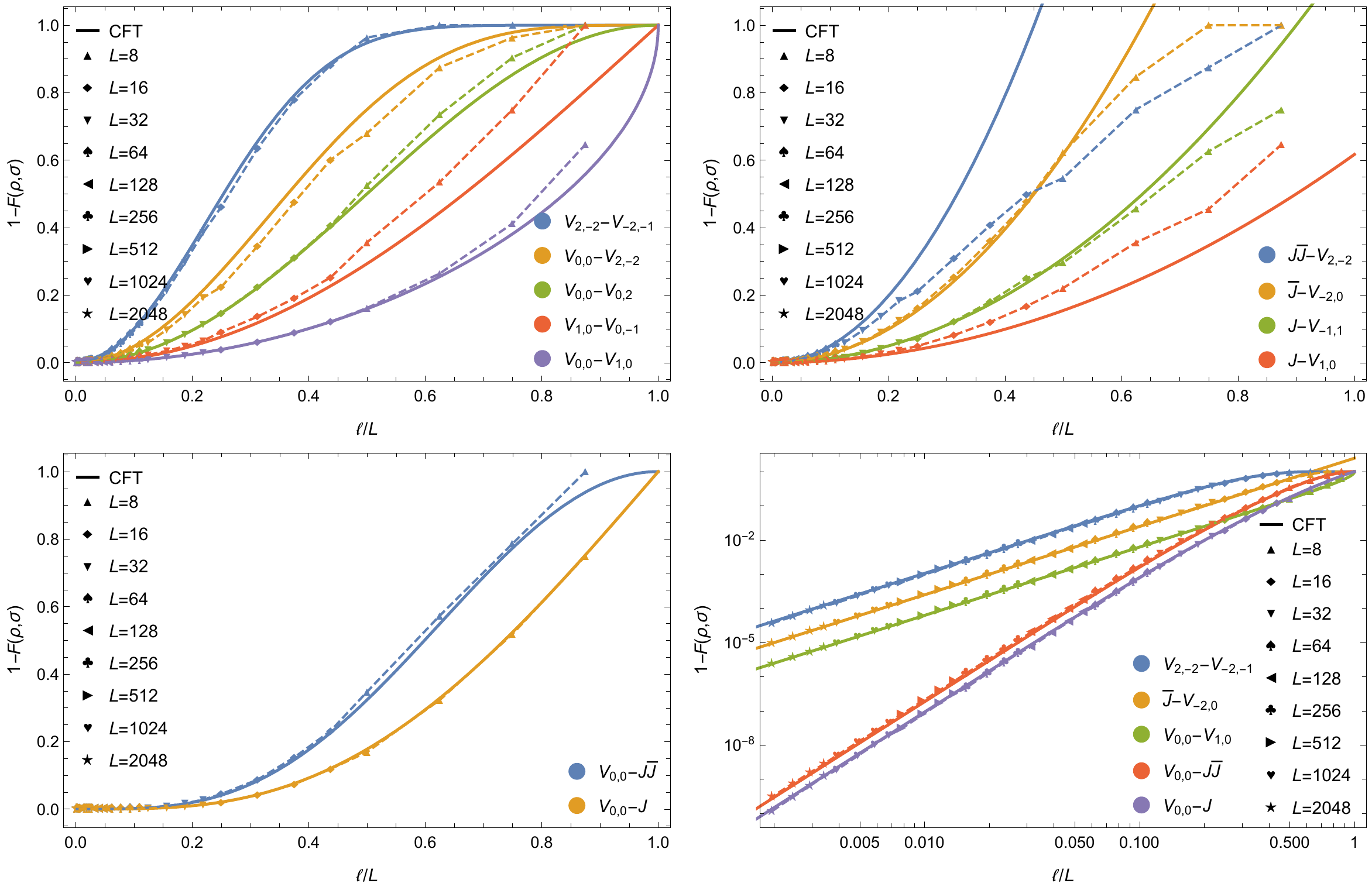}\\
  \caption{Fidelity $ F( \rho,\s)$ as a function of the ratio between the subsystem $\ell$ and the system size $L$ in the free compact boson.
  Solid lines are the CFT  predictions for short distance, Eqs.~(\ref{FDXXA1.1}) to (\ref{FDXXA2.2}).
    The symbols joined by dashed lines represent numerical data, with different symbols corresponding to different $L$.
Different colours correspond to different pairs of states $\r$ and $\s$.
    }
  \label{XXFD}
\end{figure}

The  R\'enyi relative entropy  between a generic primary operator $\phi$ and the ground state \cite{Lashkari:2014yva}
can be rewritten in terms of the function $F_\phi^{(p)}(x)$ in Eq. (\ref{Fphipell}) as
\be \label{RREofL}
S_p(\r_\phi\|\r_0) = \f{1}{p-1} \log\Big[ \Big( \f{p \sin\f{\pi\ell}{L}}{\sin\f{p\pi\ell}{L}} \Big)^{2p\D_\phi}
                                          F_\phi^{(p)}(p\ell) \Big].
\ee
Combing the last equation with \eqref{FvsS} we can get the various fidelities in terms of the function $F_\phi^{(1/2)}(1/2)$.
For example, by using $F_{V_{\a,\bar\a}}^{(p)}(\ell/L)=1$ \cite{Berganza:2011mh} in (\ref{RREofL}), we recover the result in \cite{Lashkari:2014yva}
\be \label{FDXXA1.1}
F(\r_0,\r_{\a,\bar\a}) = \Big( \cos\f{\pi\ell}{2L} \Big)^{\f{\a^2+\bar\a^2}{2}}.
\ee
However, we can get many more new results without making any calculation.
Using Eq. (\ref{FJpell}), in fact, we immediately get
\bea \label{FDXXA2.1}
&& F(\r_0,\r_J) = F(\r_0,\r_{\bar J}) = \f{\G^2(\f{3+\csc\f{\pi\ell}{2L}}{4})}
                                          {\G^2(\f{1+\csc\f{\pi\ell}{2L}}{4})}
                                        2\sin\f{\pi\ell}{L}, \nn\\
&& F(\r_0,\r_{J\bar J}) = \f{\G^4(\f{3+\csc\f{\pi\ell}{2L}}{4})}
                            {\G^4(\f{1+\csc\f{\pi\ell}{2L}}{4})}
                          4\sin^2\f{\pi\ell}{L}.
\eea
Note that the short interval expansion of (\ref{FDXXA1.1}) and (\ref{FDXXA2.1}) gives
\be
F(\r_0,\r_{\a,\bar\a}) = 1-\Delta\a\f{\pi^2\ell^2}{16 L^2} + o\Big(\f{\ell^2}{L^2}\Big),~~F(\r_0,\r_J) = 1-\Delta\a\f{3\pi^4\ell^4}{32 L^4} + o\Big(\f{\ell^4}{L^4}\Big),
\ee
and consequently
\be \label{FDXXA3}
 F(\r_J,\r_{\a,\bar\a}) = F(\r_{\bar J},\r_{\bar \a,\a}) \approx  F(\r_{J\bar J},\r_{\a,\bar\a}) = 1-\Delta\a\f{\pi^2\ell^2}{16 L^2} + o\Big(\f{\ell^2}{L^2}\Big).
\ee

Several examples of the leading order results just derived are checked against spin chains in Figure~\ref{XXFD}.
Furthermore, from the numerical results, we conjecture the more general result
\be \label{FDXXA1.2}
F(\r_{\a,\bar\a},\r_{\a',\bar\a'}) = \Big( \cos\f{\pi\ell}{2L} \Big)^{\f{(\a-\a')^2+(\bar\a-\bar\a')^2}{2}},
\ee
as well as
\bea \label{FDXXA2.2}
&& F(\r_J,\r_{\bar J}) = \f{\G^4(\f{3+\csc\f{\pi\ell}{2L}}{4})}
                           {\G^4(\f{1+\csc\f{\pi\ell}{2L}}{4})}
                         4\sin^2\f{\pi\ell}{L},\nn\\
&& F(\r_J,\r_{J\bar J}) = F(\r_{\bar J},\r_{J\bar J}) = \f{\G^2(\f{3+\csc\f{\pi\ell}{2L}}{4})}
                                                          {\G^2(\f{1+\csc\f{\pi\ell}{2L}}{4})}
                                                        2\sin\f{\pi\ell}{L},
\eea
which perfectly match the numerics.

\section{Free massless fermion} \label{secFermion}

In this section, we consider the 2D free massless fermion theory.
It is a $c=\f12$ CFT and the continuous limit of the critical Ising spin chain (which is a special case of the
 XY spin chain with transverse field reviewed in Appendix~\ref{appRXY}).
We will calculate various trace distances, relative entropies, and fidelities in the fermion theory and Ising spin chain.
The calculations in the 2D free massless fermion theory and Ising spin chain parallel those in the 2D free massless boson theory and XX spin chain.
Therefore, our discussion will be very brief.

In the 2D free massless fermion theory, besides the ground state $|0\rag$, we consider the excited states generated by the primary operators $\sigma$, $\mu$ with conformal weights $(\f1{16},\f1{16})$, $\psi$ and $\bar\psi$ with conformal weights $(\f12, 0)$ and $(0,\f12)$, respectively, and $\ve$ whose conformal weights are instead given by $(\f12,\f12)$.
We work in units such that all the primary operators are normalised to $1$.

\subsection{Trace distance}

Here  we first focus on the short interval expansion. 
Using the known scaling function for the R\'enyi entropies in the state $\s$ and $\mu$
$F_\s^{(p)}(\ell) = F_\mu^{(p)}(\ell) =1$  \cite{Berganza:2011mh},
and exploiting Eq. (\ref{xphiFphipell}), we immediately get
\be
x_\s = x_\mu= \f{1}{2^{1/4}} \approx 0.841.
\ee
For the $\ve$ state, we instead have \cite{elc-13}
\be \label{Feppell}
F_\ve^{(p)}(\ell) = \Big( \f2p \sin\f{\pi\ell}{L} \Big)^{2p}
\f{\Gamma^2(\f{1+p+p\csc\f{\pi\ell}{L}}{2})}
  {\Gamma^2(\f{1-p+p\csc\f{\pi\ell}{L}}{2})},
\ee
which, using (\ref{Fphipell}), leads to
\be \label{xep}
x_\ve = \f1\pi \approx 0.318.
\ee
As a consistency check, the bound (\ref{bound}) is satisfied with
\be
x_{\rm{max}}(1/8) = \f{\pi^{1/4}}{2^{5/8}} \sr{\f{\G(9/8)}{\G(13/8)}} \approx 0.885,
\ee
for $x_\s, x_{\mu}$, and
\be
x_{\rm{max}}(1) = \f{1}{\sr{6}} \approx 0.408\dots,
\ee
for $x_\ve$.

\begin{figure}[t]
  \centering
  \includegraphics[width=0.99\textwidth]{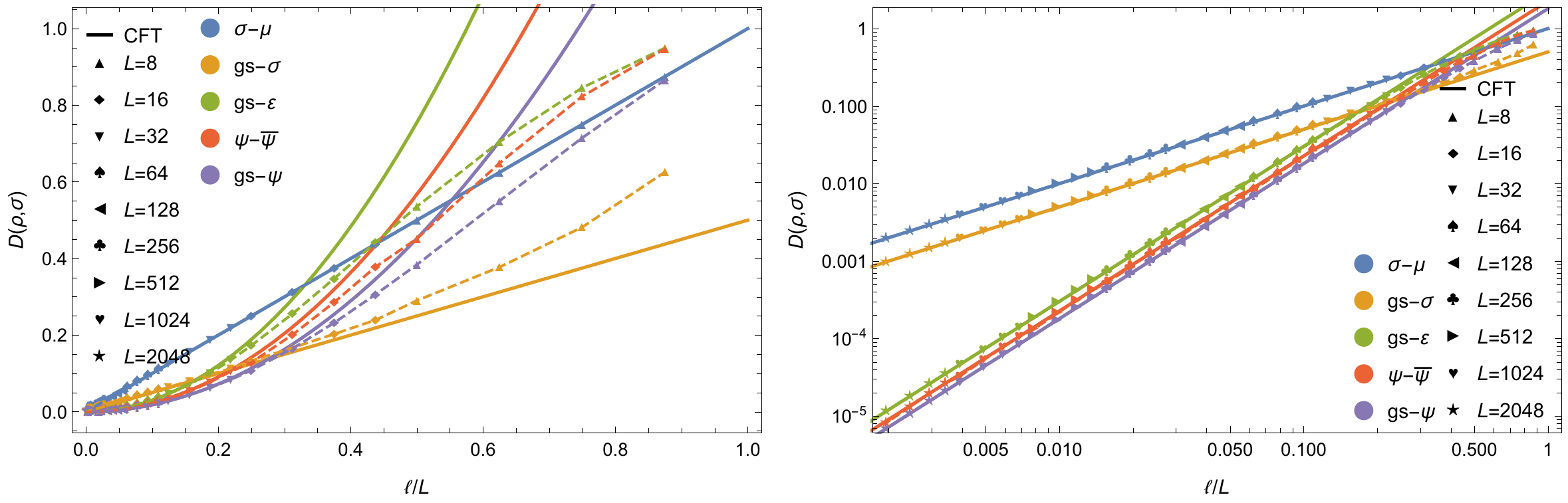}\\
  \caption{
  Trace distance $D(\r, \s)$ among the RDMs in different low-lying states as a function of the ratio between the subsystem $\ell$ and the system size $L$ in the free fermion theory.
  The solid lines denote the leading order CFT prediction in the limit of short interval, Eqs.~\eqref{TDIsingA1} and (\ref{TDIsingA3}).
    The symbols joined by dashed lines represent numerical data, with different symbols corresponding to different $L$.
    Different colours correspond to different pairs of states with ``gs'' denoting the ground state.
  }
  \label{IsingTD}
\end{figure}

Plugging the coefficient (\ref{xep})  and the expectation values
\be \label{epvevs}
\lag \ve \rag_0 = \lag \ve \rag_\psi = \lag \ve \rag_{\bar \psi} = \lag \ve \rag_\ve = 0, ~~
\lag \ve \rag_\s = - \lag \ve \rag_\m = \f{\pi}{L},
\ee
into the general formula (\ref{DrAsA}),
we obtain the leading order behaviour of the following trace distances
\bea \label{TDIsingA1}
&& D(\r_{0},\r_{\s}) = D(\r_{0},\r_{\m})= \f{\ell}{2L} + o\Big(\f{\ell}{L}\Big), \nn\\ &&
D(\r_{\s},\r_{\psi}) =  D(\r_{\s},\r_{\bar\psi}) = D(\r_{\m},\r_{\psi}) =  D(\r_{\m},\r_{\bar\psi})= \f{\ell}{2L} + o\Big(\f{\ell}{L}\Big), \nn \\
 &&  D(\r_{\s},\r_{\ve}) =D(\r_{\m},\r_{\ve}) = \f{\ell}{2L} + o\Big(\f{\ell}{L}\Big),
\eea
and
\be \label{TDIsingA3}
D(\r_{\s},\r_{\m}) = \f{\ell}{L} + o\Big(\f{\ell}{L}\Big).
\ee
Moreover, still from Eq.~(\ref{DrAsA}) and from the expectation values (\ref{epvevs}), (\ref{TXTbX}), we also get
\bea \label{tdgsps1}
&& D(\r_{0},\r_{\psi})  =
   D(\r_{0},\r_{\bar \psi}) =
   D(\r_{\psi},\r_{\ve}) =
   D(\r_{\bar \psi},\r_{\ve}) =x_T \f{ 2\pi^2\ell^2}{L^2} + o\Big(\f{\ell^2}{L^2}\Big),
\eea
with the unknown coefficients $x_T=x_{\bar T}$.

We checked several of the CFT results (\ref{TDIsingA1}) and (\ref{TDIsingA3}) against Ising spin chain numerics in Figure~\ref{IsingTD}.
From numerical spin chain results, we also get approximately (see again Figure~\ref{IsingTD})
\bea \label{TDIsingA4}
&& D(\r_{0},\r_{\psi}) =   D(\r_{0},\r_{\bar \psi}) = D(\r_{\psi},\r_\ve) =
   D(\r_{\bar \psi},\r_\ve) \approx 0.0916 \f{ 2 \pi^2\ell^2}{L^2} + o\Big(\f{\ell^2}{L^2}\Big), \nn\\
&& D(\r_{0},\r_\ve) \approx 0.153 \f{ 2 \pi^2\ell^2}{L^2} + o\Big(\f{\ell^2}{L^2}\Big),  ~~
   D(\r_{\psi},\r_{\bar \psi}) \approx 0.115 \f{ 2 \pi^2\ell^2}{L^2} + o\Big(\f{\ell^2}{L^2}\Big).
\eea
 Comparison with (\ref{tdgsps1}) leads to $x_T=x_{\bar T}\approx0.0916$, which satisfies the bound (\ref{bound}) with $x_{\rm max}(2)=1/\sr{30}= 0.183\dots$.

\subsection{An exact result}

The data in Figure \ref{IsingTD} strongly suggest that the distance  $D(\r_\s,\r_\m)$ is exactly $\ell/L$, i.e. completely fixed by the
first term in the OPE expansion. It is natural to wonder whether we can show this.
By replica trick, the stating point is always  $ {\tr(\r_\s - \r_\m)^n}$ that for free massless fermion theory can be computed by
bosonisation (see e.g. \cite{DiFrancesco:1997nk}). Using standard bosonisation rules ($\s^2=\sqrt2 \cos(\varphi/2)$ and $\m^2=\sqrt2  (\sin\varphi/2)$) and then the known
correlation functions of the vertex operators in the bosonic theory, after some long but easy algebra we get
\bea
&& \f{\tr(\r_\s - \r_\m)^n}{\tr\r_0^n} = \Big( \f{1}{4n} \sin\f{\pi\ell}{L} \Big)^{n/4}
                                         \sum_{\cS\subseteq\cS_0} \bigg\{ (-)^{|\cS|}
\Big\{
     (-)^{|\cS|}
     \sum_{\{r_i=\pm1,s_j=\pm1\}}^{\sum_i r_i + \sum_j s_j=0}
     \Big[
     \Big( \prod_{j \in {\td{\cS}}} s_j \Big) \nn\\
&& \phantom{\f{\tr(\r_\s - \r_\m)^n}{\tr\r_0^n} =} \times
     \Big( \prod_{i,i'\in{\td{\bar\cS}}}^{i<i'}\Big| \sin \f{\pi(i-i')}{n} \Big|^{r_i r_{i'}/2} \Big)
     \Big( \prod_{j,j'\in{\td{\cS}}}^{j<j'}\Big| \sin \f{\pi(j-j')}{n} \Big|^{s_j s_{j'}/2} \Big)
      \nn\\ && \phantom{\f{\tr(\r_\s - \r_\m)^n}{\tr\r_0^n} =} \times
     \Big( \prod_{i\in{\td{\bar\cS}},j\in\td\cS}\Big| \sin \f{\pi(i-j)}{n} \Big|^{r_i s_{j}/2} \Big)
     \Big]
\Big\}^{1/2} \bigg\}.
\eea
Note that the sum of the set $\mS$ is over all the subsets of $\mS_0=\{0,1,\cdots,n-1\}$, and the complement set is $\bar \cS = \cS_0/\cS$.
We also have $\td \cS = \cS \cup ( \cS + \f{\ell}{L} )$, $\td{\bar\cS} = \bar\cS \cup ( \bar\cS + \f{\ell}{L} )$.

Further simplifications of this formula appear very difficult. However it is straightforward to check numerically even for a quite large even integer $n$ that
\be
\f{\tr(\r_\s - \r_\m)^{n_e}}{\tr\r_0^{n_e}} = 2^{n_e} \prod_{j=1}^{n_e/2} \Big[ \sin\f{\pi(2j-1)\ell/L}{2n_e} \Big]^2.
\ee
This is exactly the same as the quantity in free massless boson theory in Eq. \eqref{nnorm_prod}.
Then, using the result for the analytic continuation in the previous section, we get the exact trace distance
\be
D(\r_\s,\r_\m) = \f{\ell}{L},
\ee
which in fact is exactly what the data in Fig. \ref{IsingTD} were suggesting.

\subsection{Relative entropy} \label{secFRE}

\begin{figure}[tbp]
  \centering
  \includegraphics[width=0.99\textwidth]{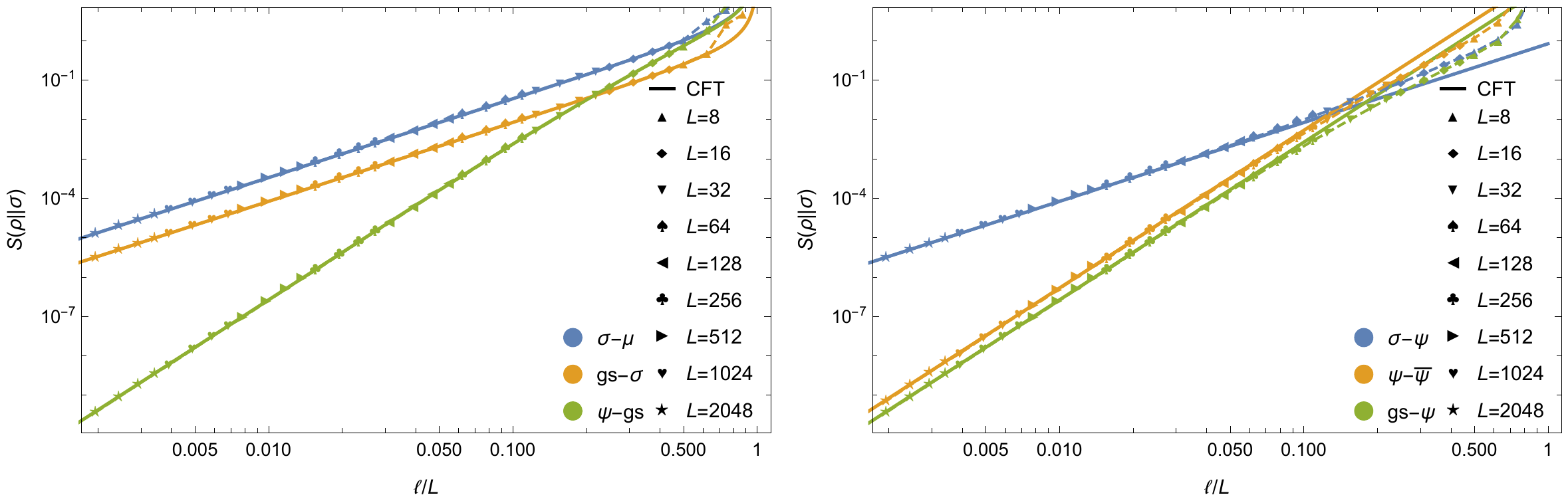}\\
  \caption{
  Relative entropy $S(\r \| \s)$ as a function of the ratio between the subsystem $\ell$ and the system size $L$ in the free fermion theory. Solid lines are the CFT short distance prediction, Eqs.~(\ref{REIsingA1}) to (\ref{REIsingA5}).
    The symbols joined by dashed lines represent numerical data, with different symbols corresponding to different $L$.
    Different colours correspond to different pairs of states $\r$ and $\s$.
  }
  \label{IsingRE}
\end{figure}

Some relative entropies in 2D free massless fermion theory have been calculated and checked against the numerical spin chain results
in \cite{Nakagawa:2017fzo}, using bosonisation and results for the free massless boson in \cite{Lashkari:2014yva,Lashkari:2015dia,Ruggiero:2016khg}.
Here, similarly, we use bosonisation, as well as the methods and results in \cite{Lashkari:2014yva,Lashkari:2015dia,Ruggiero:2016khg,Nakagawa:2017fzo},
in order to get further results.
We obtain the following relative entropies
\bea \label{REIsingA1}
&& S(\r_0\|\r_\s) = S(\r_\s\|\r_0)
 = S(\r_0\|\r_\m) = S(\r_\m\|\r_0)
 = \f14 \Big( 1- \f{\pi\ell}{L}\cot\f{\pi\ell}{L} \Big), \nn\\
&& S(\r_\s\|\r_\m) = S(\r_\m\|\r_\s) = 1- \f{\pi\ell}{L}\cot\f{\pi\ell}{L},
\eea
\be \label{REIsingA2}
 S(\r_\psi\|\r_0) = S(\r_{\bar\psi}\|\r_0) = S(\r_{\ve}\|\r_\psi) = S(\r_{\ve}\|\r_{\bar\psi}) =1- \f{\pi\ell}{L}\cot\f{\pi\ell}{L}
                                                 + \sin \f{\pi\ell}{L}
                                                 + \log\Big( 2 \sin \f{\pi\ell}{L} \Big)
                                                 + \psi\Big(\f12 \csc\f{\pi\ell}{L}\Big),
\ee
\be \label{REIsingA3}
S(\r_\ve\|\r_0) = 2 \Big( 1- \f{\pi\ell}{L}\cot\f{\pi\ell}{L} \Big)
                  + 2 \Big[ \sin \f{\pi\ell}{L}
                          + \log\Big( 2 \sin \f{\pi\ell}{L} \Big)
                          + \psi\Big(\f12 \csc\f{\pi\ell}{L}\Big) \Big],
\ee
\bea \label{REIsingA4}
&& S(\r_\psi\|\r_\s) = S(\r_\psi\|\r_\m) = S(\r_{\bar\psi}\|\r_\s) = S(\r_{\bar\psi}\|\r_\m)
 = \f54 \Big( 1- \f{\pi\ell}{L}\cot\f{\pi\ell}{L} \Big)
 + \sin \f{\pi\ell}{L} \nn\\
&& \phantom{S(\r_\psi\|\r_\s) = S(\r_\psi\|\r_\m) = S(\r_{\bar\psi}\|\r_\s) = S(\r_{\bar\psi}\|\r_\m) = }
 + \log\Big( 2 \sin \f{\pi\ell}{L} \Big)
 + \psi\Big(\f12 \csc\f{\pi\ell}{L}\Big), \\
&& S(\r_\ve\|\r_\s) = S(\r_\ve\|\r_\m) = \f{9}{4} \Big( 1- \f{\pi\ell}{L}\cot\f{\pi\ell}{L} \Big)
                    + 2 \Big[ \sin \f{\pi\ell}{L}
                            + \log\Big( 2 \sin \f{\pi\ell}{L} \Big)
                            + \psi\Big(\f12 \csc\f{\pi\ell}{L}\Big) \Big]. \nn
\eea
In particular for the relative entropies $S(\r_\s\|\r_0)$, $S(\r_\ve\|\r_0)$, $S(\r_\ve\|\r_\s)$,  we recover known results in \cite{Nakagawa:2017fzo}.
Some identities that are useful for the calculations of above relative entropies are collected in Appendix~\ref{appREF}.





In other cases, we were not able to obtain exact results. Nonetheless, using (\ref{SUHLZ}) and the expectation values (\ref{epvevs}), (\ref{TXTbX}), we can derive relative entropies at the leading order as
\bea \label{REIsingA5}
&& S(\r_0\|\r_{\psi}) = S(\r_0\|\r_{\bar \psi}) = 
S(\r_\psi\|\r_{\ve}) = S(\r_{\bar\psi}\|\r_{\ve}) =  \f{4\pi^4\ell^4}{15L^4} + o\Big( \f{\ell^4}{L^4} \Big) \nn\\.
&& S(\r_\s\|\r_{\psi}) = S(\r_\s\|\r_{\bar\psi}) =
   S(\r_\m\|\r_{\psi}) = S(\r_\m\|\r_{\bar\psi}) = \f{\pi^2\ell^2}{12L^2} + o\Big( \f{\ell^2}{L^2} \Big), \nn\\
   && S(\r_\s\|\r_{\ve}) = S(\r_\m\|\r_{\ve}) =  \f{\pi^2\ell^2}{12L^2} + o\Big( \f{\ell^2}{L^2} \Big), \nn\\
&&  S(\r_0\|\r_{\ve}) \approx S(\r_\psi\|\r_{\bar\psi}) = S(\r_{\bar\psi}\|\r_\psi) = \f{8\pi^4\ell^4}{15L^4} + o\Big( \f{\ell^4}{L^4} \Big),
\eea
We check some of the leading order relative entropies numerically in Figure~\ref{IsingRE}.

\begin{figure}[t]
  \includegraphics[width=0.95\textwidth]{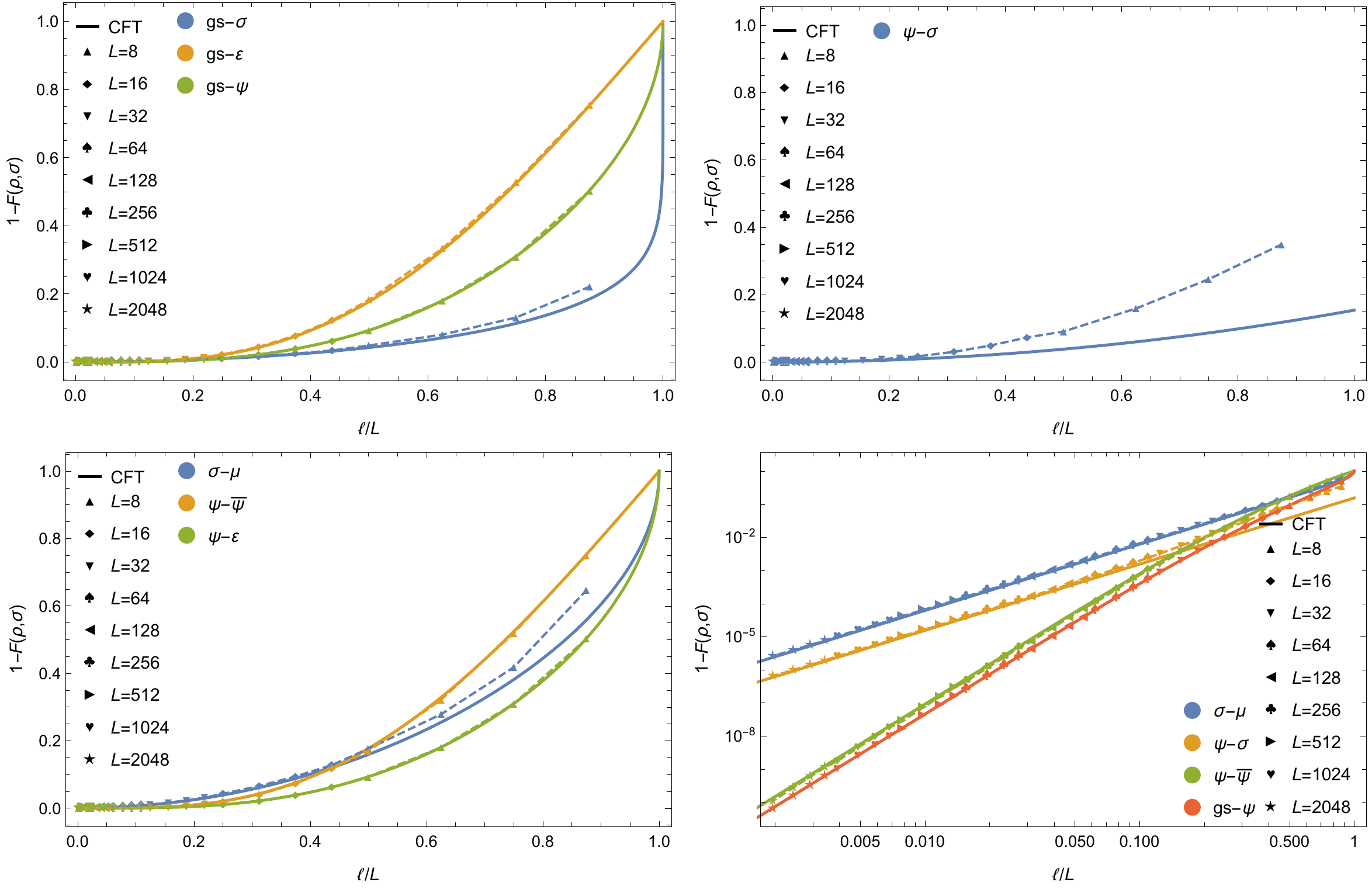}\\
  \caption{Fidelity $ F(\sigma, \rho)$ as a function of the ratio between the subsystem $\ell$ and the system size $L$ in the free fermion theory. Solid lines are the CFT predictions, Eqs.~(\ref{FDIsingA1}) to (\ref{FDIsingA3.2}).
     The symbols joined by dashed lines represent numerical data, 
      with different symbols corresponding to different $L$.
Different colours correspond to different pairs of states $\r$ and $\s$.
    }
  \label{IsingFD}
\end{figure}

\subsection{Fidelity} \label{secFFD}

As for the free boson, the fidelities in the fermion theory may be obtained plugging into Eq.~\ (\ref{RREofL}) the results
$F_\s^{(p)}(\ell) = F_\m^{(p)}(\ell) = 1$  for the R\'enyi entropies of Ref. \cite{Berganza:2011mh}, obtaining
\be \label{FDIsingA1}
F(\r_0,\r_\s) = F(\r_0,\r_\m) = \Big( \cos\f{\pi\ell}{2L} \Big)^{\f18}.
\ee
Similarly, using  $F_{\psi}^{(p)}(\ell) = F_{\bar\psi}^{(p)}(\ell) = \sr{F_\ve^{(p)}(\ell)}$ \cite{Berganza:2011mh} and (\ref{Feppell})  we get the fidelities
\bea \label{FDIsingA3.1}
&& F(\r_0,\r_\psi) = F(\r_0,\r_{\bar \psi}) = \f{\G(\f{3+\csc\f{\pi\ell}{2L}}{4})}
                                          {\G(\f{1+\csc\f{\pi\ell}{2L}}{4})}
                                        \sr{2\sin\f{\pi\ell}{L}}, \qquad 
F(\r_0,\r_{\ve}) = \f{\G^2(\f{3+\csc\f{\pi\ell}{2L}}{4})}
                                          {\G^2(\f{1+\csc\f{\pi\ell}{2L}}{4})}
                                        2\sin\f{\pi\ell}{L}.
\eea
For other pairs of states, instead, we  only get the leading order fidelities in short interval expansion
\bea \label{FDIsingA4}
&& F(\r_\psi,\r_\s) = F(\r_\psi,\r_\m) = F(\r_{\bar\psi},\r_\s) = F(\r_{\bar\psi},\r_\m) =
   1-\f{\pi^2\ell^2}{64 L^2} + o(\f{\ell^2}{L^2}), \nn\\
&& F(\r_\ve,\r_\s) = F(\r_\ve,\r_\m) = 1-\f{\pi^2\ell^2}{64 L^2} + o(\f{\ell^2}{L^2}).
\eea

We test these CFT predictions in Figure~\ref{IsingFD}.
Indeed, the numerical data allows us to conjecture the following forms
\be \label{FDIsingA2}
F(\r_\s,\r_\m) = \sqrt{\cos\f{\pi\ell}{2L}}.
\ee
and
\bea \label{FDIsingA3.2}
&& F(\r_\psi,\r_{\bar\psi}) = \f{\G^2(\f{3+\csc\f{\pi\ell}{2L}}{4})}
                                {\G^2(\f{1+\csc\f{\pi\ell}{2L}}{4})}
                              2\sin\f{\pi\ell}{L}, \nn\\
&& F(\r_\psi,\r_\ve) = F(\r_{\bar\psi},\r_\ve) = \f{\G(\f{3+\csc\f{\pi\ell}{2L}}{4})}
                                                   {\G(\f{1+\csc\f{\pi\ell}{2L}}{4})}
                                                 \sr{2\sin\f{\pi\ell}{L}},
\eea
that perfectly match the data, as shown in Figure~\ref{IsingFD}.

%
%
%

\section{Conclusion and discussion} \label{secCnD}

We  developed a systematic approach based on a replica trick to calculate the subsystem trace distance in one dimensional quantum systems and in particular 2D QFT.
We applied this method to the analytic computation of trace distances between the RDMs of one interval embedded in various low-lying energy eigenstates of  a CFT,
especially for free massless boson and fermion theories.
We obtained a full analytic result for the analytic continuation for arbitrary values of $\ell/L$ only in one case for the free bosonic theory
and another for the fermionic one.
For all other pairs of states, we have an analytic prediction only for the first term in the expansion in $\ell/L$.
We mention that, if needed, one might use known techniques for numerical analytic continuations (as e.g. in Refs. \cite{ahjk-14,dct-15}) to obtain the trace distances
from the analytically known $n$-distances for $n$ even.
We also calculated numerically the trace distances in XX and critical Ising spin chains, obtaining
perfect matches with the analytical CFT results.
We further check various analytical subsystem relative entropies and fidelities in the boson and fermion theories with the numerical spin chains results.

There is at least one aspect of our specific computations that can have important consequences also for different applications.
In fact, we have seen that there are RDMs of CFT eigenstates that have finite trace distances (and so local operators are not guaranteed to be the same in the two states),
but their (Schatten) $n$-distances, instead, vanish in the thermodynamic limit for all $n>1$.
In CFT, by means of scaling arguments, we are able to build from the $n$-norms some indicators that remain finite in the thermodynamic limit (see e.g. Eq. \eqref{Gn}),
but in a more general case (e.g. in the absence of scale invariance) it is not clear whether this is possible.
It is then natural to wonder whether some of the conclusions based on the analysis of other distances (as e.g. in Refs. \cite{fagotti2013reduced,mgdr-19})
could change if one  uses a more appropriate indicator such as the trace distance.

There are several immediate possible generalisations to the present work.
First of all one can consider other states in CFT: open systems \cite{txas-13,top-16}, disjoint intervals \cite{u-17}, finite temperature,
inhomogeneous systems \cite{Murciano:2018cfp,dsvc-17}, etc.
Secondly, one can consider subsystem trace distances in 2D massive theories \cite{Cardy:2007mb}.
Another interesting application  is related to the study of lattice entanglement Hamiltonians and their relation to
the Bisognano-Wichmann ones \cite{mgdr-19,gmcd-18,dvz-17,bw-76,ep-17,pa-18}.
Besides, one can  consider higher dimensional boson and fermion theories, trying to adapt the techniques of
Refs. \cite{Cardy:2013nua,Hung:2014npa,Chen:2016mya,Chen:2017hbk,Chen:2017yns}, at least in the small subsystem limit.

\section*{Acknowledgments}

We thank Bin Chen, Song Cheng, Jerome Dubail, and Erik Tonni for helpful discussions.
Part of this work has been carried out during the workshop  ``Entanglement in quantum systems'' at the Galileo Galilei Institute (GGI) in Florence, during
the workshop ``Quantum Information and String Theory 2019'', and during the ``It from Qubit School/Workshop'' at the Yukawa Institute for Theoretical Physics at Kyoto University.
All authors acknowledge support from ERC under Consolidator grant number 771536 (NEMO).

\appendix

\section{Review of XY spin chain} \label{appRXY}

The XY model with transverse field is defined by the Hamiltonian
\be \label{HS}
H = - \sum_{l=1}^L \Big( \f{1+\g}{4}\s_l^x\s_{l+1}^x + \f{1-\g}{4}\s_l^y\s_{l+1}^y + \f{\l}{2}\s_l^z \Big),
\ee
with $\s_l^{x,y,z}$ denoting the Pauli matrices and $L$ the total number of sites in the spin chain.
One can impose either periodic boundary conditions (PBC) as $\s_{L+1}^{x,y,z}=\s_{1}^{x,y,z}$, or anti-periodic boundary conditions (APBC)
as $\s_{L+1}^{x,y}=-\s_{1}^{x,y}$, $\s_{L+1}^{z}=\s_{1}^{z}$.
When $\g=0$ it defines the XX spin chain,  while  for $\g=1$ the Ising spin chain which is critical for $\l=1$.

The Hamiltonian (\ref{HS}) can be mapped to free fermions and exactly diagonalised \cite{Lieb:1961fr,pfeuty1970one}, as we will briefly review.
For further details, especially for the aspects of interest for this paper, see, e.g., \cite{Berganza:2011mh,calabrese2012quantum} and references therein.
We will also review the calculations of the entanglement entropies, R\'enyi entropies, and RDMs in the ground and low-lying excited states
in \cite{chung2001density,peschel2003calculation,Vidal:2002rm,Alba:2009th,Alcaraz:2011tn,Berganza:2011mh}.

The Hamiltonian (\ref{HS})  is mapped to free fermions by the Jordan-Wigner transformation
\be
 a_l = \Big(\prod_{j=1}^{l-1}\s_j^z\Big)\s_l^+, ~~
   a_l^\dag = \Big(\prod_{j=1}^{l-1}\s_j^z\Big)\s_l^-,
 \ee
where
$\s_l^\pm = \f12 ( \s_l^x \pm \ii \s_l^y )$.
By Fourier transforming, we get
\be
 b_k = \f{1}{\sr{L}}\sum_{l=1}^L\ep^{\ii l \vph_k}a_l, ~~
   b_k^\dag = \f{1}{\sr{L}}\sum_{l=1}^L\ep^{-\ii l \vph_k}a_l^\dag,
\ee
with $\vph_k = \f{2\pi k}{L}$.
We also consider two different boundary conditions for $a_l$, $a_l^\dag$: the APBC $a_{L+1}=-a_1$, $a_{L+1}^\dag=-a_1^\dag$, corresponding to the Neveu-Schwarz (NS) sector, and the PBC $a_{L+1}=a_1$, $a_{L+1}^\dag=a_1^\dag$, giving rise instead to the Ramond (R) sector.
The momenta $k$'s are half integers in the NS sector
\bea
&& {\rm odd~integer~} L: ~ k =1-\f{L}{2}, \cdots, -\f12, \f12, \cdots, \f{L}{2}-1, \f{L}{2}, \nn\\
&& {\rm even~integer~} L: ~ k =\f{1-L}{2}, \cdots, -\f12, \f12, \cdots, \f{L-1}{2},
\eea
and integers  in the R sector
\bea
&& {\rm odd~integer~} L: ~ k =\f{1-L}{2}, \cdots, -1, 0, 1, \cdots, \f{L-1}{2}, \nn\\
&& {\rm even~integer~} L: ~ k =1-\f{L}{2}, \cdots, -1, 0, 1, \cdots, \f{L}{2}-1, \f{L}{2}.
\eea
Hence, one has totally four sectors. It is useful to define the parity operator
\be \label{parity}
P = \exp\Big(\pi\ii\sum_{l=1}^L a_l^\dag a_l\Big) = \exp\Big(\pi\ii\sum_k b_k^\dag b_k\Big),
\ee
so that the four sectors are parametrised as
\bea
&& {\rm PNS~sector~with~} P = 1, \nn\\
&& {\rm APNS~sector~with~} P =-1, \nn\\
&& {\rm PR~sector~with~} P =-1, \nn\\
&& {\rm APR~sector~with~} P = 1.
\eea
For each of the four sectors, one can write the Hamiltonian as
\be \label{Hb}
H = \sum_k \Big[ (\l - \cos\vph_k) \Big( b_k^\dag b_k - \f12 \Big)
                 + \f{\ii\g}{2} ( b_k^\dag b_{-k}^\dag + b_k b_{-k} ) \Big].
\ee
In the PNS sector, one selects the states with $P=1$, and similarly in the other three sectors.

To diagonalise \eqref{Hb}, a further Bogoliubov transformation is needed. For $k\neq 0$ and $k\neq L/2$, the Bogoliubov transformation is
\be \label{BTransf}
c_k = b_k \cos\f{\th_k}{2} + \ii b_{-k}^\dag \sin\f{\th_k}{2}, ~~
c_k^\dag = b_k^\dag \cos\f{\th_k}{2} - \ii b_{-k} \sin\f{\th_k}{2}.
\ee
The parameter $\th_k \in (-\pi,\pi]$ is determined by
\bea
&& \sin\th_k = \f{\g\sin\vph_k}{\ve_k}, ~~
   \cos\th_k = \f{\l - \cos\vph_k}{\ve_k}, \nn\\
&& \ve_k = \sr{(\l - \cos\vph_k)^2+\g^2\sin^2\vph_k}.
\eea
For $k=0$ and $k=L/2$ one also has
\bea
&& c_0=b_0, ~~ c_0^\dag=b_0^\dag, ~~ \ve_0 = \l-1, \nn\\
&& c_{L/2}=b_{L/2}, ~~ c_{L/2}^\dag=b_{L/2}^\dag, ~~ \ve_{L/2} = \l+1.
\eea
After the Bogoliubov transformation, the Hamiltonian is diagonal
\be
H = \sum_k \ve_k \Big( c_k^\dag c_k - \f12 \Big).
\ee
Note that for $\g=0$, the Hamiltonian \eqref{Hb} is already diagonal and the Bogoliubov transformation is not needed.

The calculation of entanglement entropy in the ground state of the spin chain was developed in \cite{chung2001density,peschel2003calculation,Vidal:2002rm},
and was later generalised to the excited state in \cite{Alba:2009th,Alcaraz:2011tn,Berganza:2011mh}.
In the NS or R sector of the spin chain, one can define an empty state $|\es,\NS\rag$ or $|\es,\rR\rag$ that is annihilated by all the modes $c_k$
\bea
&& c_k |\es,\NS\rag = 0, ~~ k \in {\rm half ~ integers}, \nn\\
&& c_k |\es,\rR\rag = 0, ~~ k \in {\rm  integers}.
\eea
Other energy eigenstates in the spin chain can be denoted by the set of the modes $c_k^\dag$ that are excited above the empty state $|\es,\NS\rag$ or $|\es,\rR\rag$. For examples, the set $K=\{ -\f12,\f12,\f32 \}$ denotes the state $c_{-1/2}^\dag c_{1/2}^\dag c_{3/2}^\dag|\es,\NS\rag$, and the set $K=\{ -1,0 \}$ denotes the state $c_{-1}^\dag c_{0}^\dag|\es,\rR\rag$. From the complex modes $a_l$, $a_l^\dag$, one can define the Majorana modes
\be
d_{2l-1} = a_l + a_l^\dag, ~~ d_{2l} = \ii ( a_l - a_l^\dag ).
\ee
These Majorana modes $d_m$, $m=1,2,\cdots,2\ell$ are Hermitian $d_m^\dag=d_m$ and satisfy the algebra
\be
\{ d_m, d_{m'} \} = 2 \d_{mm'}.
\ee
For an interval with $\ell$ sites on the spin chain in a state $K$, one defines the correlation matrix
\be \label{GK}
\lag d_m d_{m'} \rag_K = \d_{mm'} + \G^K_{m m'},
\ee
with the $2\ell\times2\ell$ matrix  written as
\be
\G^K = \lt(\ba{cccc}
\G^K_0        & \G^K_1        & \cdots & \G^K_{\ell-1} \\
\G^K_{-1}     & \G^K_0        & \cdots & \G^K_{\ell-2} \\
\vdots        & \vdots        & \ddots & \vdots \\
\G^K_{1-\ell} & \G^K_{2-\ell} & \cdots & \G^K_0
\ea\rt),
\qquad
 \G^K_j = \lt(\ba{cc} f^K_j & g^K_j \\ -g^K_{-j} & f_j^K \ea\rt),
\ee
and
\bea
&& f^K_j = -\f{2\ii}{L} \sum_{k \in K}\sin(j \vph_k), \nn\\
&& g^K_j = -\f{\ii}{L} \sum_{k \notin K} \ep^{\ii( j\vph_k-\th_k )}+\f{\ii}{L} \sum_{k \in K} \ep^{-\ii( j\vph_k-\th_k )}.
\eea
In terms of the $2\ell$ eigenvalues $\g^K_m$, $m=1,2,\cdots,2\ell$ of $\G_K$, the entanglement entropy of the length $\ell$ interval in state $K$ is
 \cite{chung2001density,peschel2003calculation,Vidal:2002rm},
\be
S_K(\ell) = - \sum_{m=1}^{2\ell}\f{1+\g_m^K}{2}\log\f{1+\g_m^K}{2}.
\ee
The entire $2^\ell \times 2^\ell$ RDM in the state $K$ is instead given by
\be
\r_K(\ell) = \f{1}{2^\ell} \sum_{s_1,\cdots,s_{2\ell}\in\{0,1\}}
             \lag d_{2\ell}^{s_{2\ell}} \cdots d_1^{s_1} \rag_K
             d_1^{s_1} \cdots d_{2\ell}^{s_{2\ell}}.
\ee
and the multi-point correlation functions $\lag d_{2\ell}^{s_{2\ell}} \cdots d_1^{s_1} \rag_K$ can be calculated from the correlation function matrix (\ref{GK}) by the Wick theorem.
From the RDMs of various states we calculate the trace distances and other quantities.
The size of the RDMs grows exponentially, therefore we cannot reach very large $\ell$ and in this paper, in particular, we get up to $\ell=7$.
Conversely,  $L$ can be taken arbitarily large, so that we can probe a large region of the parameter $\ell/L$. 
In particular for the trace distance, for two given RDMs $\r_A$, $\s_A$, the trace distance is computed from the definition (\ref{TDDef})
\be
\label{bruteforce}
D(\r_A,\s_A) = \f12 \sum_{i = 1}^{2^\ell} | \l_i |,
\ee
with $\l_i$ being the eigenvalues of $\r_A - \s_A$.
Similarly the $n$-distances are given by $D_n(\r_A,\s_A) = (\f12 \sum_{i = 1}^{2^\ell} | \l_i |^n)^{1/n}$.

Hence, we have that by the use of Wick theorem, the correlation matrix $\G^K$ completely determines the $2^\ell \times 2^\ell$ RDM $\r_K$.
%
For a correlation matrix $\G$, we can denote the corresponding RDM as $\r_\G$.
The algebra of the RDMs studied in \cite{Fagotti:2010yr,bb-69}, obtaining
\be \label{FC}
\r_\G \r_{\G'} = \tr(\r_\G \r_{\G'}) \r_{\G\times\G'},
\ee
where the trace of two RDMs is
\be
\tr(\r_\G \r_{\G'}) = \prod_{\l \in [\rm{spectrum}(\G\G')]/2} \f{1+\l}{2},
\ee
and one defines
\be
\G\times\G' = 1 - (1-\G')(1+\G\G')^{-1}(1-\G).
\ee
The relation (\ref{FC}) can be used recursively to calculate the trace of the product of several RDMs, and therefore the even $n$-distances.

Finally we need to identify the low-lying energy eigenstates in the spin chain with the corresponding ones in CFT.
For XX spin chain and for critical Ising spin chain, this identification has been discussed, for example, in \cite{Berganza:2011mh}.
In this paper, for simplicity, we choose $L$ to be an even integer and multiple of 4.
In the XX spin chain, we only consider states in the NS sector.
Several examples of the identification of states in the spin chain and CFT are as follows
\bea
|0\rag = \prod_{k=-\f{L}{4}+\f12}^{\f{L}{4}-\f12}c_k^\dag |\es,\NS\rag                   \rm{~in~PNS~sector~}  \!\!&\lra&\!\! |0\rag         {\rm~with~} (0,0), \nn\\
c^\dag_{\f{L}{4}+\f12} |0\rag                                                            \rm{~in~APNS~sector~} \!\!&\lra&\!\!  |V_{1,0}\rag    {\rm~with~} (1/2,0), \nn\\
c_{\f{L}{4}-\f32} c_{\f{L}{4}-\f12} |0\rag                                               \rm{~in~PNS~sector~}  \!\!&\lra&\!\!  |V_{{-2},0}\rag {\rm~with~} (2,0), \nn\\
c^\dag_{-\f{L}{4}-\f32}c^\dag_{-\f{L}{4}-\f12}|0\rag                                     \rm{~in~PNS~sector~}  \!\!&\lra&\!\!  |V_{0,2}\rag    {\rm~with~} (0,2), \nn\\
c_{-\f{L}{4}+\f12} |0\rag                                                                \rm{~in~APNS~sector~} \!\!&\lra&\!\!  |V_{0,-1}\rag   {\rm~with~} (0,1/2), \nn\\
c^\dag_{-\f{L}{4}-\f32} c^\dag_{-\f{L}{4}-\f12} c^\dag_{\f{L}{4}+\f12} |0\rag            \rm{~in~APNS~sector~} \!\!&\lra&\!\!  |V_{1,2}\rag    {\rm~with~} (1/2,2), \nn\\
c_{-\f{L}{4}+\f12}c^\dag_{\f{L}{4}+\f12}c^\dag_{\f{L}{4}+\f32} |0\rag                    \rm{~in~APNS~sector~}  \!\!&\lra&\!\!  |V_{2,-1}\rag   {\rm~with~} (2,1/2), \nn\\
c_{-\f{L}{4}+\f12}c_{-\f{L}{4}+\f32}c^\dag_{\f{L}{4}+\f12}c^\dag_{\f{L}{4}+\f32} |0\rag  \rm{~in~PNS~sector~}  \!\!&\lra&\!\!  |V_{2,-2}\rag   {\rm~with~} (2,2), \nn\\
c^\dag_{-\f{L}{4}-\f12} c_{\f{L}{4}-\f12} |0\rag                                         \rm{~in~PNS~sector~}  \!\!&\lra&\!\!  |V_{{-1},1}\rag {\rm~with~} (1/2,1/2), \nn\\
c^\dag_{-\f{L}{4}-\f12} c_{\f{L}{4}-\f32}c_{\f{L}{4}-\f12} |0\rag                        \rm{~in~APNS~sector~} \!\!&\lra&\!\!  |V_{{-2},1}\rag {\rm~with~} (2,1/2), \nn\\
c_{\f{L}{4}-\f12}c^\dag_{\f{L}{4}+\f12} |0\rag                                           \rm{~in~PNS~sector~}  \!\!&\lra&\!\!  |J\rag          {\rm~with~} (1,0), \nn\\
c^\dag_{-\f{L}{4}-\f12}c_{-\f{L}{4}+\f12} |0\rag                                         \rm{~in~PNS~sector~}  \!\!&\lra&\!\!  |\bar J\rag     {\rm~with~} (0,1), \nn\\
c^\dag_{-\f{L}{4}-\f12}c_{-\f{L}{4}+\f12} c_{\f{L}{4}-\f12}c^\dag_{\f{L}{4}+\f12} |0\rag \rm{~in~PNS~sector~}  \!\!&\lra&\!\!  |J\bar J\rag    {\rm~with~} (1,1).
\eea
where the notation ``with $(h,\bar h)$'' stands for the conformal weights on the CFT side.

In the critical Ising spin chain, we consider several states both in the NS and R sectors. They include
\bea
|\es,\NS\rag                            \rm{~in~PNS~sector~}  \!\!&\lra&\!\! |0\rag        {\rm~with~} (0,0), \nn\\
c^\dag_{\f12}|\es,\NS\rag               \rm{~in~APNS~sector~} \!\!&\lra&\!\! |\psi\rag     {\rm~with~} (1/2,0), \nn\\
c^\dag_{-\f12}|\es,\NS\rag              \rm{~in~APNS~sector~} \!\!&\lra&\!\! |\bar\psi\rag {\rm~with~} (0,1/2), \nn\\
c^\dag_{-\f12}c^\dag_{\f12}|\es,\NS\rag \rm{~in~PNS~sector~}  \!\!&\lra&\!\! |\ve\rag      {\rm~with~} (1/2,1/2), \nn\\
c^\dag_{0}|\es,\rR\rag                  \rm{~in~PR~sector~}   \!\!&\lra&\!\! |\s\rag       {\rm~with~} (1/16,1/16), \nn\\
|\es,\rR\rag                            \rm{~in~APR~sector~}  \!\!&\lra&\!\! |\m\rag       {\rm~with~} (1/16,1/16).
\eea

\section{An identity in boson theory} \label{appAID}

For a subset $\mS \subseteq \mS_0$ with $\mS_0=\{0,1,\cdots,n-1\}$, in Eq.~(\ref{FnS}) we defined the function $h_n(\mS)$.
In Ref. \cite{Berganza:2011mh}, it has been shown that
\be \label{id0}
h_n(\mS_0) = 1.
\ee
More generally, for an arbitrary subset $\mS$ of $\mS_0$ and its complement $\bar\mS=\mS_0/\mS$, we have the identity
\be \label{id1}
h_n(\mS) = h_n(\bar\mS).
\ee
This relation can be simply proved by counting the poles on both sides of the equation, as described in \cite{Berganza:2011mh,Lashkari:2015dia}.
Note that $\mS\cap\bar\mS=\emptyset$, $\mS\cup\bar\mS=\mS_0$. Since $h_n(\emptyset)=1$, the identity (\ref{id0}) is a special case of (\ref{id1}). One useful corollary of the identity (\ref{id1}) is
\be
\prod_{j_1 \in \mS,j_2 \in \bar\mS}
\f{\sin^2 \f{\pi (j_1 - j_2)}{n}}{\sin\f{\pi (j_1 - j_2 + \ell/L)}{n}\sin\f{\pi (j_1 - j_2 - \ell/L)}{n}}
= h_n(\mS)^{-2}
= h_n(\bar\mS)^{-2}.
\ee

\section{Some identities in fermion theory} \label{appREF}

For a primary excited state $|\phi\rag$ with scaling dimension $\D_\phi$ in a general 2D CFT, it is easy to get the universal result
\be
\f{\tr(\r_\phi\r_0^{n-1})}{\tr\r_0^n} = \Big( \f{\sin\f{\pi\ell}{L}}{n\sin\f{\pi\ell}{n L}} \Big)^{2\D_\phi},
\label{ratio2}
\ee
leading to the universal form of the relative entropy \cite{Nakagawa:2017fzo}
\be
S(\r_\phi\|\r_0) = - S_\phi + S_0 + 2\D_\phi \Big( 1-\f{\pi\ell}{L}\cot\f{\pi\ell}{L} \Big).
\ee
Moreover, in order to compute the relative entropies in the 2D free massless fermion theory, we need the following identities
\bea
&& \f{\tr\r_\s^n}{\tr\r_0^n} = \f{\tr\r_\mu^n}{\tr\r_0^n} = 1 , \nn\\
&& \f{\tr( \r_0\r_\s^{n-1} )}{\tr\r_0^n}
 = \f{\tr( \r_0\r_\m^{n-1} )}{\tr\r_0^n}
 = \Big( \f{\sin\f{\pi\ell}{L}}{n\sin\f{\pi\ell}{n L}} \Big)^{\f14}, \nn\\
&& \f{\tr( \r_\m\r_\s^{n-1} )}{\tr\r_0^n}
 = \f{\tr( \r_\s\r_\m^{n-1} )}{\tr\r_0^n}
 = \f{\sin\f{\pi\ell}{L}}{n\sin\f{\pi\ell}{n L}}.
\eea
which can be obtained by bosonization and follow from
\bea \label{idIII}
&& \Big(\f1n\sin\f{\pi\ell}{L}\Big)^{\f{n}{2}}\f{1}{2^n}\sum_{s_0=\pm1, \cdots, s_{2n-1}=\pm1}^{s_0 + \cdots + s_{2n-1}=0}
   \Big[
      \Big(
         \prod_{0\leq j_1 < j_2 \leq n-1} \Big| \sin\f{\pi(j_1 - j_2)}{n} \Big|^{\f{s_{2j_1}s_{2j_2}+s_{2j_1+1}s_{2j_2+1}}{2}}
      \Big) \nn\\
&& ~~~~~~~~~ \times
      \Big(
         \prod_{0\leq j_1,j_2 \leq n-1} \Big| \sin\f{\pi(j_1 - j_2+\f{\ell}{L})}{n} \Big|^{\f{s_{2j_1}s_{2j_2+1}}{2}}
      \Big)
   \Big] = 1, \nn\\
&& \Big(\f1n\sin\f{\pi\ell}{L}\Big)^{\f{n-1}{2}}\f{1}{2^{n-1}}\sum_{s_2=\pm1, \cdots, s_{2n-1}=\pm1}^{s_2 + \cdots + s_{2n-1}=0}
   \Big[
      \Big(
         \prod_{1 \leq j_1 < j_2 \leq n-1} \Big| \sin\f{\pi(j_1 - j_2)}{n} \Big|^{\f{s_{2j_1}s_{2j_2}+s_{2j_1+1}s_{2j_2+1}}{2}}
      \Big) \nn\\
&& ~~~~~~~~~ \times
      \Big(
         \prod_{1 \leq j_1,j_2 \leq n-1} \Big| \sin\f{\pi(j_1 - j_2+\f{\ell}{L})}{n} \Big|^{\f{s_{2j_1}s_{2j_2+1}}{2}}
      \Big)
   \Big] = \Big( \f{\sin\f{\pi\ell}{L}}{n\sin\f{\pi\ell}{n L}} \Big)^{\f12}, \nn\\
&& \Big(\f1n\sin\f{\pi\ell}{L}\Big)^{\f{n}{2}}\f{1}{2^n}\sum_{s_0=\pm1, \cdots, s_{2n-1}=\pm1}^{s_0 + \cdots + s_{2n-1}=0}
   \Big[
      s_0 s_1
      \Big(
         \prod_{0\leq j_1 < j_2 \leq n-1} \Big| \sin\f{\pi(j_1 - j_2)}{n} \Big|^{\f{s_{2j_1}s_{2j_2}+s_{2j_1+1}s_{2j_2+1}}{2}}
      \Big) \nn\\
&& ~~~~~~~~~ \times
      \Big(
         \prod_{0\leq j_1,j_2 \leq n-1} \Big| \sin\f{\pi(j_1 - j_2+\f{\ell}{L})}{n} \Big|^{\f{s_{2j_1}s_{2j_2+1}}{2}}
      \Big)
   \Big] = - \Big( \f{\sin\f{\pi\ell}{L}}{n\sin\f{\pi\ell}{n L}} \Big)^2.
\eea
The first of the identities in (\ref{idIII}) has been proved in \cite{Berganza:2011mh}.

\section{Some formulas for the analytic continuation} \label{appAC}

For $n=1$, Eq. \eqref{anacon} simplifies to
\be\label{D1}
\log \mD_{1}[1] =  \log(\pi) -  2\int_0^{\inf} dt \f{1}{t}
\Big(
\f{1}{1+\ep^{t x/2}}-\f{\ep^{-t}}{2}
\Big).
\ee
The two integrals above are both divergent but their sum converges.
The calculation may be simplified by a sort of dimensional regularisation of each integral as
\bea
I_1(a)&=& \int_0^{\inf} dt \f{t^a}{1+\ep^{t x/2}}= 2 \left(2^a-1\right) x^{-a-1} \zeta (a+1) \Gamma (a+1)\,, \nn\\
I_2(a)&=& \int_0^{\inf} dt \f{t^a\ep^{-t}}{2}= \f{\Gamma(1+a)}2\,,
\eea
where $\zeta(a)$ is the Riemann $\zeta$ function.
The desired integral \eqref{D1} is recovered in the limit $a\to -1$ where both $I_1(a)$ and $I_2(a)$ diverges, but their difference is finite
\be
\lim_{a\to-1}(I_1(a)-I_2(a))=-\f{1}{2} \log \f{x}\pi\,,
\ee
and hence
\be
\log \mD_{1}[1] =\log x\,.
\ee

The other limit $n\to\infty$ is more cumbersome, but can be tackled with the same logic.
The starting formula is
\be
\lim_{n\to\infty}\f{\log \mD_{n}[1]}{n} =  \log(2\pi) + \int_0^{\inf} dt
\Big(\frac{e^{-t}}{t}-\frac{2}{x t^2}+\f{2}{t\left(\ep^t-1\right)}
-\frac{2 \ep^{{t x}/{2}}}{x t^2\left(\ep^t-1\right) }
+\frac{2 \ep^{t-{t   x}/{2}}}{x t^2\left(\ep^t-1\right) }
\Big).
\ee
As before each piece can be regularised in a dimensional way. There is only one problem with the term $2 /(x t^2)$ that cannot be regularised.
Anyhow, such a term cannot have a finite contribution, and so it would be sufficient to take the sum of the finite contribution of the other four integrals.
Proceeding in this way, after long but simple algebra, we arrive to the very compact form
\be
\lim_{n\to\infty}\f{\log \mD_{n}[1]}{n} =\frac{2 \left(\zeta'\left(-1,1-\frac{x}{2}\right)-\zeta'\left(-1,\frac{x}{2}\right)\right)}{x},
\label{Dinf}
\ee
where $\zeta'(z,y)\equiv \partial_z\zeta(z,y)$ denotes the derivative of the generalised $\zeta$ function with respect to the first argument.
It is possibile that such expression can be further simplified, but for our goals it is enough to write it as above.



\begin{thebibliography}{99}

\bibitem{Amico:2007ag}
L.~Amico, R.~Fazio, A.~Osterloh and V.~Vedral, \textit{{Entanglement in
  many-body systems}},
  \href{http://dx.doi.org/10.1103/RevModPhys.80.517}{Rev. Mod. Phys. {\bf 80} (2008) 517},
  [\href{https://arxiv.org/abs/quant-ph/0703044}{{\ttfamily quant-ph/0703044}}].

\bibitem{calabrese2009entanglement}
P.~Calabrese, J.~Cardy and B.~Doyon, \textit{{Entanglement entropy in extended
  quantum systems}},
  \href{http://dx.doi.org/10.1088/1751-8121/42/50/500301}{J. Phys. A  {\bf 42} (2009) 500301}.

\bibitem{Laflorencie:2015eck}
N.~Laflorencie, \textit{{Quantum entanglement in condensed matter systems}},
  \href{http://dx.doi.org/10.1016/j.physrep.2016.06.008}{Phys. Rep.  {\bf 646} (2016) 1},
  [\href{https://arxiv.org/abs/1512.03388}{{\ttfamily 1512.03388}}].

\bibitem{islam2015measuring}
R.~Islam, R.~Ma, P.~M. Preiss, M.~E. Tai, A.~Lukin, M.~Rispoli, and M. Greiner,
  \textit{Measuring entanglement entropy in a quantum many-body system},
  \href{http://dx.doi.org/10.1038/nature15750}{Nature {\bf 528} (2015) 77},
  [\href{https://arxiv.org/abs/1509.01160}{{\ttfamily 1509.01160}}].

\bibitem{kaufman2016quantum}
A.~M. Kaufman, M.~E. Tai, A.~Lukin, M.~Rispoli, R.~Schittko, P.~M. Preiss,  and M. Greiner,
\textit{Quantum thermalization through entanglement in an isolated  many-body system},
  \href{http://dx.doi.org/10.1126/science.aaf6725}{Science {\bf 353} (2016) 794},
  [\href{https://arxiv.org/abs/1603.04409}{{\ttfamily 1603.04409}}].

\bibitem{elben2018renyi}
A.~Elben, B.~Vermersch, M.~Dalmonte, J.~I. Cirac and P.~Zoller,
  \textit{R{\'e}nyi entropies from random quenches in atomic Hubbard and spin
  models},
  \href{http://dx.doi.org/10.1103/PhysRevLett.120.050406}{Phys. Rev. Lett. {\bf 120} (2018) 050406},
  [\href{https://arxiv.org/abs/1709.05060}{{\ttfamily 1709.05060}}].

\bibitem{lukin2018probing}
A.~Lukin, M.~Rispoli, R.~Schittko, M.~E. Tai, A.~M. Kaufman, S.~Choi, V. Khemani, J. Leonard, and M. Greiner,
  \textit{Probing entanglement in a many-body-localized system},
 \href{http://dx.doi.org/10.1126/science.aau0818}{Science {\bf 364}, 256 (2019)},
  [\href{https://arxiv.org/abs/1805.09819}{{\ttfamily 1805.09819}}].

\bibitem{brydges2018probing}
T.~Brydges, A.~Elben, P.~Jurcevic, B.~Vermersch, C.~Maier, B.~P. Lanyon,  P. Zoller, R. Blatt, and C. F. Roos,
  \textit{Probing entanglement entropy via randomized measurements},
 \href{http://dx.doi.org/10.1126/science.aau4963}{Science {\bf 364}, 260 (2019)},
  [\href{https://arxiv.org/abs/1806.05747}{{\ttfamily 1806.05747}}].


\bibitem{hlw-94}
C. Holzhey, F. Larsen, and F. Wilczek, {\it Geometric and renormalized entropy in conformal field theory},
\href{http://dx.doi.org/10.1016/0550-3213(94)90402-2}{Nucl. Phys. B {\bf 424}, 443 (1994)},
[\href{https://arxiv.org/abs/hep-th/9403108}{{\ttfamily hep-th/9403108}}].

\bibitem{Vidal:2002rm}
G.~Vidal, J.~I. Latorre, E.~Rico and A.~Kitaev, \textit{{Entanglement in  quantum critical phenomena}},
  \href{http://dx.doi.org/10.1103/PhysRevLett.90.227902}{Phys. Rev.
  Lett. {\bf 90} (2003) 227902},
  [\href{https://arxiv.org/abs/quant-ph/0211074}{{\ttfamily quant-ph/0211074}}];\\
J.~I. Latorre, E.~Rico and G.~Vidal, \textit{{Ground state entanglement in quantum spin chains}},
  \href{http://dx.doi.org/10.26421/QIC4.1}{Quant. Inf. Comput.
  {\bf 4} (2004) 48--92},
  [\href{https://arxiv.org/abs/quant-ph/0304098}{{\ttfamily quant-ph/0304098}}].


\bibitem{Calabrese:2004eu}
P.~Calabrese and J.~L. Cardy, \textit{{Entanglement entropy and quantum field  theory}},
  \href{http://dx.doi.org/10.1088/1742-5468/2004/06/P06002}{J. Stat. Mech. (2004) P06002},
  [\href{https://arxiv.org/abs/hep-th/0405152}{{\ttfamily hep-th/0405152}}].

\bibitem{Calabrese:2009qy}
P.~Calabrese and J.~Cardy, \textit{{Entanglement entropy and conformal field theory}},
\href{http://dx.doi.org/10.1088/1751-8113/42/50/504005}{J.  Phys. A {\bf 42} (2009) 504005},
  [\href{https://arxiv.org/abs/0905.4013}{{\ttfamily 0905.4013}}].

\bibitem{rm-04}  G. Refael and J.E. Moore, {\it Entanglement Entropy of Random Quantum Critical Points in One Dimension}
\href{https://doi.org/10.1103/PhysRevLett.93.260602}{Phys. Rev. Lett. {\bf 93},  260602 (2004)},
 [\href{https://arxiv.org/abs/cond-mat/04067373}{{\ttfamily cond-mat/04067373}}];\\
 G. Refael and J.E. Moore, {\it Criticality and entanglement in random quantum systems},
\href{https://doi.org/10.1088/1751-8113/42/50/504010}{J. of Phys. A {\bf 42}, 504010 (2009)}
 [\href{https://arxiv.org/abs/0908.1986}{{\ttfamily 0908.1986}}];\\
 N. Laflorencie, {\it Scaling of entanglement entropy in the random singlet phase},
\href{https://link.aps.org/doi/10.1103/PhysRevB.72.140408}{Phys. Rev. B {\bf 72}, 140408 (2005)},
  [\href{https://arxiv.org/abs/cond-mat/0504446}{{\ttfamily cond-mat/0504446}}];\\
 M. Fagotti, P. Calabrese, and J.E. Moore, {\it Entanglement spectrum of random-singlet quantum critical points},
 \href{https://link.aps.org/doi/10.1103/PhysRevB.83.045110}{Phys. Rev. B {\bf 83}, 045110 (2011)},
  [\href{https://arxiv.org/abs/1009.1614}{{\ttfamily 1009.1614}}].



\bibitem{top-ent}
A. Kitaev and J. Preskill, {\it Topological Entanglement Entropy},
\href{https://doi.org/10.1103/PhysRevLett.96.110404}{Phys. Rev. Lett. {\bf 96}, 110404 (2006)},
\href{http://arxiv.org/abs/hep-th/0510092}{\ttfamily [hep-th/0510092]}; \\
M. Levin and X.-G. Wen, {\it Detecting Topological Order in a Ground State Wave Function},
\href{https://doi.org/10.1103/PhysRevLett.96.110405}{Phys. Rev. Lett. {\bf 96}, 110405 (2006)},
\href{http://arxiv.org/abs/cond-mat/0510613}{\ttfamily [cond-mat/0510613]}.

\bibitem{hzs-07}
M. Haque, O. Zozulya, and K. Schoutens, {\it Entanglement entropy in fermionic Laughlin states},
\href{http://dx.doi.org/10.1103/PhysRevLett.98.060401}{Phys. Rev. Lett. {\bf 98}, 060401 (2007)},
\href{http://arxiv.org/abs/cond-mat/0609263}{\ttfamily [cond-mat/0609263]}.

\bibitem{lh-08}
H. Li and F. D. M. Haldane,
{\it Entanglement Spectrum as a Generalization of Entanglement Entropy: Identification of Topological Order in Non-Abelian Fractional Quantum Hall Effect States},
\href{http://dx.doi.org/10.1103/PhysRevLett.101.010504}{Phys. Rev. Lett. {\bf 101},  010504  (2008)},
\href{http://arxiv.org/abs/0805.0332}{\ttfamily [0805.0332]}.

\bibitem{tns}
J. I. Cirac and F. Verstraete,  {\it Renormalization and tensor product states in spin chains and lattices},
\href{http://dx.doi.org/10.1088/1751-8113/42/50/504004}{J. Phys. A {\bf 42}, 504004 (2009)},
\href{http://arxiv.org/abs/0910.1130}{\ttfamily [0910.1130]};\\
U. Schollw\"ock, {\it The density-matrix renormalization group in the age of matrix product states},
\href{http://dx.doi.org/10.1016/j.aop.2010.09.012}{Ann. Phys.  {\bf 326}, 96 (2011)},
\href{http://arxiv.org/abs/1008.3477}{\ttfamily [1008.3477]};\\
R. Orus, {\it A Practical Introduction to Tensor Networks: Matrix Product States and Projected Entangled Pair States},
\href{http://dx.doi.org/10.1016/j.aop.2014.06.013}{Ann. Phys. {\bf 349}, 117 (2014)},
\href{http://arxiv.org/abs/1306.2164}{\ttfamily [1306.2164]}.


\bibitem{cc-05}
P. Calabrese and J. Cardy,
 {\it Evolution of entanglement entropy in one-dimensional systems},
 \href{http://dx.doi.org/10.1088/1742-5468/2005/04/P04010}{J. Stat. Mech. P04010 (2005)},
  \href{https://arxiv.org/abs/cond-mat/0503393}{{\ttfamily [cond-mat/0503393]}}.

\bibitem{cramer2008exact}
M.~Cramer, C.~M. Dawson, J.~Eisert ,and T.~J. Osborne, {\it Exact relaxation  in a class of nonequilibrium quantum lattice systems},
  \href{http://dx.doi.org/10.1103/PhysRevLett.100.030602}{Phys. Rev. Lett. {\bf 100} (2008) 030602},
 \href{https://arxiv.org/abs/cond-mat/0703314}{{\ttfamily [cond-mat/0703314]}}.

\bibitem{barthel2008dephasing}
T.~Barthel and U.~Schollw{\"o}ck, \textit{Dephasing and the steady state in
  quantum many-particle systems},
  \href{http://dx.doi.org/10.1103/PhysRevLett.100.100601}{Phys. Rev. Lett. {\bf 100} (2008) 100601},
  [\href{https://arxiv.org/abs/0711.4896}{{\ttfamily 0711.4896}}].

\bibitem{deutsch2013microscopic}
J.~Deutsch, H.~Li, and A.~Sharma, \textit{Microscopic origin of thermodynamic  entropy in isolated systems},
  \href{http://dx.doi.org/10.1103/PhysRevE.87.042135}{Phys. Rev. E  {\bf 87} (2013) 042135},
  [\href{https://arxiv.org/abs/1202.2403}{{\ttfamily 1202.2403}}].

\bibitem{Gogolin:2016hwy}
C.~Gogolin and J.~Eisert, \textit{{Equilibration, thermalisation, and the
  emergence of statistical mechanics in closed quantum systems}},
  \href{http://dx.doi.org/10.1088/0034-4885/79/5/056001}{Rep. Prog. Phys. {\bf 79} (2016) 056001},
  [\href{https://arxiv.org/abs/1503.07538}{{\ttfamily 1503.07538}}].

\bibitem{calabrese2016introduction}
P.~Calabrese, F.~H. Essler and G.~Mussardo, \textit{Introduction to `quantum
  integrability in out of equilibrium systems'},
  \href{http://dx.doi.org/10.1088/1742-5468/2016/06/064001}{J. Stat. Mech. (2016) 064001}.

\bibitem{Essler:2016ufo}
F.~H.~L. Essler and M.~Fagotti, \textit{{Quench dynamics and relaxation in
  isolated integrable quantum spin chains}},
  \href{http://dx.doi.org/10.1088/1742-5468/2016/06/064002}{J. Stat.  Mech. (2016) 064002},
  [\href{https://arxiv.org/abs/1603.06452}{{\ttfamily 1603.06452}}].

\bibitem{vidmar2016generalized}
L.~Vidmar and M.~Rigol, \textit{Generalized gibbs ensemble in integrable
  lattice models},
  \href{http://dx.doi.org/10.1088/1742-5468/2016/06/064007}{J. Stat.
  Mech. (2016) 064007},
  [\href{https://arxiv.org/abs/1604.03990}{{\ttfamily 1604.03990}}].

\bibitem{alba2017entanglement}
V.~Alba and P.~Calabrese, \textit{Entanglement and thermodynamics after a
  quantum quench in integrable systems},
  \href{http://dx.doi.org/10.1073/pnas.1703516114}{PNAS {\bf 114} (2017) 7947},
  [\href{https://arxiv.org/abs/1608.00614}{{\ttfamily 1608.00614}}];\\
V.~Alba and P.~Calabrese, \textit{{Entanglement dynamics after quantum quenches
  in generic integrable systems}},
  \href{http://dx.doi.org/10.21468/SciPostPhys.4.3.017}{SciPost Phys.  {\bf 4} (2018) 017},
  [\href{https://arxiv.org/abs/1712.07529}{{\ttfamily 1712.07529}}].

%
%

\bibitem{rigol2008thermalization}
M.~Rigol, V.~Dunjko and M.~Olshanii, \textit{Thermalization and its mechanism
  for generic isolated quantum systems},
  \href{http://dx.doi.org/10.1038/nature06838}{Nature {\bf 452}
  (2008) 854}, [\href{https://arxiv.org/abs/0708.1324}{{\ttfamily 0708.1324}}].


\bibitem{Lashkari:2016vgj}
N.~Lashkari, A.~Dymarsky and H.~Liu, \textit{{Eigenstate Thermalization Hypothesis in Conformal Field Theory}},
  \href{http://dx.doi.org/10.1088/1742-5468/aab020}{J. Stat. Mech. (2018) 033101},
  [\href{https://arxiv.org/abs/1610.00302}{{\ttfamily 1610.00302}}];\\
A.~Dymarsky, N.~Lashkari and H.~Liu, \textit{Subsystem eigenstate thermalization hypothesis},
  \href{http://dx.doi.org/10.1103/PhysRevE.97.012140}{Phys. Rev. E  {\bf 97} (2018) 012140},
  [\href{https://arxiv.org/abs/1611.08764}{{\ttfamily 1611.08764}}];\\
N.~Lashkari, A.~Dymarsky and H.~Liu, \textit{{Universality of Quantum  Information in Chaotic CFTs}},
  \href{http://dx.doi.org/10.1007/JHEP03(2018)070}{JHEP {\bf 03}
  (2018) 070}, [\href{https://arxiv.org/abs/1710.10458}{{\ttfamily
  1710.10458}}].

\bibitem{Hawking:1974sw}
S.~W. Hawking, \textit{{Particle creation by black holes}},
  \href{http://dx.doi.org/10.1007/BF02345020}{Commun. Math. Phys. {\bf 43} (1975) 199};\\
S.~W. Hawking, \textit{{Breakdown of Predictability in Gravitational
  Collapse}}, \href{http://dx.doi.org/10.1103/PhysRevD.14.2460}{Phys. Rev. D {\bf 14} (1976) 2460}.

\bibitem{Mathur:2009hf}
S.~D. Mathur, \textit{{The information paradox: A Pedagogical introduction}},
  \href{http://dx.doi.org/10.1088/0264-9381/26/22/224001}{Class. Quant.
  Grav. {\bf 26} (2009) 224001},
  [\href{https://arxiv.org/abs/0909.1038}{{\ttfamily 0909.1038}}].


\bibitem{Maldacena:2013xja}
J.~Maldacena and L.~Susskind, \textit{{Cool horizons for entangled black
  holes}}, \href{http://dx.doi.org/10.1002/prop.201300020}{Fortsch.
  Phys. {\bf 61} (2013) 781},
  [\href{https://arxiv.org/abs/1306.0533}{{\ttfamily 1306.0533}}].

\bibitem{Maldacena:1997re}
J.~M. Maldacena, \textit{{The Large N limit of superconformal field theories
  and supergravity}},
  \href{http://dx.doi.org/10.4310/ATMP.1998.v2.n2.a1}{Adv. Theor. Math.
  Phys. {\bf 2} (1998) 231},
  [\href{https://arxiv.org/abs/hep-th/9711200}{{\ttfamily hep-th/9711200}}].

%
%

\bibitem{Fitzpatrick:2016ive}
A.~L. Fitzpatrick, J.~Kaplan, D.~Li, and J.~Wang, \textit{{On information loss  in AdS$_{3}$/CFT$_{2}$}},
  \href{http://dx.doi.org/10.1007/JHEP05(2016)109}{JHEP {\bf 05}
  (2016) 109}, [\href{https://arxiv.org/abs/1603.08925}{{\ttfamily
  1603.08925}}].

\bibitem{Chen:2017yze}
H.~Chen, C.~Hussong, J.~Kaplan, and D.~Li, \textit{{A Numerical Approach to Virasoro Blocks and the Information Paradox}},
  \href{http://dx.doi.org/10.1007/JHEP09(2017)102}{JHEP {\bf 09}
  (2017) 102}, [\href{https://arxiv.org/abs/1703.09727}{{\ttfamily 1703.09727}}].

\bibitem{Ryu:2006bv}
S.~Ryu and T.~Takayanagi, \textit{{Holographic derivation of entanglement
  entropy from AdS/CFT}},
  \href{http://dx.doi.org/10.1103/PhysRevLett.96.181602}{Phys. Rev. Lett. {\bf 96} (2006) 181602},
  [\href{https://arxiv.org/abs/hep-th/0603001}{{\ttfamily hep-th/0603001}}];\\
S.~Ryu and T.~Takayanagi, \textit{{Aspects of Holographic Entanglement
  Entropy}},
  \href{http://dx.doi.org/10.1088/1126-6708/2006/08/045}{JHEP  {\bf 08} (2006) 045},
  [\href{https://arxiv.org/abs/hep-th/0605073}{{\ttfamily hep-th/0605073}}];\\
T.~Nishioka, S.~Ryu and T.~Takayanagi, \textit{{Holographic Entanglement  Entropy: An Overview}},
  \href{http://dx.doi.org/10.1088/1751-8113/42/50/504008}{J. Phys. A {\bf 42} (2009) 504008},
  [\href{https://arxiv.org/abs/0905.0932}{{\ttfamily 0905.0932}}].

\bibitem{Hubeny:2007xt}
V.~E. Hubeny, M.~Rangamani and T.~Takayanagi, \textit{{A Covariant holographic
  entanglement entropy proposal}},
  \href{http://dx.doi.org/10.1088/1126-6708/2007/07/062}{JHEP  {\bf 07} (2007) 062},
  [\href{https://arxiv.org/abs/0705.0016}{{\ttfamily 0705.0016}}].


\bibitem{VanRaamsdonk:2010pw}
M.~Van~Raamsdonk, \textit{{Building up spacetime with quantum entanglement}},
  \href{http://dx.doi.org/10.1007/s10714-010-1034-0}{Gen. Rel. Grav.
  {\bf 42} (2010) 2323},
  [\href{https://arxiv.org/abs/1005.3035}{{\ttfamily 1005.3035}}].

\bibitem{Lewkowycz:2013nqa}
A.~Lewkowycz and J.~Maldacena, \textit{{Generalized gravitational entropy}},
  \href{http://dx.doi.org/10.1007/JHEP08(2013)090}{JHEP {\bf 08}
  (2013) 090}, [\href{https://arxiv.org/abs/1304.4926}{{\ttfamily 1304.4926}}].

\bibitem{Faulkner:2013ana}
T.~Faulkner, A.~Lewkowycz and J.~Maldacena, \textit{{Quantum corrections to
  holographic entanglement entropy}},
  \href{http://dx.doi.org/10.1007/JHEP11(2013)074}{JHEP {\bf 11}
  (2013) 074}, [\href{https://arxiv.org/abs/1307.2892}{{\ttfamily 1307.2892}}].

\bibitem{Faulkner:2013ica}
T.~Faulkner, M.~Guica, T.~Hartman, R.~C. Myers and M.~Van~Raamsdonk,
  \textit{{Gravitation from Entanglement in Holographic CFTs}},
  \href{http://dx.doi.org/10.1007/JHEP03(2014)051}{JHEP {\bf 03}
  (2014) 051}, [\href{https://arxiv.org/abs/1312.7856}{{\ttfamily 1312.7856}}].

\bibitem{Dong:2016fnf}
X.~Dong, \textit{{The gravity dual of R\'enyi entropy}},
  \href{http://dx.doi.org/10.1038/ncomms12472}{Nature Commun.
  {\bf 7} (2016) 12472},
  [\href{https://arxiv.org/abs/1601.06788}{{\ttfamily 1601.06788}}].

\bibitem{Dong:2016hjy}
X.~Dong, A.~Lewkowycz and M.~Rangamani, \textit{{Deriving covariant holographic
  entanglement}},
  \href{http://dx.doi.org/10.1007/JHEP11(2016)028}{JHEP {\bf 11}
  (2016) 028}, [\href{https://arxiv.org/abs/1607.07506}{{\ttfamily
  1607.07506}}].

\bibitem{Rangamani:2016dms}
M.~Rangamani and T.~Takayanagi, \textit{{Holographic Entanglement Entropy}},
  \href{http://dx.doi.org/10.1007/978-3-319-52573-0}{Lect. Notes Phys.  {\bf 931} (2017) 1},
  [\href{https://arxiv.org/abs/1609.01287}{{\ttfamily 1609.01287}}].

\bibitem{nielsen2010quantum}
M.~A. Nielsen and I.~L. Chuang, \textit{{Quantum computation and quantum
  information}}.
\newblock Cambridge University Press, Cambridge, UK, 10th anniversary~ed.,
  2010,
  \href{http://dx.doi.org/10.1017/CBO9780511976667}{10.1017/CBO9780511976667}.

\bibitem{watrous2018theory}
J.~Watrous, \textit{{The theory of quantum information}}.
\newblock Cambridge University Press, Cambridge, UK, 2018,
  \href{http://dx.doi.org/10.1017/9781316848142}{10.1017/9781316848142}.

\bibitem{fagotti2013reduced}
M.~Fagotti and F.~H. Essler, \textit{Reduced density matrix after a quantum
  quench}, \href{http://dx.doi.org/10.1103/PhysRevB.87.245107}{Phys.
  Rev. B {\bf 87} (2013) 245107},
  [\href{https://arxiv.org/abs/1302.6944}{{\ttfamily 1302.6944}}].

\bibitem{Blanco:2013joa}
D.~D. Blanco, H.~Casini, L.-Y. Hung and R.~C. Myers, \textit{{Relative Entropy
  and Holography}},
  \href{http://dx.doi.org/10.1007/JHEP08(2013)060}{JHEP {\bf 08}
  (2013) 060}, [\href{https://arxiv.org/abs/1305.3182}{{\ttfamily 1305.3182}}].



\bibitem{Balasubramanian:2014bfa}
V.~Balasubramanian, J.~J. Heckman and A.~Maloney, \textit{{Relative Entropy and
  Proximity of Quantum Field Theories}},
  \href{http://dx.doi.org/10.1007/JHEP05(2015)104}{JHEP {\bf 05}
  (2015) 104}, [\href{https://arxiv.org/abs/1410.6809}{{\ttfamily 1410.6809}}].

\bibitem{Lashkari:2014yva}
N.~Lashkari, \textit{{Relative Entropies in Conformal Field Theory}},
  \href{http://dx.doi.org/10.1103/PhysRevLett.113.051602}{Phys. Rev.
  Lett. {\bf 113} (2014) 051602},
  [\href{https://arxiv.org/abs/1404.3216}{{\ttfamily 1404.3216}}].

\bibitem{Lashkari:2015dia}
N.~Lashkari, \textit{{Modular Hamiltonian for Excited States in Conformal Field
  Theory}},
  \href{http://dx.doi.org/10.1103/PhysRevLett.117.041601}{Phys. Rev.
  Lett. {\bf 117} (2016) 041601},
  [\href{https://arxiv.org/abs/1508.03506}{{\ttfamily 1508.03506}}].



\bibitem{Sarosi:2016oks}
G.~S\'arosi and T.~Ugajin, \textit{{Relative entropy of excited states in two
  dimensional conformal field theories}},
  \href{http://dx.doi.org/10.1007/JHEP07(2016)114}{JHEP {\bf 07}
  (2016) 114}, [\href{https://arxiv.org/abs/1603.03057}{{\ttfamily
  1603.03057}}].



\bibitem{Sarosi:2016atx}
G.~S\'arosi and T.~Ugajin, \textit{{Relative entropy of excited states in
  conformal field theories of arbitrary dimensions}},
  \href{http://dx.doi.org/10.1007/JHEP02(2017)060}{JHEP {\bf 02}
  (2017) 060}, [\href{https://arxiv.org/abs/1611.02959}{{\ttfamily
  1611.02959}}].

\bibitem{Ruggiero:2016khg}
P.~Ruggiero and P.~Calabrese, \textit{{Relative Entanglement Entropies in
  1+1-dimensional conformal field theories}},
  \href{http://dx.doi.org/10.1007/JHEP02(2017)039}{JHEP {\bf 02}
  (2017) 039}, [\href{https://arxiv.org/abs/1612.00659}{{\ttfamily
  1612.00659}}].

\bibitem{Jafferis:2015del}
D.~L. Jafferis, A.~Lewkowycz, J.~Maldacena and S.~J. Suh, \textit{{Relative
  entropy equals bulk relative entropy}},
  \href{http://dx.doi.org/10.1007/JHEP06(2016)004}{JHEP {\bf 06}
  (2016) 004}, [\href{https://arxiv.org/abs/1512.06431}{{\ttfamily
  1512.06431}}].

\bibitem{Casini:2016udt}
H.~Casini, E.~Teste and G.~Torroba, \textit{{Relative entropy and the RG
  flow}}, \href{http://dx.doi.org/10.1007/JHEP03(2017)089}{JHEP
  {\bf 03} (2017) 089},
  [\href{https://arxiv.org/abs/1611.00016}{{\ttfamily 1611.00016}}].

\bibitem{Nakagawa:2017fzo}
Y.~O. Nakagawa and T.~Ugajin, \textit{{Numerical calculations on the relative
  entanglement entropy in critical spin chains}},
  \href{http://dx.doi.org/10.1088/1742-5468/aa85c1}{J. Stat. Mech. (2017) 093104},
  [\href{https://arxiv.org/abs/1705.07899}{{\ttfamily 1705.07899}}].

\bibitem{Murciano:2018cfp}
S.~Murciano, P.~Ruggiero and P.~Calabrese, \textit{{Entanglement and relative
  entropies for low-lying excited states in inhomogeneous one-dimensional
  quantum systems}},
 \href{http://dx.doi.org/10.1088/1742-5468/ab00ec}{J. Stat. Mech. (2019) 034001},
  \href{https://arxiv.org/abs/1810.02287}{{\ttfamily [1810.02287]}}.

\bibitem{kullback1951}
S.~Kullback and R.~A. Leibler, \textit{On information and sufficiency},
  \href{http://dx.doi.org/10.1214/aoms/1177729694}{Ann. Math. Statist.
  {\bf 22} (1951) 79}.

\bibitem{Zhang:2019wqo}
J.~Zhang, P.~Ruggiero and P.~Calabrese, \textit{{Subsystem Trace Distance in Quantum Field Theory}},
  \href{http://dx.doi.org/10.1103/PhysRevLett.122.141602}{Phys. Rev. Lett. {\bf 122}, 141602 (2019)},
 [ \href{https://arxiv.org/abs/1901.10993}{{\ttfamily 1901.10993}}].

\bibitem{Calabrese:2012ew}
P.~Calabrese, J.~Cardy and E.~Tonni, \textit{Entanglement negativity in
  quantum field theory},
  \href{http://dx.doi.org/10.1103/PhysRevLett.109.130502}{Phys. Rev.
  Lett. {\bf 109} (2012) 130502},
  [\href{https://arxiv.org/abs/1206.3092}{{\ttfamily 1206.3092}}].

\bibitem{Calabrese:2012nk}
P.~Calabrese, J.~Cardy and E.~Tonni, \textit{{Entanglement negativity in
  extended systems: A field theoretical approach}},
  \href{http://dx.doi.org/10.1088/1742-5468/2013/02/P02008}{J. Stat.
  Mech. (2013) P02008},
  [\href{https://arxiv.org/abs/1210.5359}{{\ttfamily 1210.5359}}].

\bibitem{ctt-13}
P Calabrese, L Tagliacozzo, and E Tonni, {\it Entanglement negativity in the critical Ising chain},
 \href{http://dx.doi.org/10.1088/1742-5468/2013/05/P05002}{J. Stat. Mech. (2013) P05002}
   [\href{https://arxiv.org/abs/1302.1113}{{\ttfamily 1302.1113}}].



\bibitem{Cardy:2007mb}
J.~L. Cardy, O.~A. Castro-Alvaredo and B.~Doyon, \textit{{Form factors of branch-point twist fields in quantum integrable models and entanglement entropy}},
  \href{http://dx.doi.org/10.1007/s10955-007-9422-x}{J.  Stat. Phys. {\bf 130} (2008) 129},
  [\href{https://arxiv.org/abs/0706.3384}{{\ttfamily 0706.3384}}].



\bibitem{Headrick:2010zt}
M.~Headrick, \textit{{Entanglement R\'enyi entropies in holographic theories}},
  \href{http://dx.doi.org/10.1103/PhysRevD.82.126010}{Phys. Rev. D
  {\bf 82} (2010) 126010},
  [\href{https://arxiv.org/abs/1006.0047}{{\ttfamily 1006.0047}}].

\bibitem{Calabrese:2010he}
P.~Calabrese, J.~Cardy and E.~Tonni, \textit{{Entanglement entropy of two  disjoint intervals in conformal field theory II}},
  \href{http://dx.doi.org/10.1088/1742-5468/2011/01/P01021}{J. Stat. Mech. (2011) P01021},
  [\href{https://arxiv.org/abs/1011.5482}{{\ttfamily 1011.5482}}].

\bibitem{Rajabpour:2011pt}
M.~Rajabpour and F.~Gliozzi, \textit{{Entanglement Entropy of Two Disjoint
  Intervals from Fusion Algebra of Twist Fields}},
  \href{http://dx.doi.org/10.1088/1742-5468/2012/02/P02016}{J. Stat.
  Mech. (2012) P02016},
  [\href{https://arxiv.org/abs/1112.1225}{{\ttfamily 1112.1225}}].

\bibitem{Chen:2013kpa}
B.~Chen and J.-j. Zhang, \textit{{On short interval expansion of R\'enyi entropy}},
\href{http://dx.doi.org/10.1007/JHEP11(2013)164}{JHEP  {\bf 11} (2013) 164},
  [\href{https://arxiv.org/abs/1309.5453}{{\ttfamily 1309.5453}}].

\bibitem{Chen:2016lbu}
B.~Chen, J.-B. Wu and J.-j. Zhang, \textit{{Short interval expansion of R\'enyi  entropy on torus}},
  \href{http://dx.doi.org/10.1007/JHEP08(2016)130}{JHEP {\bf 08}
  (2016) 130}, [\href{https://arxiv.org/abs/1606.05444}{{\ttfamily
  1606.05444}}].

\bibitem{rtc-18}
P. Ruggiero, E. Tonni, and P. Calabrese, {\it Entanglement entropy of two disjoint intervals and the recursion formula for conformal blocks},
\href{http://dx.doi.org/10.1088/1742-5468/aae5a8}{J. Stat. Mech. (2018) 113101},
 [\href{https://arxiv.org/abs/1805.05975}{{\ttfamily  1805.05975}}].

\bibitem{Lin:2016dxa}
F.-L. Lin, H.~Wang and J.-j. Zhang, \textit{{Thermality and excited state
  R\'enyi entropy in two-dimensional CFT}},
  \href{http://dx.doi.org/10.1007/JHEP11(2016)116}{JHEP {\bf 11}
  (2016) 116}, [\href{https://arxiv.org/abs/1610.01362}{{\ttfamily
  1610.01362}}].

\bibitem{He:2017vyf}
S.~He, F.-L. Lin and J.-j. Zhang, \textit{{Subsystem eigenstate thermalization
  hypothesis for entanglement entropy in CFT}},
  \href{http://dx.doi.org/10.1007/JHEP08(2017)126}{JHEP {\bf 08}
  (2017) 126}, [\href{https://arxiv.org/abs/1703.08724}{{\ttfamily  1703.08724}}];\\
S.~He, F.-L. Lin and J.-j. Zhang, \textit{{Dissimilarities of reduced density
  matrices and eigenstate thermalization hypothesis}},
  \href{http://dx.doi.org/10.1007/JHEP12(2017)073}{JHEP {\bf 12}
  (2017) 073}, [\href{https://arxiv.org/abs/1708.05090}{{\ttfamily 1708.05090}}].


\bibitem{Ohmori:2014eia}
K.~Ohmori and Y.~Tachikawa, \textit{{Physics at the entangling surface}},
  \href{http://dx.doi.org/10.1088/1742-5468/2015/04/P04010}{J. Stat. Mech. (2015) P04010},
  [\href{https://arxiv.org/abs/1406.4167}{{\ttfamily 1406.4167}}].

\bibitem{Cardy:2016fqc}
J.~Cardy and E.~Tonni, \textit{{Entanglement hamiltonians in two-dimensional conformal field theory}},
  \href{http://dx.doi.org/10.1088/1742-5468/2016/12/123103}{J. Stat. Mech. (2016) 123103},
  [\href{https://arxiv.org/abs/1608.01283}{{\ttfamily 1608.01283}}].

\bibitem{act-17}
V. Alba, P. Calabrese, and E. Tonni,
Entanglement spectrum degeneracy and Cardy formula in 1+1 dimensional conformal field theories,
 \href{http://dx.doi.org/10.1088/1751-8121/aa9365}{J. Phys. A {\bf 51}, 024001 (2018)},
\href{http://arxiv.org/abs/1707.07532}{\ttfamily [1707.07532]}.

\bibitem{Alcaraz:2011tn}
F.~C. Alcaraz, M.~I. Berganza and G.~Sierra, \textit{{Entanglement of low-energy excitations in Conformal Field Theory}},
  \href{http://dx.doi.org/10.1103/PhysRevLett.106.201601}{Phys. Rev.  Lett. {\bf 106} (2011) 201601},
  [\href{https://arxiv.org/abs/1101.2881}{{\ttfamily 1101.2881}}].

\bibitem{Berganza:2011mh}
M.~I. Berganza, F.~C. Alcaraz and G.~Sierra, \textit{{Entanglement of excited  states in critical spin chains}},
  \href{http://dx.doi.org/10.1088/1742-5468/2012/01/P01016}{J. Stat.
  Mech. (2012) P01016},
  [\href{https://arxiv.org/abs/1109.5673}{{\ttfamily 1109.5673}}].







\bibitem{p-14}
T. Palmai,
{\it Excited state entanglement in one dimensional quantum critical systems: Extensivity and the role of microscopic details},
\href{http://dx.doi.org/10.1103/PhysRevB.90.161404}{Phys. Rev. B {\bf 90}, 161404 (2014)},
 [\href{https://arxiv.org/abs/1406.3182}{{\ttfamily 1406.3182}}].


\bibitem{p-16}
T. Palmai, {\it Entanglement Entropy from the Truncated Conformal Space},
\href{http://dx.doi.org/10.1016/j.physletb.2016.06.012}{Phys. Lett. {\bf B}, 445 (2016)},
 [\href{https://arxiv.org/abs/1605.00444}{{\ttfamily 1605.00444}}].


\bibitem{txas-13}
L. Taddia, J. C. Xavier, F. C. Alcaraz, and G. Sierra,
{\it Entanglement Entropies in Conformal Systems with Boundaries},
\href{http://dx.doi.org/10.1103/PhysRevB.88.075112}{Phys. Rev. B {\bf 88}, 075112 (2013)},
 [\href{https://arxiv.org/abs/1302.6222}{{\ttfamily 1302.6222}}].


\bibitem{top-16}
L. Taddia, F. Ortolani, and T. Palmai,
{\it R\'enyi entanglement entropies of descendant states in critical systems with boundaries: conformal field theory and spin chains},
\href{http://dx.doi.org/10.1088/1742-5468/2016/09/093104}{J. Stat. Mech. (2016) 093104},
 [\href{https://arxiv.org/abs/1606.02667}{{\ttfamily 1606.02667}}].

\bibitem{rrs-14}
G. Ramirez, J. Rodriguez-Laguna, and G. Sierra, {\it Entanglement in low-energy states of the random-hopping model},
\href{http://dx.doi.org/10.1088/1742-5468/2016/07/P07003}{J. Stat. Mech. (2014) P07003},
 [\href{https://arxiv.org/abs/1402.5015}{{\ttfamily 1402.5015}}].






\bibitem{k-91}
J. Kurchan, {\it Replica trick to calculate means of absolute values: applications to stochastic equations},
\href{https://doi.org/10.1088/0305-4470/24/21/011}{J. Phys. A {\bf 24}, 4969 (1991)}.

\bibitem{g-vari}
D. M. Gangardt and A. Kamenev, {\it Replica Treatment of the Calogero Sutherland Model},
\href{https://doi.org/10.1016/S0550-3213(01)00326-1}{Nucl. Phys. B {\bf 610}, 578 (2001)}
 [\href{https://arxiv.org/abs/cond-mat/0102405}{{\ttfamily cond-mat/0102405}}]; \\
S. M. Nishigaki, D. M. Gangardt, and A. Kamenev, {\it Correlation functions of the BC Calogero Sutherland model},
\href{https://doi.org/10.1088/0305-4470/36/12/316}{J. Phys. A {\bf 36}, 3137 (2003)}
 [\href{https://arxiv.org/abs/cond-mat/0207301}{{\ttfamily cond-mat/0207301}}];\\
D.~M.~Gangardt, {\it Universal correlations of trapped one-dimensional impenetrable bosons},
\href{https://doi.org/10.1088/0305-4470/37/40/002}{J. Phys. A {\bf 37}, 9335 (2004)}
 [\href{https://arxiv.org/abs/cond-mat/0404104}{{\ttfamily cond-mat/0404104}}].



\bibitem{cs-08}
P.~Calabrese and R.~Santachiara,
{\it Off-diagonal correlations in one-dimensional anyonic models: a replica approach},
 \href{http://dx.doi.org/10.1088/1742-5468/2009/03/P03002}{J. Stat. Mech. (2009) P03002},
 [\href{https://arxiv.org/abs/0811.2991}{{\ttfamily 0811.2991}}];\\
G. Marmorini, M. Pepe, and P. Calabrese,
{\it One-body reduced density matrix of trapped impenetrable anyons in one dimension},
 \href{http://dx.doi.org/10.1088/1742-5468/2016/07/073106}{J. Stat. Mech. (2016) 073106},
[\href{https://arxiv.org/abs/1605.00838}{{\ttfamily 1605.00838}}].

\bibitem{dei-18}
T. Dupic, B. Estienne, and Y. Ikhlef, {\it Entanglement entropies of minimal models from null vectors},
\href{https://scipost.org/10.21468/SciPostPhys.4.6.031}{SciPost Phys. { \bf 4}, 031 (2018)},
 [\href{https://arxiv.org/abs/1709.09270}{{\ttfamily 1709.09270}}].


\bibitem{elc-13}
F. H. L. Essler, A. M. L\"auchli, and P. Calabrese, {\it Shell-Filling Effect in the Entanglement Entropies of Spinful Fermions},
\href{http://dx.doi.org/10.1103/PhysRevLett.110.115701}{Phys. Rev. Lett. {\bf 110}, 115701 (2013)},
 [\href{https://arxiv.org/abs/1211.2474}{{\ttfamily 1211.2474}}]; \\
%
P. Calabrese, F. H. L. Essler, and A. L\"auchli, \emph{Entanglement entropies of the quarter filled Hubbard model},
\href{http://dx.doi.org/10.1088/1742-5468/2014/09/P09025}{J. Stat. Mech. (2014) P09025},
[\href{https://arxiv.org/abs/1406.7477}{{\ttfamily 1406.7477}}].






\bibitem{DiFrancesco:1997nk}
P.~Di~Francesco, P.~Mathieu and D.~S\'en\'echal, \textit{{Conformal Field
  Theory}}.
\newblock Springer, New York, USA, 1997,
  \href{http://dx.doi.org/10.1007/978-1-4612-2256-9}{10.1007/978-1-4612-2256-9}.

\bibitem{Blumenhagen:2009zz}
R.~Blumenhagen and E.~Plauschinn, \textit{{Introduction to conformal field
  theory}}, \href{http://dx.doi.org/10.1007/978-3-642-00450-6}{Lect. Notes Phys. {\bf 779} (2009) 1}.

\bibitem{no-anomaly}
  Z.~Li and J.~j.~Zhang,
  {\it On one-loop entanglement entropy of two short intervals from OPE of twist operators},
  \href{http://dx.doi.org/10.1007/JHEP05(2016)130}{JHEP {\bf 05}, 130 (2016)}
[\href{https://arxiv.org/abs/1604.02779}{1604.02779}]; \\
  W.~Z.~Guo, F.~L.~Lin and J.~Zhang,
  {\it Note on ETH of descendant states in 2D CFT},
   \href{http://dx.doi.org/10.1007/JHEP01(2019)152}{JHEP {\bf 01}, 152 (2019)}
 [\href{https://arxiv.org/abs/1810.01258}{1810.01258}].




\bibitem{chung2001density}
M.-C. Chung and I.~Peschel, \textit{Density-matrix spectra of solvable  fermionic systems},
  \href{http://dx.doi.org/10.1103/PhysRevB.64.064412}{Phys. Rev. B {\bf 64} (2001) 064412},
  [\href{https://arxiv.org/abs/cond-mat/0103301}{{\ttfamily  cond-mat/0103301}}].

\bibitem{peschel2003calculation}
I.~Peschel, \textit{{Calculation of reduced density matrices from correlation functions}},
  \href{http://dx.doi.org/{10.1088/0305-4470/36/14/101}}{J. Phys. A {\bf 36} (2003) L205},
  [\href{https://arxiv.org/abs/cond-mat/0212631}{{\ttfamily cond-mat/0212631}}];\\
I. Peschel and V. Eisler, {\it Reduced density matrices and entanglement entropy in free lattice models},
\href{http://dx.doi.org/10.1088/1751-8113/42/50/504003}{J. Phys. A {\bf 42}, 504003 (2009)},
 [\href{https://arxiv.org/abs/0906.1663}{{\ttfamily 0906.1663}}];
;\\
I. Peschel, {\it Entanglement in solvable many-particle models},
\href{http://dx.doi.org/10.1007/s13538-012-0074-1}{Braz. J. Phys. 42, 267 (2012)},
 [\href{https://arxiv.org/abs/1109.0159}{{\ttfamily 1109.0159}}].

\bibitem{Alba:2009th}
V.~Alba, M.~Fagotti and P.~Calabrese, \textit{{Entanglement entropy of excited states}},
  \href{http://dx.doi.org/10.1088/1742-5468/2009/10/P10020}{J. Stat.  Mech. (2009) P10020},
  [\href{https://arxiv.org/abs/0909.1999}{{\ttfamily 0909.1999}}].



\bibitem{Fagotti:2010yr}
M.~Fagotti and P.~Calabrese, \textit{{Entanglement entropy of two disjoint
  blocks in XY chains}},
  \href{http://dx.doi.org/10.1088/1742-5468/2010/04/P04016}{J. Stat. Mech. (2010) P04016},
  [\href{https://arxiv.org/abs/1003.1110}{{\ttfamily 1003.1110}}].


\bibitem{Li:2016qbo}
Z.~Li and J.-j. Zhang, \textit{{Holographic R\'enyi entropy for two-dimensional
  $\mathcal{N}=(2,2)$ superconformal field theory}},
  \href{http://dx.doi.org/10.1103/PhysRevD.95.126009}{Phys. Rev. D {\bf 95} (2017) 126009},
  [\href{https://arxiv.org/abs/1611.00546}{{\ttfamily 1611.00546}}].

\bibitem{ccen-10}
P. Calabrese, M. Campostrini, F. Essler and B. Nienhuis,
{\it Parity effect in the scaling of block entanglement in gapless spin chains},
\href{http://dx.doi.org/10.1103/PhysRevLett.104.095701}{Phys. Rev. Lett {\bf 104}, 095701 (2010)},
  [\href{https://arxiv.org/abs/0911.4660}{{\ttfamily 0911.4660}}].


\bibitem{ce-10} P. Calabrese and F. H. L. Essler, {\it Universal corrections to scaling for block entanglement in spin-1/2 XX chains},
\href{http://dx.doi.org/10.1088/1742-5468/2010/08/P08029}{J. Stat. Mech. P08029 (2010)},
  [\href{https://arxiv.org/abs/1006.3420}{{\ttfamily 1006.3420}}].

\bibitem{fc-11} M. Fagotti and P. Calabrese,
{\it Universal parity effects in the entanglement entropy of XX chains with open boundary conditions},
\href{http://dx.doi.org/10.1088/1742-5468/2011/01/P01017}{J. Stat. Mech. P01017 (2011)},
 [\href{https://arxiv.org/abs/1010.5796}{{\ttfamily 1010.5796}}].

\bibitem{cmv-11}
P. Calabrese, M. Mintchev, and E. Vicari, {\it Entanglement Entropy of One-Dimensional Gases},
\href{https://doi.org/10.1103/PhysRevLett.107.020601}{Phys. Rev. Lett. {\bf 107}, 020601 (2011)},
[\href{https://arxiv.org/abs/1105.4756}{{\ttfamily 1105.4756}}];
\\
P. Calabrese, M. Mintchev, and E. Vicari,
 {\it The entanglement entropy of one-dimensional systems in continuous and homogeneous space},
\href{http://dx.doi.org/10.1088/1742-5468/2011/09/P09028}{J. Stat. Mech. P09028 (2011)},
[\href{https://arxiv.org/abs/1107.3985}{{\ttfamily 1107.3985}}].

\bibitem{cc-10}
J. Cardy and P. Calabrese,
\emph{Unusual Corrections to Scaling in Entanglement Entropy},
\href{http://dx.doi.org/10.1088/1742-5468/2010/04/P04023}{J. Stat. Mech. (2010) P04023},
[\href{https://arxiv.org/abs/1002.4353}{{\ttfamily 1002.4353}}].



\bibitem{Wilde:2014eda}
M.~M. Wilde, A.~Winter, and D.~Yang, \textit{{Strong Converse for the Classical
  Capacity of Entanglement Breaking and Hadamard Channels via a Sandwiched
  R{\'e}nyi Relative Entropy}},
  \href{http://dx.doi.org/10.1007/s00220-014-2122-x}{Commun. Math. Phys. {\bf 331} (2014) 593},
  [\href{https://arxiv.org/abs/1306.1586}{{\ttfamily 1306.1586}}].

\bibitem{muller2013quantum}
M.~M{\"u}ller-Lennert, F.~Dupuis, O.~Szehr, S.~Fehr and M.~Tomamichel,
  \textit{{On quantum R{\'e}nyi entropies: A new generalization and some
  properties}}, \href{http://dx.doi.org/10.1063/1.4838856}{J. Math. Phys. {\bf 54} (2013) 122203},
  [\href{https://arxiv.org/abs/1306.3142}{{\ttfamily 1306.3142}}].


\bibitem{ahjk-14}
C. M. Agon, M. Headrick, D. L. Jafferis, S. Kasko, {\it Disk entanglement entropy for a Maxwell field},
\href{https://doi.org/10.1103/PhysRevD.89.025018}{Phys. Rev. D {\bf 89}, 025018 (2014)},
[\href{https://arxiv.org/abs/1310.4886}{{\ttfamily 1310.4886}}].


\bibitem{dct-15}
C. De Nobili, A. Coser, and E. Tonni, {\it Entanglement entropy and negativity of disjoint intervals in CFT: Some numerical extrapolations},
\href{http://dx.doi.org/10.1088/1742-5468/2015/06/P06021}{J. Stat. Mech. (2015) P06021},
 [\href{https://arxiv.org/abs/1501.04311}{{\ttfamily 1501.04311}}].

\bibitem{mgdr-19}
T. Mendes-Santos, G. Giudici, M. Dalmonte, M. A. Rajabpour,
{\it Entanglement Hamiltonian of quantum critical chains and conformal field theories},
 [\href{https://arxiv.org/abs/1906.00471 }{{\ttfamily 1906.00471}}].

 \bibitem{u-17}
T. Ugajin, {\it Mutual information of excited states and relative entropy of two disjoint subsystems in CFT},
\href{http://dx.doi.org/10.1007/JHEP10(2017)184}{JHEP {\bf 10} (2017) 184},
\href{https://arxiv.org/abs/1611.03163}{{\ttfamily [1611.03163]}}.

\bibitem{dsvc-17} J. Dubail, J.-M. St\'ephan, J. Viti, and P. Calabrese,
{\it Conformal field theory for inhomogeneous one-dimensional quantum systems: the example of non-interacting Fermi gases},
\href{http://dx.doi.org/10.21468/SciPostPhys.2.1.002}{Scipost Phys. {\bf 2}, 002 (2017)},
\href{https://arxiv.org/abs/1606.04401}{{\ttfamily [1606.04401]}}.


\bibitem{dvz-17}
M. Dalmonte, B. Vermersch, and P. Zoller, {\it Quantum Simulation and Spectroscopy of Entanglement Hamiltonians},
\href{https://doi.org/10.1038/s41567-018-0151-7}{Nature Phys. {\bf 14}, 827 (2018)},
\href{https://arxiv.org/abs/1707.04455}{\ttfamily [1707.04455]}.

\bibitem{gmcd-18}
G. Giudici, T. Mendes-Santos, P. Calabrese, and M. Dalmonte,
{\it Entanglement Hamiltonians of lattice models via the Bisognano-Wichmann theorem},
\href{http://dx.doi.org/10.1103/PhysRevB.98.134403}{Phys. Rev. B {\bf 98}, 134403 (2018)},
 \href{https://arxiv.org/abs/1807.01322}{{\ttfamily  [1807.01322]}}.

 \bibitem{bw-76}
J. Bisognano and E. Wichmann, {\it On the duality condition for quantum fields},
\href{http://dx.doi.org/10.1063/1.522898}{J. Math. Phys. {\bf 17}, 303 (1976);}\\
J. Bisognano and E. Wichmann, {\it On the Duality Condition for a Hermitian Scalar Field},
\href{http://dx.doi.org/10.1063/1.522605}{J. Math. Phys. 16, 985 (1975);}\\
P. Hislop and R. Longo,  {\it Modular structure of the local algebras associated with the free massless scalar field theory},
\href{http://dx.doi.org/10.1007/BF01208372}{Comm. Math. Phys 84, 71 (1982).}

\bibitem{ep-17}
V. Eisler and I. Peschel, {\it Analytical results for the entanglement Hamiltonian of a free-fermion chain},
\href{http://dx.doi.org/10.1088/1751-8121/aa76b5}{J. Phys. A {\bf 50} 284003 (2017)},
 \href{https://arxiv.org/abs/1703.08126}{{\ttfamily  [1703.08126]}};\\
V. Eisler and I. Peschel, {\it Properties of the entanglement Hamiltonian for finite free-fermion chains},
\href{http://dx.doi.org/10.1088/1742-5468/aace2b}{ J. Stat. Mech. (2018) 104001},
 \href{https://arxiv.org/abs/1805.00078 }{{\ttfamily  [1805.00078]}};\\
V. Eisler, E. Tonni, and I. Peschel, {\it On the continuum limit of the entanglement Hamiltonian},
 \href{https://arxiv.org/abs/1902.04474}{{\ttfamily  [1902.04474]}}.

\bibitem{pa-18}
F. Parisen Toldin and F. F. Assaad, {\it  Entanglement Hamiltonian of Interacting Fermionic Models},
\href{https://doi.org/10.1103/PhysRevLett.121.200602}{Phys. Rev. Lett. {\bf 121}, 200602 (2018)},
\href{https://arxiv.org/abs/1804.03163}{{\ttfamily  [1804.03163]}}.



\bibitem{Cardy:2013nua}
J.~Cardy, \textit{{Some results on the mutual information of disjoint regions
  in higher dimensions}},
  \href{http://dx.doi.org/10.1088/1751-8113/46/28/285402}{J. Phys. A {\bf 46} (2013) 285402},
  [\href{https://arxiv.org/abs/1304.7985}{{\ttfamily 1304.7985}}].

\bibitem{Hung:2014npa}
L.-Y. Hung, R.~C. Myers and M.~Smolkin, \textit{{Twist operators in higher dimensions}},
\href{http://dx.doi.org/10.1007/JHEP10(2014)178}{JHEP  {\bf 10} (2014) 178},
  [\href{https://arxiv.org/abs/1407.6429}{{\ttfamily 1407.6429}}].

\bibitem{Chen:2016mya}
B.~Chen and J.~Long, \textit{{R\'enyi mutual information for a free scalar field  in even dimensions}},
  \href{http://dx.doi.org/10.1103/PhysRevD.96.045006}{Phys. Rev. D {\bf 96} (2017) 045006},
  [\href{https://arxiv.org/abs/1612.00114}{{\ttfamily 1612.00114}}].

\bibitem{Chen:2017hbk}
B.~Chen, L.~Chen, P.-x. Hao and J.~Long, \textit{{On the Mutual Information in Conformal Field Theory}},
  \href{http://dx.doi.org/10.1007/JHEP06(2017)096}{JHEP {\bf 06}
  (2017) 096}, [\href{https://arxiv.org/abs/1704.03692}{{\ttfamily 1704.03692}}].

\bibitem{Chen:2017yns}
B.~Chen, Z.-Y. Fan, W.-M. Li and C.-Y. Zhang, \textit{{Holographic Mutual  Information of Two Disjoint Spheres}},
  \href{http://dx.doi.org/10.1007/JHEP04(2018)113}{JHEP {\bf 04}
  (2018) 113}, [\href{https://arxiv.org/abs/1712.05131}{{\ttfamily 1712.05131}}].



\bibitem{Lieb:1961fr}
E.~H. Lieb, T.~Schultz and D.~Mattis, \textit{{Two soluble models of an antiferromagnetic chain}},
  \href{http://dx.doi.org/10.1016/0003-4916(61)90115-4}{Ann. Phys.  {\bf 16} (1961) 407}.

\bibitem{pfeuty1970one}
P.~Pfeuty, \textit{{The one-dimensional Ising model with a transverse field}},
  \href{http://dx.doi.org/10.1016/0003-4916(70)90270-8}{Ann. Phys.  {\bf 57} (1970) 79}.

\bibitem{calabrese2012quantum}
P.~Calabrese, F.~H. Essler and M.~Fagotti, \textit{Quantum quench in the
  transverse field ising chain: I. time evolution of order parameter correlators},
  \href{http://dx.doi.org/10.1088/1742-5468/2012/07/P07016}{J. Stat. Mech. (2012) P07016},
  [\href{https://arxiv.org/abs/1204.3911}{{\ttfamily 1204.3911}}].



\bibitem{bb-69}
R. Balian and E. Brezin, {\it Nonunitary Bogoliubov transformations and extension of Wick's theorem},
\href{https://link.springer.com/article/10.1007/BF02710281}{Il Nuovo Cimento B {\bf 64}, 37 (1969).}



\end{thebibliography}
\end{document}